\documentclass[prd,aps,floats,twocolumn]{revtex4}
\usepackage{commath}
\pdfoutput=1
\usepackage[usenames, dvipsnames]{color}
\usepackage{graphicx, array}
\usepackage{graphics}
\usepackage{graphics,epsfig}
\usepackage{amssymb}
\usepackage{amsmath}
\usepackage{ulem}
\usepackage{multirow}
\usepackage{subfigure}
\usepackage{bm}
\usepackage{txfonts}
\usepackage{cancel}
\usepackage{url}
\usepackage{hyperref}
\usepackage{ulem}
\newcolumntype{C}[1]{>{\centering\arraybackslash}m{#1}}
\def\mnras{MNRAS}

 \def\be{\begin{equation}}
\def\ee{\end{equation}}
 \def\bi{\begin{itemize}}
 \def\ei{\end{itemize}}
  \def\ben{\begin{enumerate}}
\def\een{\end{enumerate}}
  \def\bt{\begin{tabular}}
\def\et{\end{tabular}}
\def\bc{\begin{center}}
\def\ec{\end{center}}

\def\bea{\begin{eqnarray}}
\def\eea{\end{eqnarray}}

\def\ba{\begin{eqnarray}}
\def\ea{\end{eqnarray}}

\begin{document}

\input{epsf}

\title{Convolution Lagrangian perturbation theory for biased tracers beyond general relativity}
\author {Georgios Valogiannis and Rachel Bean.}
\affiliation{Department of Astronomy, Cornell University, Ithaca, New York 14853, USA.}
\label{firstpage}

\begin{abstract}

We compare analytic predictions for real and Fourier space two-point statistics for biased tracers from a variety of  Lagrangian Perturbation Theory approaches against those from state of the art N-body simulations in $f(R)$ Hu-Sawicki and the nDGP braneworld modified gravity theories.
 
We show that the novel physics of gravitational collapse in scalar tensor theories with the chameleon or the Vainshtein screening mechanism can be effectively factored in with bias parameters analytically predicted using the Peak-Background Split formalism  when updated to include  the environmental sensitivity of modified gravity theories as well as changes to the halo mass function. 
 
We demonstrate that Convolution Lagrangian Perturbation Theory (CLPT) and Standard Perturbation Theory (SPT) approaches provide accurate analytic methods to predict the correlation function and power spectra, respectively, for biased tracers in modified gravity models and are able to characterize both the BAO, power-law and small scale regimes needed for upcoming galaxy surveys such as DESI, Euclid, LSST and WFIRST.
\end{abstract}

\maketitle

\section{Introduction}
\label{sec:intro}
About 20 years after the discovery of cosmic acceleration \citep{Perlmutter:1998np, Riess:2004nr}, the $\Lambda$CDM model of the universe is rightfully called the standard model of cosmology, after having successfully passed a broad range of tight observational tests \citep{Eisenstein:2005su, Percival:2007yw, Percival:2009xn, Kazin:2014qga, Spergel:2013tha, Ade:2013zuv, Ade:2015xua}. This model augments the cosmic baryonic and leptonic matter components in the Standard Model, with two additional ingredients that dominate the cosmic energy budget, cold dark matter (CDM) and a cosmological constant, $\Lambda$, with the latter playing the role of vacuum energy that is responsible for the observed late-time acceleration. Central to the above paradigm is also the assumption that gravitational evolution is governed by Einstein's General Theory of Relativity (GR), which led to the observed inhomogeneous pattern of the Large-Scale Structure (LSS) of the Universe through the process of gravitational instability. The value of the vacuum energy predicted by quantum field theory, however, is orders of magnitude larger than the best-fit one that is necessary to explain cosmic acceleration, so $\Lambda$ needs to be fine-tuned; the so-called cosmological constant problem \citep{Weinberg:1988cp}. Such an unfortunate mismatch, together with the need to fully explore the space of all theoretical alternatives, has generated growing interest in testing gravity models that self-accelerate though a large-scale modification to GR, instead of dark energy, the so-called Modified Gravity (MG) models \citep{Koyama:2015vza, Ishak:2018his}.

Modifying the Einstein-Hilbert action, however, introduces in principle an additional degree of freedom that is conformally coupled to matter and can produce significant deviations from the predictions of GR, which have passed a wide array of precise observational tests, especially in the Solar System \citep{Will:2005va}. Furthermore, the recent simultaneous detection of gravitational waves and EM counterparts by the LIGO/Virgo collaboration \citep{TheLIGOScientific:2017qsa,Goldstein:2017mmi,Savchenko:2017ffs,Monitor:2017mdv,GBM:2017lvd}, has placed additional constraints \citep{Sakstein:2017xjx,Ezquiaga:2017ekz,Creminelli:2017sry,Baker:2017hug} into the form of the most general expression of a scalar-tensor theory that produces second order equations of motion, described by the Horndeski action \citep{Horndeski1974,PhysRevD.84.064039}. 

In order to be able to confront such tight constraints successfully, while at the same time provide a stable self-accelerative cosmic mechanism, viable candidates contain a restoring property, called ``screening" \citep{Khoury:2010xi,Khoury:2013tda}, which is a dynamical mechanism that weakens the additional fifth forces in high-density environments through the corresponding scalar field self-interactions. In the Vainshtein mechanism \citep{VAINSHTEIN1972393,Babichev:2013usa}, GR is recovered thanks to the second derivative terms in the scalar field Lagrangian, that become large in high density environments and effectively weaken the coupling to the matter sources. The Vainshtein mechanism is very efficient in the vicinity of a massive source and contains a rich phenomenology, which makes it particularly attractive. Another popular class of screening consists of the chameleons \citep{PhysRevD.69.044026,PhysRevLett.93.171104}, where in regions of high potential the scalar fields become massive and cannot propagate, resulting thus in suppression of the fifth forces. Despite the fact that chameleons cannot produce self-acceleration \citep{Wang:2012kj}, their very interesting phenomenology makes them serve as ideal testbeds for gravity and considerable efforts have been put into their study in the past decade \citep{Burrage:2017qrf}. Other screening classes include the symmetrons \citep{PhysRevLett.104.231301,Olive:2007aj}, that employ spontaneous symmetry breaking and share qualitative similarities with the chameleons and the K-Mouflage \citep{Dvali:2010jz,Babichev:2013usa}, in which deviations are suppressed when scalar field gradients exceed a certain value. 

The observed inhomogeneous LSS of the Universe, is the outcome of the subsequent nonlinear gravitational evolution of the primordial density fluctuations, partially modulated by the late-time acceleration. As a result, it provides us with an observational window into the fundamental physics that shaped this process, including sensitive tests of the underlying large-scale gravitational law, making it particularly valuable for constraining the various MG models. Indeed, as we are entering the era of ``precision cosmology", multiple spectroscopic and photometric surveys of the LSS, both already operating, such as the DES \citep{Abbott:2005b} and also about to be commissioned in the next decade, like DESI \citep{Levi:2013gra}, Euclid \citep{Laureijs:2011gra}, the LSST \citep{Abell:2009aa} and WFIRST \citep{Spergel:2013tha}, will provide us with particularly precise maps of the LSS that will shed light on the mysterious nature of the dark sector. Taking full advantage of this wealth of cosmological information poses a great challenge for experiment and theory alike. 

From a theoretical standpoint, models of structure formation rely upon accurately tracing the nonlinear evolution of dark matter perturbations. In the linear regime and when gravity is governed by GR, different modes evolve independently, with the time evolution encapsulated in the scale-independent growth factor. At  nonlinear scales, however, the dynamics of the self-gravitating dark matter system can only be tracked accurately by full-blown N-body simulations, which are highly computationally expensive . Additional complexity arises in accounting for the fact that the observed galaxies do not perfectly trace the underlying dark matter density field, but are biased tracers of it \citep{1984ApJ...284L...9K}. While this effect can be easily captured by introducing a multiplicative bias factor in the large scales \citep{1988MNRAS.235..715E}, in the regime of nonlinear dynamics, empirical modeling needs to be combined with sophisticated simulations in order to predict the spatial distribution of galaxies inside gravitationally collapsed dark matter halos \citep{0004-637X-575-2-587,0004-637X-609-1-35,0004-637X-633-2-791}. Furthermore, when MG configurations are considered, the fifth forces introduce an additional layer of complexity, scale-dependent growth is generally present even at the linear level and in the nonlinear scales, one needs to solve the scalar field Klein-Gordon (KG) equation that is highly nonlinear and adds to the computational costs significantly. The intermediate, quasi-linear, scales can fortunately be analytically accessed by higher order Perturbation Theory (PT) \citep{Bernardeau:2001qr,2009PhRvD..80d3531C} approaches or hybrid methods \citep{Tassev:2013pn,Valogiannis:2016ane}. 

The Lagrangian Perturbation Theory approach \citep{Zeldovich:1969sb,1989A&A...223....9B,Bouchet:1994xp,Hivon:1994qb,Taylor:1996ne,Matsubara:2007wj,Matsubara:2008wx,2013MNRAS.429.1674C,Matsubara:2015ipa,Desjacques:2016bnm} to structure formation is one the oldest and most popular analytical frameworks in the literature, that has been been particularly successful at describing the Baryon Acoustic Oscillation (BAO) peak \citep{Eisenstein:2005su}, observed at a comoving scale of $\sim$110 Mpc/$h$, and the power-law correlation function, on comoving scales $\sim$20-90 Mpc/$h$, in $\Lambda$CDM  \citep{Vlah:2014nta}. Combined with a model for halo bias \citep{1984ApJ...284L...9K,Mo:1996cn,Sheth:1999mn,2009JCAP...08..020M,PhysRevD.88.023515,Matsubara:2008wx}, it can be used to predict the 2-point statistics for halos in the real and redshift space \citep{2013MNRAS.429.1674C,Matsubara:2008wx,Vlah:2016bcl}, which serves as a crucial step to the theoretical description of galaxy clustering. Additional contributions from small-scale physics can also be included using techniques inspired by effective field theory theory \citep{Vlah:2015sea,Vlah:2016bcl}. In the context of MG cosmologies, extensive studies have been performed in the framework of Eulerian Standard Perturbation Theory (SPT) \citep{PhysRevD.79.123512,Taruya:2013quf,PhysRevD.88.023527,PhysRevD.90.123515,PhysRevD.92.063522,Fasiello:2017bot,Bose:2017dtl,Bose:2018zpk}. LPT was first found to work very well within the COLA hybrid framework for chameleon and Vainshtein MG cosmologies \citep{Valogiannis:2016ane}, while third order LPT for dark matter was recently developed in the case of scalar-tensor theories in \citep{Aviles:2017aor}.

In this work, we perform a comprehensive study of how LPT can be used to make predictions for biased tracers in modified gravity theories, and how well the predictions compare with full numerical simulations for a variety of modified gravity models. We  study chameleon and Vainsthein MG theories, focusing on the $f(R)$ Hu-Sawicki \citep{Hu:2007nk} and $n$DGP braneworld models \citep{Dvali:2000hr} as popular, representative examples for each category. The underlying dark matter clustering is described using the formalism in \citep{Aviles:2017aor}. We then consider the Convolution Lagrangian Perturbation Theory (CLPT)   \citep{2013MNRAS.429.1674C} for biased tracers, as well as the variants, using the particular resummation scheme in \citep{Vlah:2015sea,Vlah:2016bcl}. We extend   the peak-background split formalism (PBS) \citep{1984ApJ...284L...9K,1986ApJ...304...15B,PhysRevD.88.023515}  in which the Lagrangian bias factors are calculated as responses, to account for modifications to the halo mass function in modified gravity theories and the environmental dependence of screening effects. We show how this formalism should be extended in the case of each screening mechanism. 
Finally, we cross-check and validate our results in terms of the 2-point statistics against state-of-the-art cosmological N-body simulations that allow us to assess the LPT predictions in the nonlinear, quasi non-linear and baryon acoustic oscillation peak regimes. Comparison with simulations is required to ensure that the predictions will be sufficiently robust for upcoming surveys such as DESI, Euclid, LSST and WFIRST.

 We note that a recent paper, submitted to the arXiv while this work was being finalized, also studied biased tracers in MG using CLPT \citep{Aviles:2018saf} but used a different bias scheme and results were not compared against full simulations.  

Our paper is structured as follows: in Section \ref{sec:Models}, we present the MG models we studied and the details of the N-body simulations employed to test our LPT implementation.  We summarize the formalism for developing biased tracer statistics for GR in Section \ref{sec:LPTDM}.  In Section \ref{sec:Results:LPT}, we discuss  the modifications required to the perturbative schemes to predict the real and Fourier space 2-point statistics and the associated bias parameters for tracers of different masses  for models beyond GR.   The LPT predictions are compared to statistics derived from simulations in Section \ref{sec:Results:Compsims}. Finally, we conclude, and discuss implications for future work, in Section \ref{sec:conclusions}. More detail on the derivations in the paper are given in the Appendix.

\section{Formalism}
\label{sec:Formalism}

\subsection{Modified gravity models}
\label{sec:Models}

\subsubsection{The $f(R)$ model}
Despite the latest surge in the field of MG, deformations to GR, together with associated experimental tests, have been around for almost as long as GR itself \citep{Will:2005va}. One of the oldest attempts consisted of adding a nonlinear function $f(R)$ of the Ricci scalar $R$ to the standard Einstein-Hilbert action, the so-called $f(R)$ theories \citep{DeFelice:2010aj}, with a resulting action $S$ of the form:
\begin{equation}\label{actFr}
S=\int d^4x \sqrt{-g} \left[\frac{R+f(R)}{16 \pi G} + \mathcal{L}_m \right],
\end{equation}
with $\mathcal{L}_m$ denoting the matter sector Lagrangian and G the gravitational constant. Since such an action frees up an additional degree of freedom, the latest interest in this class of theories comes from the idea that a modification of this type is responsible for cosmic acceleration, rather than dark energy \citep{Carroll:2003wy}. 

Such a model is the Hu-Sawicki $f(R)$ model \citep{Hu:2007nk}, which we study in this paper and the functional form of which is given by:
\begin{equation}\label{fRHu}
f(R)=-m^2\frac{c_1\left(R/m^2\right)^n}{c_2\left(R/m^2\right)^n+1}.
\end{equation}
In equation (\ref{fRHu}), $\Omega_{m0}$ denotes the matter fractional energy density and $H_0$ the Hubble Constant, both evaluated today, $m=H_0\sqrt{\Omega_{m0}}$, which has dimensions of mass and $n$, $c_1$ and $c_2$ are free parameters. 

In order to match the $\Lambda$CDM expansion history and for sufficiently small values of $|f_{R_0}|$, the background value of the Ricci scalar, $\bar{R}$, becomes equal to:
\begin{align}
\bar{R}= 3 \Omega_m H_0^2 \left(1+ 4 \frac{\Omega_{\Lambda0}}{\Omega_{m0}} \right).
\end{align}
where $\Omega_{\Lambda0}$ is the dark energy fractional density evaluated at the present time. 

The derivative $f_{R}=\frac{df(R)}{dR}$ becomes a functional of the cosmological parameters when evaluated today, through the relationship
\begin{equation}
\bar{f}_{R_0} =-n\frac{c_1}{c_2^2}\left(\frac{\Omega_{m0}}{3(\Omega_{m0}+\Omega_{\Lambda0})}\right)^{n+1}.
\end{equation}
The above mapping allows us to reduce the number of free parameters and the Hu-Sawicki model is commonly parametrized by quoting the values chosen for $n$ and $|f_{R_0}|$, with the latter being the background value of the fifth potential evaluated today. 

The reason the Hu-Sawicki model is so popular is that it can be cast into the form of a scalar-tensor theory that realizes the chameleon screening mechanism \citep{PhysRevD.69.044026,PhysRevLett.93.171104}, through a conformal transformation \citep{Brax:2008hh}, with the quantity $f_{R}$ identified as the scalar field that is coupled to matter. Through the interplay between matter and the self-interaction potential, the scalar chameleon field becomes massive near high over-densities and the associated fifth forces get exponentially suppressed due to the Yukawa effect. A stronger screening effect is manifested in lower values of the parameter $|f_{R_0}|$, with the limit of $|f_{R_0}|\rightarrow0$ exactly recovering GR. Following the literature and also because of the available simulations (as we will discuss below in \ref{sim}), we choose $n=1$ and consider three different $f(R)$ models with $\abs{\bar{f}_{R_0}}=\{10^{-6},10^{-5}, 10^{-4}\}$, which we shall refer to, from now on, as F6, F5 and F4.

When considering perturbations around a homogeneous and isotropic FRW metric in the conformal Newtonian gauge, the resulting system of the Poisson and KG equations becomes \citep{Hu:2007nk}:
\begin{equation}
\begin{aligned}\label{eq:poisson}
\nabla^2 \Phi_N &= 4 \pi G  a^2 \delta \rho_m - \frac{1}{2} \nabla^2 f_R, \\
\nabla^2 f_R &= -\frac{a^2}{3} \delta R - \frac{8 \pi G a^2}{3} \delta \rho_m,
\end{aligned}
\end{equation}
where $\Phi_N$ is the Newtonian potential and $\delta \rho_m$ the matter density perturbation. $\delta R$, the perturbation to the Ricci scalar, can be written as a function of the scalar field $f_R$, which will play a central role in our screening implementation in Section \ref{sec:LPTDM}.

\subsubsection{The $n$DGP model}
\label{nDGPsec}

The second MG model under consideration comes from the realm of higher-dimensional braneworld cosmology. The simplest example of such a configuration is the so-called Dvali-Gabadadze-Porrati (DGP) model \citep{Dvali:2000hr}, which is described by an action of the following form:
\begin{equation}\label{actDGP}
S=\int d^4x \sqrt{-g} \left[\frac{R}{16 \pi G} + \mathcal{L}_m \right] + \int d^5x \sqrt{-g_5} \left(\frac{R_5}{16 \pi G r_c}\right).
\end{equation}
In this model, the spacetime consists of 5 dimensions, rather than the usual 4, but the standard model fields are restricted to a 4-dimensional (4D) brane and the free parameter, $r_c$, denotes the length-scale below which gravity becomes 4D. $R_5$ and $g_5$ denote the corresponding 5D versions of the Ricci scalar and metric tensor determinant, respectively. The resulting Friedman equation from (\ref{actDGP}) is:
\begin{equation}\label{FridDGP}
\epsilon \frac{H}{r_c}=H^2 - \frac{\rho_m}{24 \pi G}, 
\end{equation}
where $\epsilon=\pm 1$. Each of the two values of $\epsilon$ represents a particular branch of the model, with $\epsilon=+1$ producing the self-accelerating solution, which, however, suffers from undesirable "ghost" instabilities \citep{Koyama:2007zz} and is thus an unphysical model to consider. The value of $\epsilon=-1$ corresponds to the so-called normal branch, hereafter called $n$DGP, which is well behaved, but does not self-accelerate and can match a $\Lambda$CDM expansion history only in the presence of dark energy. 

When focusing on the normal branch and in the quasi-static limit for sub-horizon scales, the perturbations in the conformal Newtonian gauge give the modified Poisson system of equations \citep{PhysRevD.79.123512}:
\begin{equation}
\begin{aligned}\label{eq:poisson_dgp}
\nabla^2 \Phi_N &= 4 \pi G  a^2 \delta \rho_m + \frac{1}{2} \nabla^2 \varphi, \\
\nabla^2 \varphi &= \frac{8 \pi G a^2}{3 \beta} \delta \rho_m - \frac{r_c^2}{3 \beta a^2}\Big[(\nabla^2 \varphi)^2- (\nabla_i \nabla_j \varphi)^2 \Big],
\end{aligned}
\end{equation}
with the coupling $\beta$ given by
\begin{align}
\label{eq:dgpbeta}
\beta(a) = 1 + 2 H(a) r_c \left(1 + \frac{\dot{H(a)}}{3 H(a)^2} \right).
\end{align}

The $n$DGP model is a typical example of a scalar-tensor theory that realizes the Vainshtein screening mechanism \citep{VAINSHTEIN1972393,Babichev:2013usa}, in which the modifications to gravity are suppressed in the existence of large second derivatives of the scalar field. In the second equation of (\ref{eq:poisson_dgp}), in particular, it can be seen how, once the second derivatives of the scalar field become large in high densities, the second term on the right hand side (r.h.s.) becomes significant and effectively weakens the source strength, resulting in strong screening of  the fifth forces.

We consider the $n$DGP model for two choices of the free parameter $r_c$, corresponding to $n\equiv H_0 r_c=1$ and $n=5$, which we shall call, from now on, N1 and N5, respectively.

\subsubsection{N-body Simulations}
\label{sim}

Accurate realizations of structure formation in the nonlinear regime can only be achieved by performing N-body simulations. In this paper, we test our results against two groups of state-of-the-art N-body simulations, that serve complementary purposes to each other, as explained below. 

The first group of simulations, to which we shall refer as Group I from now on, are the {\sc ELEPHANT} simulations, presented in \citep{Cautun:2017tkc}. These span the parameter space of both MG models we study: the F4, F5 and F6 $f(R)$ models and N1 and N5 $n$DGP models.  The $f(R)$ simulations were performed using the {\sc ECOSMOG} code \citep{1475-7516-2012-01-051,Bose:2016wms}, while the $n$DGP ones \citep{Hellwing:2017pmj} using the {\sc ECOSMOG-V} version \citep{Li:2013nua,Barreira:2015xvp}, both of which were based on the GR code {\sc RAMSES} \citep{Teyssier:2001cp} and where suitably extended to integrate the scalar field KG equation for the corresponding models using adaptive-mesh-refinement techniques. The parameters describing the background $\Lambda$CDM cosmology are the best-fit ones given by the 9-year WMAP release \citep{0067-0049-208-2-19} and have the following values: $\Omega_b=0.046$, $\Omega_{cdm}=0.235$, $\Omega_m=0.281$, $\Omega_L=0.719$, $h=0.7$, $n_s=0.971$ and $\sigma_8=0.82$. $N_p=1024^3$ equal mass particles were placed in a simulation box with a side $L_{box}=1024$ Mpc/h and the density field was resolved in a $1024^3$ resolution grid. Furthermore, the simulations were initialized at redshift $z_i=49$ using the Zel'dovich approximation \citep{Zeldovich:1969sb} and evolved through $z_f=0$. 

Gravitationally bound dark matter halos were identified using the {\sc ROCKSTAR} halo finder \citep{2013ApJ...762..109B}. Finally, so as to get an estimate of the variance, each model was simulated for 5 random realizations, corresponding to different random phases in the initial density field.

The $1024^3$ $(Mpc/h)^3$ volume simulations' results in Group I become noisy at scales $r>100$ Mpc/h.  
To probe the BAO scales, where LPT has been previously found to perform very well for GR \citep{Vlah:2014nta}, we also test our results against the largest volume $f(R)$ simulations performed to date for the modified gravity lightcone simulation project \citep{Arnold:2018nmv}.
 In these simulations, which we will call Group II from now on, the box side is $L_{box}=1536$ Mpc/h with $2048^3$ equal mass particles used, for GR and the $\abs{\bar{f}_{R_0}}$=$10^{-5}$ model. The parameters describing the background $\Lambda$CDM cosmology are the best-fit ones given by the Planck collaboration \citep{2016A&A...594A..13P} and have the following values: $\Omega_b=0.0486$, $\Omega_m=0.3089$, $\Omega_L=0.6911$, $h=0.6774$, $n_s=0.9667$ and $\sigma_8=0.8159$. The simulations were performed using the MG code {\sc MG-GADGET} \citep{doi:10.1093/mnras/stt1575}, which is a MG extension to the code {\sc P-GADGET3}, an improved version of the code {\sc GADGET-2} \citep{2005MNRAS.364.1105S}, created for GR simulations. Dark matter halo catalogues were produced using the {\sc SUBFIND} algorithm \citep{2001MNRAS.328..726S}. Each model has been simulated for one random realization. 

For more detailed discussions on the N-body implementations, we refer interested readers to the corresponding publications.

\subsection{Convolution Lagrangian Perturbation Theory for biased tracers in MG}
\label{sec:LPTDM}

\subsubsection{LPT for dark matter}\label{LPTdarkmatter}

The Lagrangian Perturbation Theory approach to structure formation has been extensively studied \citep{Zeldovich:1969sb,1989A&A...223....9B,Bouchet:1994xp,Hivon:1994qb,Taylor:1996ne,Matsubara:2007wj,Matsubara:2008wx,2013MNRAS.429.1674C,Matsubara:2015ipa,Desjacques:2016bnm} in the context of $\Lambda$CDM scenarios. Opposite to the Eulerian picture, in which one monitors the evolution of the desired quantities at a given, fixed, position, in LPT one instead tracks down the evolution of a given fluid element over time. Starting from an initial, Eulerian, comoving position $\bold{q}$ at a desired early time $t_0$, each mass element is mapped to its comoving Lagrangian position $\bold{x}(\bold{q},t)$ at time $t$, through the relationship
\begin{equation}\label{Lagpos}
\bold{x}(\bold{q},t) = \bold{q} + \bold{\Psi}(\bold{q},t).
\end{equation}
The Lagrangian displacement $\bold{\Psi}(\bold{q},t)$, taken to be 0 at the initial time $t_0$, is the fundamental quantity of interest in LPT.  Furthermore, enforcing mass conservation, through the continuity equation, between the initial and final infinitesimal volume elements centered around $\bold{q}$ and $\bold{x}$, respectively, gives $\rho_{m}(\bold{x},t)d^3x=\rho_{m}(\bold{q},t_0)d^3q$. Assuming $t_{0}$ refers to an epoch early enough that the density perturbations around the background density $\bar{\rho}$ are negligible, meaning $\rho_{m}(\bold{q},t_0)=\bar{\rho}_m$, allows us to obtain the dark matter fractional overdensity, $\delta_m$, in the Lagrangian picture:
\begin{equation}\label{delJac}
1 + \delta_m(\bold{x},t) = \int d^3q \delta_{D}\left[\bold{x}-\bold{q}-\bold{\Psi}(\bold{q},t)\right] = \frac{1}{J(\bold{q},t)},
\end{equation}
with $\delta_{D}$ being the Dirac delta function and $J(\bold{q},t)$ the determinant of the deformation matrix 
\begin{equation}\label{Jacobian}
J_{ij} = \frac{\partial x^i}{\partial q^j} = \delta_{ij} + \frac{\partial \Psi^i}{\partial q^j}.
\end{equation}
For an irrotational flow, which is a good approximation for cold dark matter and assuming that the gravitational evolution is governed by GR, perturbations around a flat FRW metric give the geodesic and Poisson equations, in the quasi-static approximation and for sub-horizon scales, as:
\begin{equation}
\begin{aligned}\label{eq:geopoisson}
\ddot{\bold{x}} + 2H\dot{\bold{x}} &= -\frac{1}{a^2}\nabla_{\bold{x}}\psi(\bold{x},t), \\
\frac{1}{a^2}\nabla^2_{\bold{x}}\psi(\bold{x},t) &=  4 \pi G\bar{\rho}_m \delta(\bold{x},t).
\end{aligned}
\end{equation}
We should point out that in (\ref{eq:geopoisson}) $\psi(\bold{x},t)$ denotes the metric perturbation, which should not be confused with the Lagrangian displacement field $\bold{\Psi}(\bold{q},t)$.
\newline In the LPT picture, we perturbatively expand $\bold{\Psi}$ as
\begin{equation}\label{eq:psiexp}
\bold{\Psi}(\bold{q},t) = \sum_{n=1}^{\infty}\bold{\Psi}^{(n)}(\bold{q},t) = \bold{\Psi}^{(1)}(\bold{q},t)+\bold{\Psi}^{(2)}(\bold{q},t)+\bold{\Psi}^{(3)}(\bold{q},t)...
\end{equation}
and equations (\ref{Lagpos})-(\ref{eq:psiexp}) form a closed system, that is recursively solved for the various orders of $\bold{\Psi}$. The first order solution is the so-called Zel'dovich approximation \citep{Zeldovich:1969sb}.

In MG theories, as also explained in Section \ref{sec:Models}, an additional degree of freedom is present, that directly couples to matter and causes particles to deviate from the nominal geodesics of GR. Consequently, equations (\ref{eq:geopoisson}) are, in principle, modified for a scalar-tensor theory and so is the LPT framework presented above. In \citep{Valogiannis:2016ane}, the LPT approach was expanded for chameleons and symmetrons, including the first order contribution to the Klein-Gordon equation and was shown to perform very well in the context of the COLA hybrid framework. The LPT approach for scalar-tensor theories up to third order was presented in \citep{Aviles:2017aor}, the main results of which we summarize in this section.

In \citep{PhysRevD.92.063522}, which developed an SPT framework for studying MG theories with a screening mechanism in the nonlinear regime, based on the closure theory approximation in \citep{Taruya:2007xy}, the scalar field KG equation for a Brans-Dicke-like (BD) theory with interactions of a scalar field, $\phi$, was written as:
\begin{equation}\label{BransDicke}
(3 + 2 \omega_{BD})\frac{1}{a^2} k_x^2\phi(\bold{k}_x,t) = 8 \pi G \bar{\rho}_m \delta(\bold{k}_x,t) - \mathcal{I}(\phi),
\end{equation}
where $\mathcal{I}(\phi)$ denotes the perturbative form of the field self-interaction term:
\begin{eqnarray}
\label{interaction}
 \mathcal{I}(\phi) &=& M_1(\bold{k},t)\phi \nonumber
\\
&& + \frac{1}{2}\int \frac{d^3 k_1 d^3 k_2}{\left(2 \pi\right)^3} \delta_D(\bold{k}-\bold{k}_{12})M_2(\bold{k}_{1},\bold{k}_{2})\phi(\bold{k}_{1})\phi(\bold{k}_{2})\nonumber\\
&& + \frac{1}{6}\int \frac{d^3 k_1 d^3 k_2 d^3 k_3}{\left(2 \pi\right)^6} \delta_D(\bold{k}-\bold{k}_{123})\nonumber \\&& \   \hspace{1.5cm} \times M_3(\bold{k}_{1},\bold{k}_{2},\bold{k}_{3})\phi(\bold{k}_{1})\phi(\bold{k}_{2})\phi(\bold{k}_{3}),
\end{eqnarray}
where we adopted the standard notation $\bold{k}_{ijk}=\bold{k}_{i}+\bold{k}_{j}+\bold{k}_{k}$ and $M_1(\bold{k},t)$, $M_2(\bold{k}_{1},\bold{k}_{2})$ and $M_3(\bold{k}_{1},\bold{k}_{2},\bold{k}_{3})$ are mass terms. The higher order piece in the Fourier space representation of the interaction term in equation (\ref{interaction}), incorporates the screening effect, up to third order,  that is responsible for recovering GR at small scales. The mapping between a given scalar-tensor theory and the BD form above can be easily performed through assigning appropriate values to the mass terms and the BD coupling $\omega_{BD}$ above, as we will later show for our two models of study. 

The perturbed modified Einstein equations have the form \citep{Aviles:2017aor}
\begin{equation}
\begin{aligned}\label{eq:modpoisson}
\nabla_{\bold{x}}\mathcal{\hat{T}}\bold{\Psi} &= -\frac{1}{a^2}\nabla^2_{\bold{x}}\psi(\bold{x},t), \\
\frac{1}{a^2}\nabla^2_{\bold{x}}\psi(\bold{x},t) &=  4 \pi G\bar{\rho}_m \delta(\bold{x},t) -\frac{1}{2 a^2} \nabla^2\phi - \frac{1}{2 a^2}\left(\nabla^2_{\bold{x}}\phi -\nabla^2\phi \right),
\end{aligned}
\end{equation}
where we use (\ref{Lagpos}) and introduce the time derivative operator $\mathcal{\hat{T}}=\frac{d^2}{dt^2} + 2H\frac{d}{dt}$, as in \citep{Matsubara:2007wj}.
The last term in the second line of (\ref{eq:modpoisson}), called $\it{frame}$-$\it{lagging}$ in \citep{Aviles:2017aor}, is a geometrical term that occurs due to the fact that, in LPT, the KG equation should be expressed in Lagrangian $\left(\nabla^2\right)$, rather than Eulerian coordinates $\left(\nabla^2_{\bold{x}}\right)$. Taking this into account, equations (\ref{BransDicke})-(\ref{eq:modpoisson}) are combined to give
\begin{equation}
\begin{aligned}\label{eq:finpsi}
& \left(J^{-1}\right)_{ij}\mathcal{\hat{T}}\Psi_{i,j}(\bold{k}) = -A(k)\delta(\bold{k}) + \frac{k^2}{a^2 3\Pi(k)}\delta\mathcal{I}(\bold{k}) \\
& + \frac{M_1(k)}{3\Pi(k)}\frac{1}{2 a^2}\left(\nabla^2_{\bold{x}}\phi -\nabla^2\phi \right)(\bold{k}),
\end{aligned}
\end{equation}
where, following the definitions in \citep{PhysRevD.92.063522}, we have
\begin{equation}
\begin{aligned}\label{eq:sources}
A(k) &=  4 \pi G\bar{\rho}_m\left(1 + \frac{k^2}{a^2 3\Pi(k)}\right), \\
\Pi(k) &= \frac{1}{3a^2}\left[\left(3+2\omega_{BD}\right)k^2 + M_1a^2\right]
\end{aligned}
\end{equation}
and all the quantities are Fourier transforms in the Lagrangian $\bold{q}$-space. The inverse Jacobean in (\ref{eq:finpsi}) reflects the derivative transformation to the $\bold{q}$-space, where the Einstein notation is adopted. Furthermore, $\delta(\bold{k})$, $\delta\mathcal{I}(\bold{k})$ and $\left(\nabla^2_{\bold{x}}\phi -\nabla^2\phi \right)(\bold{k})$ are the Lagragian Fourier transformations of the Lagrangian-transformed overdensity (\ref{delJac}), the higher order interaction Kernels in (\ref{interaction}) and the frame-lagging Kernel, correspondingly. The expression for the latter is given in \citep{Aviles:2017aor}. Equation (\ref{eq:finpsi}) forms a closed system with (\ref{delJac}), (\ref{Jacobian}), (\ref{BransDicke}) and (\ref{interaction}) that is solved, perturbatively, to obtain the MG solution up to various orders in $\bold{\Psi}$, as in (\ref{eq:psiexp}). 

Solving for the first order solution, one gets \citep{Valogiannis:2016ane}:
\begin{equation}\label{eq:Zeldisp}
\bold{k}\cdot\bold{\Psi}^{(1)} = i D^{(1)}(k,t)\delta^{(1)}(\bold{k},t=0),
\end{equation}
which can be easily solved for the displacement field, as:
\begin{equation}\label{eq:Zeldispsol}
{\Psi}^{j}(k,t) = \frac{i k^j}{k^2} D^{(1)}(k,t)\delta^{(1)}(\bold{k},t=0).
\end{equation}
We see that the r.h.s of (\ref{eq:Zeldispsol}) can be conveniently decomposed into a product of the first order density mode at very early times, $\delta^{(1)}(\bold{k},t=0)$, early enough to be gaussian and a space-time dependent growth factor $D^{(1)}(k,t)$, given by:
\begin{equation}\label{growth1st}
\mathcal{\hat{T}}D^{(1)}(k,t)=A(k)D^{(1)}(k,t).
\end{equation}
In the GR limit, $A(k)=A(k=0)=4 \pi G\bar{\rho}_m$, $D^{(1)}$ becomes scale independent and is nothing else than the first order growing mode for GR; the Zel'dovich approximation.

Moving on to the second order piece, we have:
\begin{equation}\label{eq:Zeldisp2}
\bold{k}\cdot\bold{\Psi}^{(2)} = \frac{i}{2} \int \frac{d^3 k_1 d^3 k_2}{\left(2 \pi\right)^3} \delta_D(\bold{k}-\bold{k}_{12}) D^{(2)}(\bold{k}_1,\bold{k}_2)\delta^{(1)}_1\delta^{(1)}_2,
\end{equation}
where, for compactness, we adopted the notation $\delta^{(1)}_1=(\bold{k}_1,0)$ and the second order growth factor, $D^{(2)}(\bold{k}_1,\bold{k}_2)$, is given by \citep{Aviles:2017aor}:
\begin{eqnarray}\label{eq:secgrowth}
D^{(2)}(\bold{k}_1,\bold{k}_2) &=& D^{(2)}_a(\bold{k}_1,\bold{k}_2)-D^{(2)}_b(\bold{k}_1,\bold{k}_2)\frac{\bold{k}_1\cdot\bold{k}_2}{k^2_1k^2_1} \nonumber
\\ && +D^{(2)}_{FL}(\bold{k}_1,\bold{k}_2)-D^{(2)}_{\delta \mathcal{I}}(\bold{k}_1,\bold{k}_2).
\end{eqnarray}
The four individual components are given by:
\begin{equation}
\begin{aligned}\label{eq:D2sources}
\left(\mathcal{\hat{T}}-A(k)\right)D^{(2)}_a(\bold{k}_1,\bold{k}_2) &=  A(k)D^{(1)}(k_1)D^{(1)}(k_2), \\
\left(\mathcal{\hat{T}}-A(k)\right)D^{(2)}_b(\bold{k}_1,\bold{k}_2) &=  \left(A(k_1)+A(k_2)-A(k)\right)D^{(1)}(k_1)D^{(1)}(k_2), \\
\left(\mathcal{\hat{T}}-A(k)\right)D^{(2)}_{\delta \mathcal{I}}(\bold{k}_1,\bold{k}_2) &=  \left(\frac{2A_0}{3}\right)^2\frac{k^2}{a^2}\frac{M_{2}(\bold{k}_1,\bold{k}_2)D^{(1)}(k_1)D^{(1)}(k_2)}{6\Pi(k)\Pi(k_1)\Pi(k_2)}, \\
\left(\mathcal{\hat{T}}-A(k)\right)D^{(2)}_{FL}(\bold{k}_1,\bold{k}_2) &=  \left(\frac{M_1}{3\Pi(k)}\right)K^{(2)}_{FL}(\bold{k}_1,\bold{k}_2)D^{(1)}(k_1)D^{(1)}(k_2).
\end{aligned}
\end{equation}
The two last terms represent the second order contributions to the growth factor, given by the screening and frame-lagging effects, correspondingly, while the expression for $K^{(2)}_{FL}$ is given in \citep{Aviles:2017aor}. Despite its lengthier expression, when taking the GR limit we get $D^{(2)}_{FL}=D^{(2)}_{\delta \mathcal{I}}=0$ and $D^{(2)}_{a}=D^{(2)}_{b}$, allowing $D^{(2)}$ to become scale-independent, reducing to the known GR result, which can be well approximated by $D^{(2)}(t)=-\frac{3}{7}\left(D^{(1)}(t)\right)^2$ for $\Lambda$CDM cosmologies \citep{Bouchet:1994xp}. 

Solving for the third order piece in (\ref{eq:finpsi}) results in a lengthy differential equation for the third order MG growth factor, $D^{(3)}(\bold{k}_1,\bold{k}_2,\bold{k}_3)$, that also needs to be symmetrized. The result is given by equation (\ref{D3MG}) in the appendix \ref{polyderiv}. It should be also noted that equations (\ref{growth1st}), (\ref{eq:D2sources}) 
and f(\ref{D3MG}) can be either solved by inverting the linear operator $\left(\mathcal{\hat{T}}-A(k)\right)$ using its Green function, as done in \citep{Matsubara:2015ipa,Aviles:2017aor}, or by numerically solving the corresponding differential equations. Even though both methods give results that agree with each other well, we chose to proceed with the latter because it is computationally faster. The differential equations were solved using a $5^{th}$ order Runge-Kutta scheme, implemented in $\it{Mathematica}$ \citep{Mathematica}.

We finish this section by showing the particular expressions for the mass terms in (\ref{interaction}) and the sources in (\ref{eq:sources}) for the $f(R)$ and $n$DGP models we study. For the $f(R)$ Hu-Sawicki model, the Brans-Dicke scalar is simply $\omega_{BD}=0$, while the mass terms are given by the expansion \citep{PhysRevD.79.123512}:
\begin{equation}\label{eq:fRmasses}
M_{n} = \frac{d^{n}\bar{R}(f_{R_0})}{df_{R_0}^{n}},
\end{equation}
which, using (\ref{fRHu}), gives \citep{Aviles:2017aor}:
\begin{equation}
\begin{aligned}\label{eq:fRmasses2}
M_{1}(a) &= \frac{3 H_0^2}{2 |f_{R_0}|}\frac{\left(\Omega_m a^{-3}+4 \Omega_{\Lambda}\right)^3}{\left(\Omega_m+4 \Omega_{\Lambda}\right)^2},\\
M_{2}(a) &= \frac{9 H_0^2}{4 |f_{R_0}|}\frac{\left(\Omega_m a^{-3}+4 \Omega_{\Lambda}\right)^5}{\left(\Omega_m+4 \Omega_{\Lambda}\right)^4},\\
M_{3}(a) &= \frac{45 H_0^2}{8 |f_{R_0}|}\frac{\left(\Omega_m a^{-3}+4 \Omega_{\Lambda}\right)^7}{\left(\Omega_m+4 \Omega_{\Lambda}\right)^6}.
\end{aligned}
\end{equation}

In the case of the $n$DGP braneworld model, a similar procedure, informed by (\ref{eq:poisson_dgp}), gives the relevant expressions \citep{PhysRevD.79.123512,Aviles:2018saf}:
\begin{equation}
\begin{aligned}\label{eq:nDGPmasses}
&M_{1}(a) = 0,\\
&M_{2}(\bold{k}_1,\bold{k}_2,a) = 2\frac{n^2}{H_0^2 a^4}\left(k_1^2k_2^2-(\bold{k_1}\cdot\bold{k_2})^2\right),\\
&M_{3}(\bold{k}_1,\bold{k}_2,\bold{k}_3,a) = 18\frac{n^2 \beta(a)}{H_0^2 a^6 4 \pi G\bar{\rho}_m}\Biggl((\bold{k_1}\cdot\bold{k_2})k_1^2k_3^2+2(\bold{k_1}\cdot\bold{k_2})^2k_3^2\\
& -2(\bold{k_1}\cdot\bold{k_2})(\bold{k_1}\cdot\bold{k_3})(\bold{k_2}\cdot\bold{k_3})-(\bold{k_1}\cdot\bold{k_2})(\bold{k_1}\cdot\bold{k_3})^2\Biggr), \\
\end{aligned}
\end{equation}
with $\beta$ defined in (\ref{eq:dgpbeta}). It is interesting to notice that, even though the interaction term in (\ref{eq:poisson_dgp}) contains only second order derivatives, in (\ref{eq:nDGPmasses}) a $3^{rd}$ order mass contribution is now present, that arises when transforming the Eulerian derivatives to the Lagrangian space through (\ref{Jacobian}).

\subsubsection{2-point statistics for biased tracers in GR}\label{sec:twopointGR}

The perturbative theory of galaxy clustering \citep{Desjacques:2016bnm}, which aims to describe the statistics of biased tracers in the quasi-linear regime, consists of a perturbative description for the evolution of the underlying dark matter density field, combined with an analytical description for the bias parameters at each given order. 
In the case of cold dark matter, the calculation of the 2-point statistics, even in LPT, is a more straightforward process, since one just needs to plug the ${\bf q}$-space Lagrangian overdensity, mapped to the Eulerian frame through (\ref{delJac}), into the common expressions for the autocorrelation function:
\begin{equation}\label{xi}
\xi(r) = \langle \delta_m(\bold{x})\delta_m(\bold{x+r})\rangle
\end{equation}
and its Fourier space counterpart, the matter power spectrum
\begin{equation}\label{Pk}
\left(2\pi\right)^3\delta_D(\bold{k}+\bold{k'})P(k) = \langle \tilde{\delta}_m(\bold{k})\tilde{\delta}_m(\bold{k'})\rangle.
\end{equation}

When studying biased tracers, like for example dark matter halos, we need an analytical model to describe their statistical prevalance with respect to the underlying density field.   Following \citep{Matsubara:2008wx,2013MNRAS.429.1674C}, we employ a model of a local in matter density Lagrangian bias in which the positions of biased tracers  are purely specified by a distribution of the  underlying CDM density field $\delta(\bold{q},t=0)\equiv\delta(\bold{q})$, encoded through a function $F\left[\delta_R(\bold{q})\right]$, as
\begin{equation}\label{biasF}
\rho_{X}(\bold{q}) =\bar{\rho}_{X} F\left[\delta_R(\bold{q})\right],
\end{equation}
where, consistent with the literature, we use the subscript $X$ to indicate biased tracers. $\delta_R(\bold{q})$ denotes the primordial density field smoothed over some scale $R$, while $\bar{\rho}_{X}$ is the mean density of tracers.   Density conservation  provides the equivalent of equation (\ref{delJac}) for tracers,
\begin{equation}\label{delJacX}
\delta_X(\bold{x},t) = \int d^3q F\left[\delta_R(\bold{q})\right]\delta_{D}\left[\bold{x}-\bold{q}-\bold{\Psi}(\bold{q},t)\right] -1. 
\end{equation}
This model of local Lagrangian bias, which corresponds to a non-local bias in the Eulerian space, can be extended to include a biasing scheme that is non-local in the Lagrangian space \citep{PhysRevD.83.083518}. Combining (\ref{delJacX}) and (\ref{xi}) and after some transformations one gets the general expression for the 2-point correlation function for biased tracers in LPT,
\begin{equation}\label{xiX}
1+\xi_{X}(r) = \int d^3q\int \frac{d^3k}{\left(2 \pi\right)^3}e^{i\bold{k}\cdot(\bold{q}-\bold{r})} \int \frac{d\lambda_1}{\left(2 \pi\right)}\frac{d\lambda_2}{\left(2 \pi\right)} L(\bold{q},\bold{k},\lambda_1,\lambda_2),
\end{equation}
with
\begin{equation}\label{LX}
L(\bold{q},\bold{k},\lambda_1,\lambda_2)=\tilde{F}_1 \tilde{F}_2\underbrace{\langle e^{i\left[\lambda_1 \delta_1+\lambda_2 \delta_2+\bold{k}\cdot(\bold{\Delta}))\right]} \rangle}_\text{$K(\bold{q},\bold{k},\lambda_1,\lambda_2)$}, 
\end{equation}
where $\tilde{F}_1$, $\tilde{F}_2$ are the Fourier space representations of $F$, with corresponding wavemodes $\lambda_1$ and $\lambda_2$ and $\bold{\Delta}=\bold{\Psi}_2-\bold{\Psi}_1$. The notation $\delta(\bold{q}_1)\equiv \delta_1$ has been adopted for all quantities. 
The ensemble average K in (\ref{LX}) can be cast into an exponent of a power series in cumulants, through the cumulant expansion theorem, $\langle e^{iX} \rangle = \exp\left[\sum_{N=1}^{\infty}\frac{i^N}{N!} \langle X^{N} \rangle_c \right]$, which, combined with a multinomial expansion, gives:
\begin{equation}\label{multinom}
K(\bold{q},\bold{k},\lambda_1,\lambda_2)=\exp \left[\sum \frac{i^{n+m+r}}{m!n!r!}\lambda_1^m\lambda_2^n k_{i_1}..k_{i_r}\langle \delta_1^{m} \delta_1^{n} \Delta_{i_1}..\Delta_{i_r}\rangle_c\right], 
\end{equation}
in terms of a series of correlators
\begin{equation}
\begin{aligned}\label{eq:correl}
\sigma_R^2 &= \langle \delta^2 \rangle_c \\
\xi_L(\vec{q}) &= \langle \delta_1 \delta_2 \rangle_c,  \\
A_{ij}^{m n}(\vec{q}) &= \langle \delta_i^{m} \delta_j^{n} \Delta_i \Delta_j\rangle_c,  \\ 
W_{ijk}^{m n}(\vec{q}) &= \langle \delta_i^{m} \delta_j^{n} \Delta_i \Delta_j \Delta_k\rangle_c,  \\ 
U_{i}^{m n}(\vec{q}) &= \langle \delta_1^{m} \delta_2^{n} \Delta_i \rangle_c,
\end{aligned}
\end{equation}
where we adopted the commonly used notation for the Lagrangian cumulants in (\ref{eq:correl}).

Keeping all terms in (\ref{multinom}) that contain cumulants up to third order, which is the equivalent of the one-loop correction to the linear power spectrum, results in a highly oscillatory integrand that presents challenges when ensuring the integral is fully converged. \citep{Matsubara:2007wj} proposed expanding  all contributions to the exponent but the scale-independent ``zero-lag" piece of $A_{ij}^{0 0}$, which results in a non-perturbative resummation scheme that is simpler to handle analytically. Building upon this result, \citep{2013MNRAS.429.1674C} proposed keeping all the terms of $A_{ij}^{0 0}$ in the exponent, in their Convolution Lagrangian Perturbation Theory (CLPT) scheme. Keeping only the linear component of $A_{ij}^{0 0}$ exponentiated, as done in \citep{Vlah:2015sea,Vlah:2016bcl}, and performing the $\lambda$ and $k$ integrations in (\ref{xiX}) gives a CLPT expression for the 2-point real space correlation function,
\begin{eqnarray}
\label{xiXfinal}
1 + \xi_{X}(r) &=& \int d^3q \frac{e^{-\frac{1}{2} (q_i - r_i) (A^{-1}_{L})_{ij}(q_j - r_j)}} {\left(2\pi\right)^{3/2} |A_{L}|^{1/2}} \times    \Biggl( 1 - \frac{1}{2}G_{ij} A^{loop}_{ij} \nonumber\\
 && + \frac{1}{6}\Gamma_{ijk} W_{ijk}- b_1 \left(2U_i g_i + A^{10}_{ij}G_{ij}\right) \nonumber \\
&&-b_2\left(U^{(1)}_iU^{(1)}_jG_{ij}+U^{20}_ig_i\right)\nonumber\\
&&+b_1^2\left(\xi_L-U^{(1)}_iU^{(1)}_jG_{ij}-U^{11}_ig_i\right) \nonumber\\
&& +\frac{1}{2}b_2^2\xi_L^2 -2b_1b_2\xi_LU^{(1)}_ig_i \Biggr),
\end{eqnarray}
with 
\begin{eqnarray}
\label{determs}
 g_i &\equiv& (A^{-1}_{L})_{ij}(q_j - r_j), \nonumber \\
 G_{ij} &\equiv& (A^{-1}_L)_{ij}-g_ig_j, \nonumber\\
 \Gamma_{ijk} &\equiv& (A^{-1}_{L})_{ij} g_k +(A^{-1}_{L})_{ki} g_j +(A^{-1}_{L})_{jk} g_i - g_i g_j g_k. 
\end{eqnarray}
Furthermore, in (\ref{xiXfinal}) we define $U^{10}_i\equiv U_i$, $W^{000}_{ijk}\equiv W_{ijk}$ and use superscript numbers in brackets to indicate the various orders of contribution.  The $1^{st}$ and $2^{nd}$ order Lagrangian bias factors, $b_1$ and $b_2$, are the expectation values of the $1^{st}$ and $2^{nd}$ order derivatives of the Lagrangian bias function $F$, respectively, through the identity \citep{Matsubara:2008wx,PhysRevD.83.083518},
\begin{equation}
\begin{aligned}\label{biasfn}
b_n\equiv \int \frac{d \lambda}{2\pi}\tilde{F}(\lambda)e^{-\frac{1}{2}\lambda^2 \sigma^2_R}\left(i \lambda \right)^n = \bigg \langle \frac{d^nF}{d\delta^n} \bigg \rangle.
\end{aligned}
\end{equation}
In the case of dark matter, we have $F=1$ and $\tilde{F}(\lambda)=2\pi \delta_D(\lambda)$ \citep{2013MNRAS.429.1674C}, and we recover $b_1=b_2=0$ for the unbiased, dark matter distribution.

The Fourier transform gives the CLPT power spectrum for biased tracers \citep{Vlah:2015sea,Vlah:2016bcl}:
\begin{eqnarray}
\label{PkXfinal}
P_{X}(k) &=& \int d^3q e^{i\bold{k}\cdot\bold{q}}e^{-\frac{1}{2} k_ik_j A^{L}_{ij}} \times \Biggl( 1 - \frac{1}{2}k_ik_j A^{loop}_{ij}  - \frac{i}{6}k_ik_jk_k W_{ijk} \nonumber\\ 
&&+ b_1 \left(2i k_i U_i  - k_ik_jA^{10}_{ij}\right) + b_2\left(i k_i U^{20}_i - k_ik_jU^{(1)}_iU^{(1)}_j\right) \nonumber\\
&&+b_1^2\left(\xi_L + ik_iU^{11}_i -k_ik_jU^{(1)}_iU^{(1)}_j\right)\nonumber \\
&& +\frac{1}{2}b_2^2\xi_L^2 + 2b_1b_2\xi_L i k_iU^{(1)}_i\Biggr).
\end{eqnarray}

In addition to the one-loop expressions for the two-point statistics, we also calculate the Zel'dovich ($1^{st}$ order LPT) approximation \citep{White:2014gfa} for biased tracers in the configuration and Fourier space, which can be identified as the subset of terms in (\ref{xiXfinal}) and (\ref{PkXfinal}) that are linear in $P_{L}(k)$. These are the terms that depend on combinations of $\xi_L$, $U^{(1)}$ and $A^{L}_{ij}$.

While, for GR, CLPT does a very good job at modeling the configuration space $\xi(r)$, it is known to perform less well in reconstructing clustering in the Fourier space \citep{Vlah:2014nta}. Expanding the resummed exponent in (\ref{PkXfinal})
and performing the resulting integrals gives the Eulerian one-loop Standard Perturbation Theory (SPT) power spectrum for biased tracers  in GR \citep{Matsubara:2008wx},
\begin{eqnarray}
\label{PkXSPTGR}
P^{SPT}_{X}(k) &=&  \left(1-k^2 \sigma_L^2\right)\left(1+b_1\right)^2P_{L}(k) + \frac{9}{98}Q_1(k) + \frac{3}{7}Q_2(k)  \nonumber \\
&&+ \frac{1}{2}Q_3(k)+ b_1\left(\frac{6}{7}Q_5(k)+2Q_7(k)\right) + b_2\left(\frac{3}{7}Q_8(k)+Q_9(k)\right)\nonumber\\
&& + b_1^2 \left(Q_9(k)+Q_{11}(k)\right) + 2b_1b_2Q_{12}(k)  + \frac{1}{2}b_2^2 Q_{13}(k) \nonumber\\
&&+ \frac{6}{7}\left(1+b_1\right)^2\left[R_1(k)+R_2(k)\right] -\frac{8}{21}\left(1+b_1\right)R_1(k),
\end{eqnarray}
where 
\begin{equation}\label{sigmalin}
\sigma_L^2=\frac{1}{6 \pi^2}\int_0^{\infty}dkP_L(k),
\end{equation}
 is the 1D variance of the Lagrangian displacement and the functions $Q_n$ and $R_n$ 
were defined in \citep{Matsubara:2008wx} for GR. 
The SPT power spectrum has been shown to follow the power spectrum much better than the CLPT prediction in GR \citep{Vlah:2014nta}. 

The correlation function obtained from Fourier transforming  (\ref{PkXSPTGR}) is, unfortunately, known to be ill-behaved \citep{Matsubara:2007wj}. However, if one performs an alternative resummation proposed in \citep{Matsubara:2007wj,Matsubara:2008wx}, 
known as Lagrangian Resummation Theory (LRT), the resulting power spectrum, 
\begin{eqnarray}
\label{PkXLRTGR}
P^{LRT}_{X}(k) &=& e^{-k^2 \sigma_L^2} \Biggl[\left(1+b_1\right)^2P_{L}(k) + \frac{9}{98}Q_1(k) + \frac{3}{7}Q_2(k) + \frac{1}{2}Q_3(k)\nonumber  \\
&&+ b_1\left(\frac{6}{7}Q_5(k)+2Q_7(k)\right) + b_2\left(\frac{3}{7}Q_8(k)+Q_9(k)\right)\nonumber\\
&& + b_1^2 \left(Q_9(k)+Q_{11}(k)\right) + 2b_1b_2Q_{12}(k)\nonumber \\
&& + \frac{1}{2}b_2^2 Q_{13}(k)+ \frac{6}{7}\left(1+b_1\right)^2\left[R_1(k)+R_2(k)\right] \nonumber \\
&&-\frac{8}{21}\left(1+b_1\right)R_1(k) \Biggr],
\end{eqnarray}
which differs from (\ref{PkXSPTGR}) only by the exponential prefactor, can be Fourier transformed to the configuration space and is found to characterize the BAO scales well for both dark matter \citep{Matsubara:2007wj,Vlah:2014nta} and biased tracers \citep{Matsubara:2008wx,2013MNRAS.429.1674C}. It  decays sharply for large values of $k$ however.

\subsubsection{Calculation of bias parameters in GR}\label{sec:biasparametersGR}

In this section, we present an analytical model for the calculation of the bias parameters (\ref{biasfn}) in GR, which will be extended to include MG in section \ref{sec:biasparameters}. It should be noted though, that even in the complete absence of an analytical model, one or both of the bias parameters in CLPT can be treated as free parameters, to be fitted over simulations, as for example done in \citep{2013MNRAS.429.1674C,doi:10.1111/j.1365-2966.2011.19379.x}. With regards to analytical models for bias, arguably the most popular one is the halo approach \citep{Mo:1995cs,Mo:1996cn,Sheth:1998xe,Sheth:1999mn,Ma:2000ik,Peacock:2000qk,Scoccimarro:2000gm,Cooray:2002dia}, that is based on the extended Press-Schechter (PS) formalism \citep{1974ApJ...187..425P,1991ApJ...379..440B}, in combination with the Peak-Background Split (PBS) approach \citep{1984ApJ...284L...9K}. A discussion of the accuracy of such approaches can be found in \citep{2010MNRAS.402..589M,Hoffmann:2015mma}. In what follows, we briefly summarize the main ingredients of this prescription in GR. 

Let $M_0$ be the mass of a collapsed region at a redshift of interest $z$, that is enclosed in a Lagrangian region of radius $R_0$, which, given the mean matter density $\rho_{m0}$, will be given by
\begin{equation}\label{radiusdef}
R_0 = \left(\frac{3 M_0}{4 \pi \rho_{m0}} \right)^{\frac{1}{3}}.
\end{equation}
The variance of matter density fluctuations in this region is,
\begin{equation}\label{variance}
\sigma^2(M_0) = \int \frac{dk k^2}{2 \pi^2} W^2\left(k R_0\right) P_L(k,z=0),
\end{equation}
with $P_L(k,z=0)$ the linear matter power spectrum evaluated today and $W\left(k R_0\right)$ the top-hat smoothing Kernel,
\begin{equation}\label{tophat}
W\left(k R_0\right) = \frac{3\left[\sin(k R_0)-k R_0\cos(k R_0)\right]}{\left(k R_0\right)^3}.
\end{equation}
For GR, density perturbations are evolved in time, relative to present time using the GR linear growth factor $D(z)$.

Based on the PS theory and its variants \citep{1974ApJ...187..425P,1991ApJ...379..440B}, the comoving mean number density of halos per logarithmic mass bin $d\ln M$, $\bar{n}_h$, can be analytically modeled as:
\begin{equation}\label{PSfunction}
\bar{n}_h(M) = \frac{\partial^2\bar{N}_h}{\partial V \partial \ln M} = \frac{\bar{\rho}_m}{M}\nu_c(M)f\left[\nu_c(M)\right] \frac{d\ln \nu_c(M)}{dM}, 
\end{equation}
where $\bar{N}_h$ is the mean number of halos with mass $M$, in a bin of width $dM$, enclosed in a comoving volume $V$.

The quantity $\nu_c(M)$, the peak significance, is given by 
\begin{equation}\label{peak}
\nu_c(M)=\frac{\delta_{cr}}{\sigma(M,z)}=\frac{\delta_{cr}}{D(z)\sigma(M)}. 
\end{equation}
where $D(z)$ is the linear growth factor at the time of collapse $z$, normalized so that $D(z=0)=1$. 
In (\ref{peak}), $\sigma(M,z)=D(z)\sigma(M)$ is the variance at redshift $z$, with $\sigma(M)$ the variance ($\ref{variance}$) evaluated today and $\delta_{cr}$ is the critical overdensity for collapse at redshift $z$. For an Einstein De-Sitter (EDS) cosmology, the latter is always $\delta_{cr}=1.686$, which turns out to be a very good approximation for $\Lambda$CDM cosmologies and will be adopted here. 

$f\left[\nu_c(M)\right]$ is the multiplicity function, that in the original PS theory is given by:
\begin{equation}\label{psmult}
\nu_cf\left[\nu_c\right]=\sqrt{\frac{2}{\pi}}\nu_ce^{\frac{-\nu_c^2}{2}}.
\end{equation}
The prescription (\ref{PSfunction}), often referred to as the universal mass function, is exact in an EDS universe with a power law power spectrum. While (\ref{psmult}) has been used to describe the halo mass function for a broad range of cosmologies, it  lacks the necessary accuracy for precision predictions. For this reason, Sheth and Tormen (ST) \citep{Sheth:1999mn},  introduced an alternative function:
\begin{equation}\label{stmult}
\nu_cf\left[\nu_c\right]=\sqrt{\frac{2}{\pi}}A(p)\left[1+\frac{1}{(q \nu_c^2)^p}\right]\sqrt{q}\nu_ce^{\frac{-q\nu_c^2}{2}},
\end{equation}
where $A(p)=\left[1+\pi^{-\frac{1}{2}}2^{-p}\Gamma(0.5-p)\right]^{-1}$ and $q,p$ are free parameters that can be fitted over N-body simulations. The best fit pair was initially proposed to be $(q,p)=(0.707,0.3)$ which was later updated to $(q,p)=(0.75,0.3)$. These are considered to be the ``standard" ST parameters \citep{Sheth:1999mn,Sheth:1999su}. For the PS function,  $q=1,p=0$. 

Based on the PBS argument \citep{1984ApJ...284L...9K}, a large-wavelength  density perturbation $\Delta$ (that is effectively  constant on small scales) has the same effect on the formation of biased tracers as a modification to the mean background density by this offset. If by $\bar{n}_h(M, \Delta)$ we denote the halo mass function's response to such a perturbation, also sometimes called the conditional mass function, then the fractional overdensity of halos will be given by \citep{Mo:1995cs,Mo:1996cn,Sheth:1998xe}:
\begin{equation}\label{halodensity}
 1 + \delta_h(M)= \frac{\bar{n}_h(M,\Delta)}{\bar{n}_h(M,0)}, 
\end{equation}
where $\bar{n}_h(M,0)=\bar{n}_h(M)$, is the standard, unconditional halo mass function. It also worth noticing that equation (\ref{halodensity}) also defines $F\left[\delta_R(\bold{q})\right]$, through (\ref{biasF}). The Lagrangian bias of order $n$, is then given by
\begin{equation}\label{biasrig}
b_{n}^L(M)= \frac{1}{\bar{n}_h(M,0)}\frac{d^n\bar{n}_h(M,\Delta)}{d \Delta^n}\Biggr|_{\substack{\Delta=0}}, 
\end{equation}
where the time argument in the above is assumed and omitted for simplicity. Equation (\ref{biasrig}) is the rigorous definition of the bias parameters, that is exact even in the absence of an analytical description for the halo mass function and can be calculated numerically, for example employing separate universe simulations \citep{PhysRevD.88.023515}. In the presence of a universal mass function, the conditional mass function is given by the same expression (\ref{PSfunction}), but with the modified peak significance $\nu'_c(M)=\frac{\delta_{cr}-\Delta}{D(z)\sigma(M)}$, in which case the the bias factors are easily calculated \citep{Sheth:1998xe} by (\ref{halodensity}), as: 
\begin{equation}\label{biasgeneral}
b_{n}^L(M)= \frac{(-1)^n}{\left(D(z)\sigma(M)\right)^n}\frac{1}{\nu_cf\left[\nu_c\right]}\frac{d^n \left(\nu_cf\left[\nu_c\right]\right)}{d\nu_c^n}, 
\end{equation}
which are the so-called PBS Lagrangian biases. Applied on the ST function (\ref{stmult}), the PBS biases are:
\begin{equation}
\begin{aligned}\label{pbsbiasGR}
b_{1}^L (M)&= \frac{-1}{\delta_{cr}}\left[q\nu_c^2-1+\frac{2p}{1+\left(q\nu_c^2\right)^p}\right], \\
b_{2}^L (M)&= \frac{1}{\delta^2_{cr}}\left[q^2\nu_c^4-3q\nu_c^2+\frac{2p\left(2q\nu_c^2+2p-1\right)}{1+\left(q\nu_c^2\right)^p}\right].\\
\end{aligned}
\end{equation}
These PBS biases are identical to the Lagrangian bias factors defined within the context of CLPT through (\ref{biasfn})  \citep{Matsubara:2008wx,PhysRevD.83.083518}.

\section{Results}
\label{sec:Results}

\subsection{Lagrangian Perturbation Theory for Biased Tracers in MG}
\label{sec:Results:LPT}

\subsubsection{2-point statistics for Biased Tracers in MG}\label{sec:twopointMG}

In section \ref{sec:twopointGR}, we showed the expressions for the calculation of 2-point statistics in CLPT and its variants, under the assumption that gravitational evolution is governed by GR. Here we explain how each of these relationships have to be modified in the case of MG theories. We note  that these results are consistent with those recently presented in \citep{Aviles:2018saf}.

The two-point statistics for biased tracers in MG are given, as in GR, by the definitions \ref{xi} and \ref{Pk}, in the configuration and Fourier space, respectively. Considering biased tracers in the Lagrangian space, through (\ref{biasF}), the overdensity of biased tracers in LPT is given by
\begin{equation}\label{delJacXMG}
\delta_X(\bold{x},t) = \int d^3q F\left[\delta_R(\bold{q})\right]\delta_{D}\left[\bold{x}-\bold{q}-\bold{\Psi}(\bold{q},t)\right] -1,
\end{equation}
where we used density conservation. The above equation is similar to (\ref{delJacX}) for GR, but differs in that the Lagrangian field $\bold{\Psi}(\bold{q})$ follows the MG evolution presented in Section \ref{LPTdarkmatter}. In particular, if we work in terms of the Fourier transform of $\bold{\Psi}(\bold{q})$, labeled as $\tilde{\bold{\Psi}}(\bold{p})$, the $n^{th}$ order LPT solutions in MG will be given by, 
\begin{eqnarray}
\label{psifourmain}
\tilde{\Psi}^{(n)}_j(\bold{p}) &=& \frac{i}{n!}\int\frac{d^3p_1}{(2\pi)^3}..\frac{d^3p_{n}}{(2\pi)^3} \delta_D^3\left(\sum_{j=1}^{n}p_j-p\right) \nonumber \\
&& \times \ L_j^{(n)}(\bold{p}_1,..,\bold{p}_n)\tilde{\delta}_{L}(\bold{p}_1)..\tilde{\delta}_{L}(\bold{p}_n), 
\end{eqnarray}
where $\tilde{\delta}_{L}(\bold{p}_n)$ are the linear-density Fourier transformed fields at the time of evaluation and the Kernels $ L_j^{(n)}(\bold{p}_1,..,\bold{p}_n)$ are given by \citep{Aviles:2017aor}:
\begin{equation}
\begin{aligned}\label{LPTKernelsmain}
L_j^{(1)}(\bold{p}) &= \frac{p^j}{p^2}, \\
L_j^{(2)}(\bold{p}_1,\bold{p}_2) &= \frac{p^j}{p^2} \frac{D^{(2)}(\bold{p_1},\bold{p_2})}{D^{(1)}(p_1)D^{(1)}(p_2)},  \\
 (L_j^{(3)})_{sym}(\bold{p}_1,\bold{p}_2,\bold{p}_3) &= i\frac{p^j}{p^2} \frac{D^{(3)}_{sym}(\bold{p_1},\bold{p_2},\bold{p}_3)}{D^{(1)}(p_1)D^{(1)}(p_2)D^{(1)}(p_3)}.
 \end{aligned}
\end{equation}
The MG growth factors in (\ref{LPTKernelsmain}), are the ones given by (\ref{growth1st}), (\ref{eq:secgrowth}) and (\ref{D3MG}). Plugging the overdensity (\ref{delJacXMG}) into (\ref{xi}) and (\ref{Pk}) and working as in (\ref{xiX})-(\ref{eq:correl}), we arrive at equations (\ref{xiXfinal}) and (\ref{PkXfinal}), that give the 2-point statistics for biased tracers in MG and depend on the MG Lagrangian correlators
\begin{equation}
\begin{aligned}\label{eq:correlMG}
\sigma_R^2 &= \langle \delta^2 \rangle_c \\
\xi_L(\vec{q}) &= \langle \delta_1 \delta_2 \rangle_c,  \\
A_{ij}^{m n}(\vec{q}) &= \langle \delta_i^{m} \delta_j^{n} \Delta_i \Delta_j\rangle_c,  \\ 
W_{ijk}^{m n}(\vec{q}) &= \langle \delta_i^{m} \delta_j^{n} \Delta_i \Delta_j \Delta_k\rangle_c,  \\ 
U_{i}^{m n}(\vec{q}) &= \langle \delta_1^{m} \delta_2^{n} \Delta_i \rangle_c. 
\end{aligned}
\end{equation}
For the MG correlators (\ref{eq:correlMG}), we use the same definition and index structure as in GR, but these functions differ from their GR counterparts, because the quantities in the cumulants follow the MG LPT solutions (\ref{LPTKernelsmain}). In particular, plugging (\ref{psifourmain}) and (\ref{LPTKernelsmain}) into (\ref{eq:correlMG}), we get the Lagrangian correlators in MG:
\begin{eqnarray}
\label{qfuncsmain}
V_1^{(112)}(q)&=&\frac{1}{2\pi^2}\int \frac{dk}{k}  \left(-\frac{3}{7}\right)\left[R_1(k)\right]_{MG} j_1(kq), \nonumber\\
V_3^{(112)}(q)&=&\frac{1}{2\pi^2}\int \frac{dk}{k}  \left(-\frac{3}{7}\right)\left[Q_1(k)\right]_{MG} j_1(kq), \nonumber\\
S^{(112)}(q)&=&\frac{3}{14\pi^2}\int \frac{dk}{k} \left[2\left[R_1\right]_{MG}+4R_2 +\left[Q_1\right]_{MG} +2Q_2\right]\frac{j_2(kq)}{kq}, \nonumber\\
T^{(112)}(q)&=&\frac{-3}{14\pi^2}\int \frac{dk}{k} \left[2\left[R_1\right]_{MG}+4R_2 +\left[Q_1\right]_{MG} +2Q_2\right]j_3(kq), \nonumber\\
U^{(1)}(q)&=&\frac{1}{2\pi^2}\int dk k \left(-1\right)P_L(k) j_1(kq), \nonumber\\
U^{(3)}(q)&=&\frac{1}{2\pi^2}\int dk k \left(-\frac{5}{21}\right)R_1(k) j_1(kq), \nonumber\\
U_{20}^{(2)}(q)&=&\frac{1}{2\pi^2}\int dk k \left(-\frac{6}{7}\right)Q_8(k) j_1(kq), \nonumber\\
U_{11}^{(2)}(q)&=&\frac{1}{2\pi^2}\int dk k \left(-\frac{6}{7}\right)\left[R_1(k)+R_2(k)\right]_{MG} j_1(kq), \nonumber\\
X_{10}^{(12)}(q)&=&\frac{1}{2\pi^2}\int dk \frac{1}{14}\Biggl(2\left(\left[R_1\right]_{MG}-R_2(k)\right) +3\left[R_1\right]_{MG} j_0(kq) \nonumber\\
-&3&\left[ \left[R_1\right]_{MG} + 2R_2+2\left[R_1(k)+R_2(k)\right]_{MG} +2Q_5 \right]\frac{j_1(kq)}{kq} \Biggr), \nonumber\\
Y_{10}^{(12)}(q)&=&\frac{1}{2\pi^2}\int dk \left(-\frac{3}{14}\right)\Biggl( \left[R_1\right]_{MG} + 2R_2  \nonumber\\
&& +2\left[R_1(k)+R_2(k)\right]_{MG} +2Q_5\Biggr)\times \left[j_0(kq)-3\frac{j_1(kq)}{kq}\right] , \nonumber\\
X(q)&=&\frac{1}{2\pi^2}\int dk\ a(k) \left[\frac{2}{3}-2\frac{j_1(kq)}{kq}\right], \nonumber\\
Y(q)&=&\frac{1}{2\pi^2}\int dk\ a(k) \left[-2j_0(kq)+6\frac{j_1(kq)}{kq}\right], 
\end{eqnarray}
where we additionally performed the decompositions
\begin{equation}
\begin{aligned}
A_{ij}^{mn}(q)&=X_{mn}(q)\delta_{ij}+Y_{mn}\hat{q}_{i}\hat{q}_{j}\\
W_{ijk}(q)&=V_{1}(q)\hat{q}_{i}\delta_{jk}+V_{2}(q)\hat{q}_{j}\delta_{ki}+V_{3}(q)\hat{q}_{k}\delta_{ij}+T(q)\hat{q}_{i}\hat{q}_{j}\hat{q}_{k}
\end{aligned}
\end{equation}
and defined $a(k)=P_L(k)+\frac{9}{98}Q_1(k)+\frac{10}{21}R_1(k)$.
The functions $Q_n(k)$ and $R_n(k)$ that appear in the r.h.s of (\ref{qfuncsmain}), are defined, as in GR, to be:
\begin{eqnarray}
Q_n(k) &=& \frac{k^3}{4 \pi^2 }\int_0^{\infty} drP_L(kr) \nonumber \\
&& \times \int_{-1}^{1}dx P_L(k\sqrt{1+r^2-2rx})\tilde{Q}_n(r,x)  \nonumber
\\
\label{QRformmain}
R_n(k) &= &\frac{k^3}{4 \pi^2 }P_L(k) \int_0^{\infty} drP_L(kr)\int_{-1}^{1}dx \tilde{R}_n(r,x).
\end{eqnarray}
The scale and redshift dependency of the growth factors alters the evaluation of these expressions relative to GR:
\begin{equation}
\begin{aligned}\label{Qsmain}
&\tilde{Q}_1 = r^2\left(\bar{D}_a^{(2)}-\bar{D}_b^{(2)}\frac{x^2+r^2-2rx}{1+r^2-2rx}+\bar{D}_{FL}^{(2)}-\bar{D}_{\delta \mathcal{I}}^{(2)}\right)^2 \\
&\tilde{Q}_2 = \frac{rx(1-rx)}{1+r^2-2rx}\left(\bar{D}_a^{(2)} - \bar{D}_b^{(2)}\frac{x^2+r^2-2rx}{1+r^2-2rx}+\bar{D}_{FL}^{(2)}-\bar{D}_{\delta \mathcal{I}}^{(2)}\right) \\
&\tilde{Q}_3 = \frac{x^2(1-rx)^2}{(1+r^2-2rx)^2} \\
&\tilde{Q}_5 = rx\left(\bar{D}_a^{(2)}-\bar{D}_b^{(2)}\frac{x^2+r^2-2rx}{1+r^2-2rx}+\bar{D}_{FL}^{(2)}-\bar{D}_{\delta \mathcal{I}}^{(2)}\right) \\
&\tilde{Q}_7 = \frac{x^2(1-rx)}{(1+r^2-2rx)} \\
&\tilde{Q}_8 = r^2\left(\bar{D}_a^{(2)}-\bar{D}_b^{(2)}\frac{x^2+r^2-2rx}{1+r^2-2rx}+\bar{D}_{FL}^{(2)}-\bar{D}_{\delta \mathcal{I}}^{(2)}\right) \\
&\tilde{Q}_9 = \frac{rx(1-rx)}{1+r^2-2rx} \\
&\tilde{Q}_{11} = x^2 \\
&\tilde{Q}_{12} = rx \\
&\tilde{Q}_{13} = r^2 \\
&\left[\tilde{Q}_1\right]_{MG} = \frac{r^2 (1-x^2)}{1+r^2-2rx}\left(\bar{D}_a^{(2)}-\bar{D}_b^{(2)}\frac{x^2+r^2-2rx}{1+r^2-2rx}+\bar{D}_{FL}^{(2)}-\bar{D}_{\delta \mathcal{I}}^{(2)}\right).\\ 
\end{aligned}
\end{equation}
and
\begin{equation}
\begin{aligned}\label{Rsmain}
&\tilde{R}_{1} = \frac{21}{10} r^2 \frac{D^{(3)}_{sym}(\bold{k},-\bold{p},\bold{p})}{D^{(1)}(\bold{k})\left(D^{(1)}(\bold{p})\right)^2} \\
&\tilde{R}_{2} = \frac{rx(1-rx)}{1+r^2-2rx}\left(\bar{D}_a^{(2)}-\bar{D}_b^{(2)}x^2+\bar{D}_{FL}^{(2)}-\bar{D}_{\delta \mathcal{I}}^{(2)}\right) \\
&\left[\tilde{R_1}(k)+\tilde{R_2}(k)\right]_{MG} = \frac{r^2(1-rx)}{1+r^2-2rx}\left(\bar{D}_a^{(2)}-\bar{D}_b^{(2)}x^2+\bar{D}_{FL}^{(2)}-\bar{D}_{\delta \mathcal{I}}^{(2)}\right)\\
&\left[\tilde{R}_1\right]_{MG} = \frac{r^2(1-x^2)}{1+r^2-2rx}\left(\bar{D}_a^{(2)}-\bar{D}_b^{(2)}x^2+\bar{D}_{FL}^{(2)}-\bar{D}_{\delta \mathcal{I}}^{(2)}\right).\\ 
\end{aligned}
\end{equation}
The functions $Q_1$-$Q_{13}$ in (\ref{Qsmain}) and $R_1$, $R_{2}$ in (\ref{Rsmain}) differ from GR as they depend on the MG growth factors (\ref{growth1st}), (\ref{eq:secgrowth}) and (\ref{D3MG}). In addition, the functions $\left[Q_1\right]_{MG}$, $\left[R_1\right]_{MG}$ and $\left[R_1(k)+R_2(k)\right]_{MG}$ are new ones that arise in MG. In the GR limit, that corresponds to $\bar{D}_a^{(2)}=\bar{D}_b^{(2)}=1$ and $\bar{D}_{FL}^{(2)}=\bar{D}_{\delta \mathcal{I}}^{(2)}=0$, $\left[Q_1\right]_{MG}=Q_1$, $\left[R_1\right]_{MG}=R_1$, $\left[R_1(k)+R_2(k)\right]_{MG}=R_1(k)+R_2(k)$ and the functions $Q_n$ and $R_n$ reduce to their GR expressions in \citep{Matsubara:2007wj}. In that limit, furthermore, the correlators (\ref{qfuncsmain}) recover their GR forms presented in \citep{2013MNRAS.429.1674C}. The derivations of the above, along with a more detailed discussion, are presented in Appendices \ref{polyderiv} and \ref{correlderiv}.

\begin{figure}[!tb]
\bc
{\includegraphics[width=0.5\textwidth]{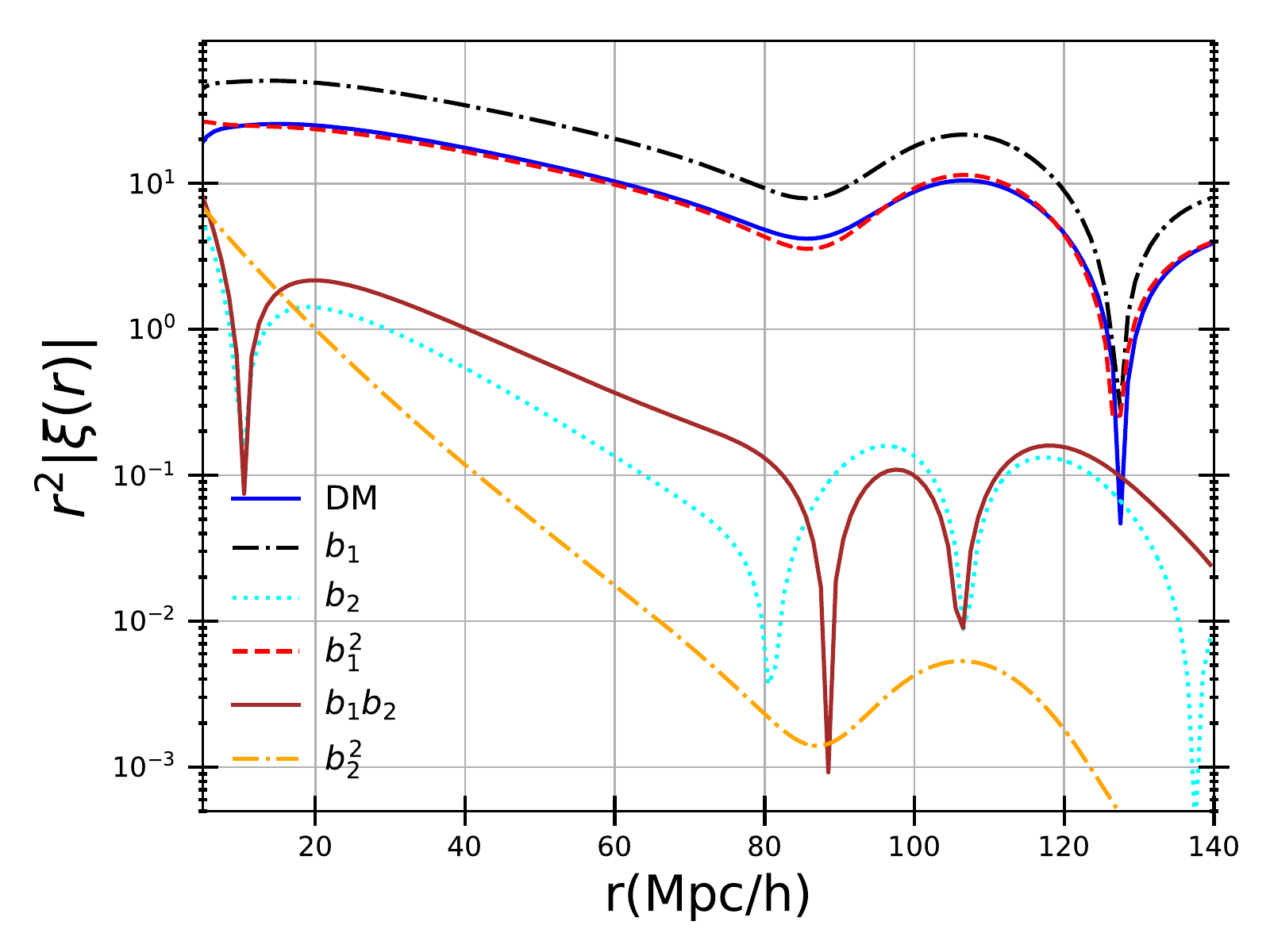}

}
\caption{Contributions to the CLPT correlation function prediction, $\xi$, given in (\ref{xiXfinal}), for the F4 model at z=0.5, by the various terms in the expansion: dark matter (no bias prefactors) [blue solid], $b_1$ term [black dash-dot],  $b_2$ term [cyan dotted], $b_1^2$ term [red dash], $b_2^2$ term [orange dash-dot] and $b_1b_2$ term [brown solid].}
\label{CLPTstuff}
\ec
\end{figure}

In Figure \ref{CLPTstuff}, we show the contributions of the different terms in (\ref{xiXfinal}) as a function of $r$ for the F4 model (which predicts the largest deviations from GR), evaluated at $z=0.5$.

As in the GR case, proceeding to expand the resummed exponent in (\ref{PkXfinal}) and performing the resulting integrals, as shown in Appendix \ref{spectraderiv}, gives the equivalent of the Eulerian one-loop SPT power spectrum for biased tracers in MG:
\begin{equation}
\begin{aligned}\label{PkXSPT}
&P^{SPT}_{X}(k) =  \\
&  \left(1-k^2 \sigma_L^2\right)\left(1+b_1\right)^2P_{L}(k) + \frac{9}{98}Q_1(k) + \frac{3}{7}Q_2(k) + \frac{1}{2}Q_3(k)  \\
&+ b_1\left(\frac{6}{7}Q_5(k)+2Q_7(k)\right) + b_2\left(\frac{3}{7}Q_8(k)+Q_9(k)\right)\\
& + b_1^2 \left(Q_9(k)+Q_{11}(k)\right) + 2b_1b_2Q_{12}(k) \\
& + \frac{1}{2}b_2^2 Q_13(k)+ \frac{6}{7}\left(b_1^2+b_1\right)\left[R_1(k)+R_2(k)\right]_{MG} \\
&+ \frac{6}{7}\left(1+b_1\right)\left[R_1(k)+R_2(k)\right] -\frac{8}{21}\left(1+b_1\right)R_1(k).
\end{aligned}
\end{equation}

Equation (\ref{PkXSPT}), that depends on the functions (\ref{QRformmain}), is the MG version of (\ref{PkXSPTGR}). Appendix \ref{polyderiv} provides a more thorough discussion. While we refer to this as the ``SPT" expression, we note  that, unlike in GR, where equation (\ref{PkXSPTGR}) has been shown to be identical to the SPT expression, in MG, additional terms that appear when transforming the 
Klein-Gordon equation from Eulerian to Lagrangian coordinates need to be considered to show the equivalence \cite{Aviles:2018saf}.

The LRT power spectrum for MG theories is obtained, just like in GR, by keeping the zero-lag term exponentiated:
\begin{eqnarray}
\label{PkXLRT}
P^{LRT}_{X}(k) &=&  e^{-k^2 \sigma_L^2} \Biggl[\left(1+b_1\right)^2P_{L}(k) + \frac{9}{98}Q_1(k) + \frac{3}{7}Q_2(k) + \frac{1}{2}Q_3(k)  \nonumber \\
&&+ b_1\left(\frac{6}{7}Q_5(k)+2Q_7(k)\right) + b_2\left(\frac{3}{7}Q_8(k)+Q_9(k)\right)\nonumber\\
&& + b_1^2 \left(Q_9(k)+Q_{11}(k)\right) + 2b_1b_2Q_{12}(k)\nonumber \\
&& + \frac{1}{2}b_2^2 Q_13(k)+ \frac{6}{7}\left(b_1^2+b_1\right)\left[R_1(k)+R_2(k)\right]_{MG}\nonumber \\
&&+ \frac{6}{7}\left(1+b_1\right)\left[R_1(k)+R_2(k)\right] -\frac{8}{21}\left(1+b_1\right)R_1(k) \Biggr].
\end{eqnarray}
The derivation is discussed in Appendix \ref{spectraderiv}. The configuration space counterpart, $\xi^{LRT}_{X}(r)$, is obtained by Fourier transforming (\ref{PkXLRT}),
\begin{equation}\label{xiXLRT}
\begin{aligned}
 \xi^{LRT}_{X}(r) &=\int \frac{d^3k}{\left(2\pi\right)^3} e^{i\bold{k}\cdot\bold{r}}P^{LRT}_{X}(k)  \\
& = \int \frac{dk}{2 \pi^2} k^2P^{LRT}_{X}(k)j_0(kr), \\
\end{aligned}
\end{equation}
with $j_0(kr)$ the zeroth-order Bessel function.

To evaluate the expressions (\ref{xiXfinal}), (\ref{PkXfinal}), (\ref{PkXSPT}) and (\ref{PkXLRT}) we modified a publicly available code released by \citep{Vlah:2016bcl} in \footnote{\url{https://github.com/martinjameswhite/CLEFT_GSM}}, 
that efficiently performs the 2D integrals in (\ref{xiXfinal}), (\ref{PkXfinal}) using Haskel transformations, as well those in (\ref{qfuncsmain}) and (\ref{QRformmain}). On top of the functions $Q_1$-$Q_{13}$ and $R_1$, $R_{2}$, the code was extended to evaluate the new functions  $\left[Q_1\right]_{MG}$, $\left[R_1\right]_{MG}$ and $\left[R_1(k)+R_2(k)\right]_{MG}$, as well as the modified correlators (\ref{qfuncsmain}). We make this code publicly available in \footnote{\url{https://github.com/CornellCosmology/bias_MG_LPT_products}}. Our modified version accepts the modified gravity model growth factors, $D_{MG}(k,z)$ as input along with the linear power spectrum given by:
\begin{equation}\label{Plin}
P_{MG}^{L}(k,z) = \left(\frac{D^1_{MG}(k,z)}{D^1_{GR}(k,0)}\right)^2P_{GR}^{L}(k,0).
\end{equation}
The linear power spectrum for the background $\Lambda$CDM cosmology is calculated using the publicly available code CAMB \citep{Lewis:1999bs}. 

After calculating the necessary MG growth factors using our $\sc{Mathematica}$ notebook, we feed our modified version of the code with tabulated values of the growth factors for the various values of $k$, $r$ and $x$ needed at a given cosmological redshift $z$. The PYTHON module computes the various $Q_n(k)$ and $R_n(k)$ functions through equations (\ref{Qsmain}) and (\ref{Rsmain}), which are then used to calculate the various components of the CLPT power spectrum $P_{X}(k)$. The k functions can then be simply combined to give the SPT and LRT power spectra, by equations (\ref{PkXSPT}) and (\ref{PkXLRT}), respectively. Finally, the modified C$++$ counterpart follows a similar procedure to compute the configuration space two-point correlation function given by CLPT, through (\ref{xiXfinal}).
The procedure is explained in more detail in the Appendix \ref{correlderiv}.

We finish this section by noting that, even though we restrict our model to the case of a local Lagrangian bias, our framework can be extended to include a curvature bias and/or corrections from EFT, as in \citep{Vlah:2016bcl}. For modified gravity theories with scale-dependent growth, a general expansion bias is not possible in principle, though for some theories, such as the $f(R)$, the effects of the fifth force can be perturbatively absorbed in terms of higher-order derivatives \citep{Desjacques:2016bnm, Aviles:2018saf}.

\subsubsection{Calculation of bias parameters in MG}\label{sec:biasparameters}

In this section we turn to the final necessary ingredient to describe biased tracers, an analytical model for the calculation of the bias parameters in MG.

Central to the GR derivation in Section \ref{sec:biasparametersGR}, is the assumption that spherical collapse is independent of the exterior spacetime, which, in the case of GR evolution, is given by Birkhoff's theorem. In MG theories, however, the additional degree of freedom violates Birkhoff's theorem, which will have important consequences for the modeling of the halo mass function, as well as on its response to an external density perturbation. 

The Press-Schechter formalism (\ref{PSfunction}) relies upon the assumption that the linearly evolved critical overdensity, $\delta_{cr}$, is always constant at the time of collapse, for example $\delta_{cr}=1.686$ for an EDS evolution. This is not the case for MG
due to the presence of the additional scalar field that generates the fifth forces. Following \citep{doi:10.1111/j.1365-2966.2011.20404.x,Lombriser:2013wta}, if we define $1+\delta_h = y_h^{-3}$, the evolution of a spherically symmetric halo density perturbation,  will be given by: 
\begin{equation}\label{collapseh}
y''_h -\left(2-\frac{3}{2}\Omega(a)\right) y'_h +\frac{1}{2}\Omega(a)\frac{G_{eff}}{G}\left(y_h^{-3}-1\right)y_h = 0, 
\end{equation}
where $'$ denotes derivatives with respect to $\ln(a)$ and $G_{eff}$, the modified Newton's constant, is given by 
\begin{equation}
G_{eff} =  \left(1+E\right)G,
\label{eq:Geff}
\end{equation}
 
For the $n$DGP braneworld model,\citep{2010PhRvD..81f3005S}
\begin{equation}
E=  \frac{2}{3 \beta(a)}\frac{\sqrt{1+\chi^3}-1}{\chi^{-3}},
\end{equation}
where $\beta(a)$ was defined in (\ref{eq:nDGPmasses}) and
\begin{equation}\label{chinDGP}
\chi^{-3} = \frac{\Omega_m n^2}{1.10894 a^3 \beta^2(a)}\frac{y_h^3-1}{y_h^3},
\end{equation}
with the $n$DGP parameter, $n=H_0r_c$. Note that the fifth force modification does not depend on either the mass or the environment, a property of the Vainshtein mechanism, which means that the collapse barrier for this model is redshift and scale independent \citep{2010PhRvD..81f3005S,Barreira:2013xea}. Thanks to this property, the halo biases for this model can be easily calculated by the corresponding GR expressions (\ref{pbsbiasGR}), with a different value for the $\it{constant}$ threshold $\delta_{cr}$. For the background cosmology of the Group I simulations, at $z=0.5$, we integrate equation (\ref{collapseh}) and find $\delta_{cr}$ to have values (linearly extrapolated at $z=0.5$) of $\delta_{cr}=1.571$ and $\delta_{cr}=1.657$ for the N1 ($n=1$) and N5 ($n=5$) cases, respectively. 

For $f(R)$ models, for a collapsing sphere of mass $M$ and radius $R_{th}$, $E$ is given by
\begin{equation}\label{Fforce}
E = 2\beta^2\left[3\frac{\Delta R}{R_{th}}-3\left(\frac{\Delta R}{R_{th}}\right)^2+\left(\frac{\Delta R}{R_{th}}\right)^3\right],
\end{equation}
with $\beta=\frac{1}{\sqrt{6}}$. Finally, $\frac{\Delta R}{R_{th}}$ is given by
\begin{equation}
\begin{aligned}\label{Shell}
& \frac{\Delta R}{R_{th}} = \frac{|f_{R0}| a^3}{\Omega_m y_h^{-3}H_0^2 R_{th}^2} \\
& \times \left[\left(\frac{1+4\frac{\Omega_{\Lambda}}{\Omega_m}}{\left(y_{env} a\right)^{-3}+4\frac{\Omega_{\Lambda}}{\Omega_m}}\right)^{n+1} -  \left(\frac{1+4\frac{\Omega_{\Lambda}}{\Omega_m}}{\left(y_{h} a\right)^{-3}+4\frac{\Omega_{\Lambda}}{\Omega_m}}\right)^{n+1}\right]. \\
\end{aligned}
\end{equation}
Here the overdensity related to $y_h$ is embedded on a longer wavelength environment with over density, $1+\delta_{env} = y_{env}^{-3}$. In models that realize the chameleon screening mechanism, like the $f(R)$ Hu-Sawicki, only a thin shell of a massive sphere contributes to the fifth force, with a fractional thickness $\frac{\Delta R}{R_{th}}$, that causes an enhancement given by $E$. It is this factor $E$ that causes the environmental dependence of spherical collapse in MG, through (\ref{collapseh}). When set equal to zero, we recover the standard GR solution, which is the one that describes the evolution of the environment, since on such a long wavelength perturbations $G_{eff}\approx G$ and we have:
\begin{equation}\label{collapseenv}
y''_{env} -\left(2-\frac{3}{2}\Omega(a)\right) y'_{env} +\frac{1}{2}\Omega(a)\left(y_{env}^{-3}-1\right)y_{env} = 0.
\end{equation}
%

\begin{figure}[!tb]
\bc
{\includegraphics[width=0.48\textwidth]{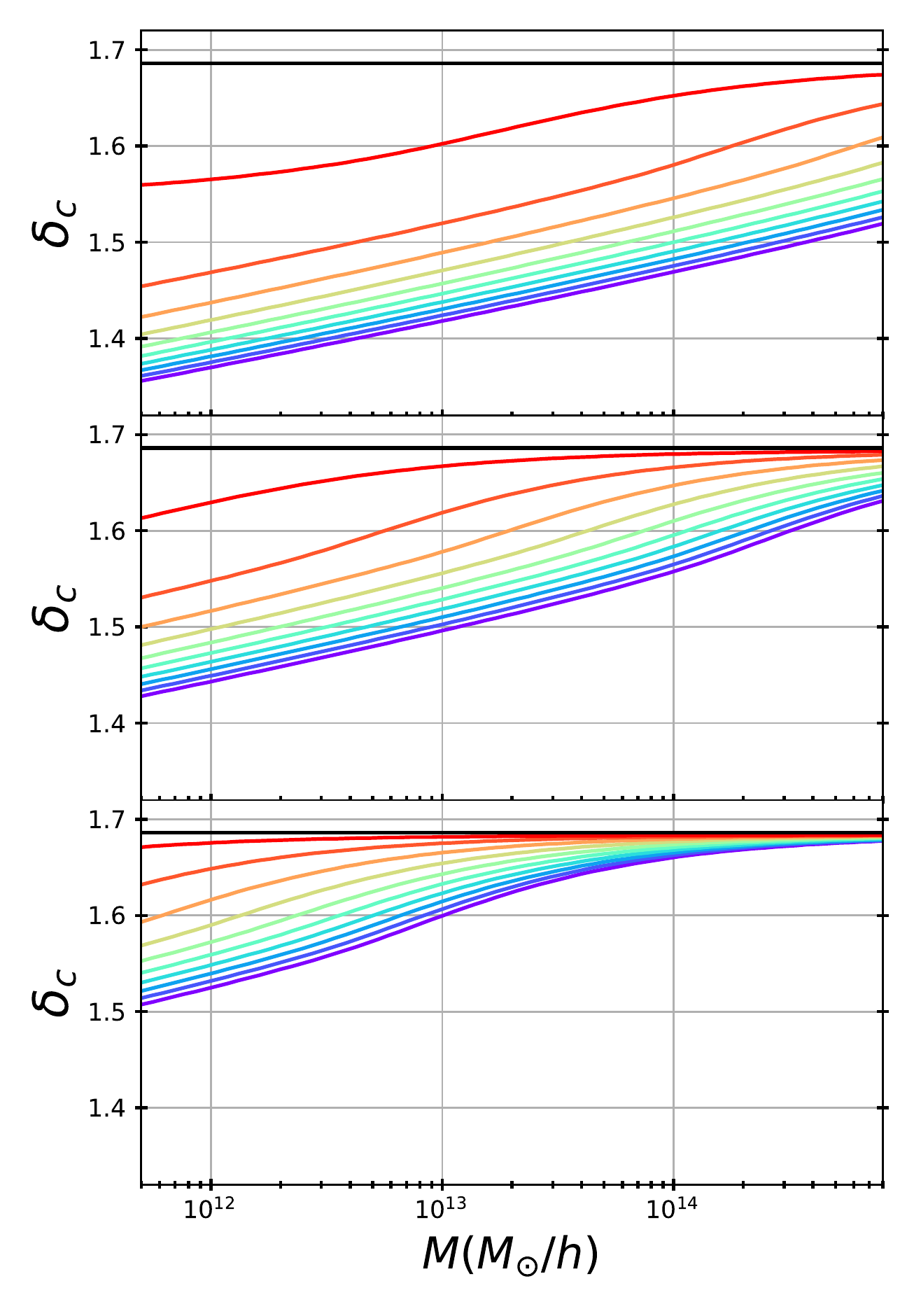}
}
\caption{Critical overdensity for collapse, $\delta_{cr}$, as a function of halo mass for the F4 [top], the F5 [middle] and the F6 [bottom] MG models at $z=0.5$ and for various environments. The different color curves correspond to the solutions for the following values of the environment overdensity, $\delta_{env}=\left[-1 (\mathrm{purple}),-0.72,-0.43,-0.15,0.13,0.42,0.7,0.98,1.27,1.55 (\mathrm{red})\right]$. The horizontal black line shows the $\delta_{cr}=1.686$ value for GR. The background cosmology is that of the Group I simulations.}
\label{collapsevsM}
\ec
\end{figure}

In light of the coupling between a collapsing halo and the background on which it evolves, as seen through (\ref{collapseh}), it is clear that in MG the collapse barrier, $\delta_{cr}$, that is constant in GR, should now be promoted to a function of both the mass $M$ and the environment overdensity $\delta_{env}$. For each choice of $M$ and $\delta_{env_i}$, equations (\ref{collapseh}) and (\ref{collapseenv}) form a system of coupled differential equations that we solve simultaneously, as in \citep{doi:10.1111/j.1365-2966.2011.20404.x,Lombriser:2013wta}. The critical overdensity is identified as the smallest value of the boundary condition $\delta_{h}(a_i)$ at the initial scale factor $a_i=0.002$, that causes a singularity i.e. signifying the onset of nonlinear collapse, at the scale factor of interest, which is then evolved to that scale factor $a$ through \citep{doi:10.1111/j.1365-2966.2011.20404.x,Lombriser:2013wta}:
\begin{equation}
\begin{aligned}\label{delhevolv}
\delta_{h}(a) = \frac{D^{(1)}(a)}{D^{(1)}(a_i)}\delta_{h}(a_i),
\end{aligned}
\end{equation}
with $D^{(1)}$ the linear GR growth factor 

In Figure \ref{collapsevsM}, we show the critical density, $\delta_{cr}(z, M,\delta_{env})$ as a function of mass, $M$, for the three $f(R)$ models at $z=0.5$, as it was obtained for a variety of environmental densities, $\delta_{env}$. Because the gravitational strength,  $G_{eff}$, is greater in the modified models this allows haloes to form more easily, translating into a lower value of $\delta_{cr}$ than the value in $\Lambda$CDM. $G_{eff}$ is scale dependent and tends towards the GR value on large scales (which would collapse into large mass haloes). GR is also recovered for highly screened models and high density environments. For this reason the critical threshold tends towards the GR value $\delta_{cr}=1.686$ for smaller values of $f_{R0}$,  increasing values of $M$ and more positive values of $\delta_{env}$ (regions with larger screening) 
as found in \citep{doi:10.1111/j.1365-2966.2011.20404.x,Lombriser:2013wta}. The 
deviations from GR become progressively less pronounced as we move from weaker (F4, top panel) to stronger (F6, lower panel) screening, as one would expect.

Having calculated the function $\delta_{cr}(z, M,\delta_{env})$ for the three $f(R)$ models, the MG halo mass function can be again given by the universal prescription (\ref{PSfunction}), but now with a modified peak significance
\begin{equation}\label{peakMG}
\nu_{cMG}(z, M,\delta_{env})=\frac{\delta_{cr}(z, M,\delta_{env})}{D^{(1)}(z)\sigma(M)}. 
\end{equation}
The growth factors in MG are scale-dependent, meaning that in equation (\ref{peakMG}) one should in principle use $D^{(1)}_{MG}(k,z)$, however, it has been shown \citep{doi:10.1111/j.1365-2966.2011.20404.x,Lombriser:2013wta}, that it is sufficient to use the GR growth factor, $D^{(1)}(z)$ to define the $\nu_c$ parameters. Modifications beyond this assumption are accounted for later in the free parameters of the halo mass function.

The dependence on the mass and the environment alters, and makes environment dependent, not only the unconditional halo mass function in MG, but also the conditional one. In the presence of a long-wavelength density perturbation, $\Delta$, the conditional halo mass function will be again described by (\ref{PSfunction}), but now with a modified $\nu'_{cMG}$ given by:
\begin{equation}
\begin{aligned}\label{peakMGprime}
\nu'_{cMG}(M,\delta_{env})=\frac{\delta_{cr}(z, M,\delta_{env}+\Delta)-\Delta}{D^{(1)}(z)\sigma(M)}.
\end{aligned}
\end{equation}
Application of the bias definition (\ref{biasrig}) in that case, gives the first and second order bias parameters 
\begin{equation}
\begin{aligned}\label{pbsbiasMG}
& b_{MG}^1(M,\delta_{env}) =  \frac{\frac{d\delta_{cr}(M,\delta_{env})}{d \delta_{env}}-1}{\delta_{cr}(M,\delta_{env})}\left[q\nu_{cMG}^2-1+\frac{2p}{1+\left(q\nu_{cMG}^2\right)^p}\right], 
\\
& b_{MG}^2(M,\delta_{env}) = \\
& \frac{\left(\frac{d\delta_{cr}(M,\delta_{env})}{d \delta_{env}}-1\right)^2}{\delta^2_{cr}(M,\delta_{env})}\left[q^2\nu_{cMG}^4-3q\nu_{cMG}^2+\frac{2p\left(2q\nu_{cMG}^2+2p-1\right)}{1+\left(q\nu_{cMG}^2\right)^p}\right]  \\
& + \frac{d^2\delta_{cr}(M,\delta_{env})}{d \delta_{env}^2}\frac{1}{\delta_{cr}(M,\delta_{env})}\left[q\nu_{cMG}^2-1+\frac{2p}{1+\left(q\nu_{cMG}^2\right)^p}\right].\\
\end{aligned}
\end{equation}
In the case of a sample of halos in a mass range of width $dM$ around a single value $M$, the conditional and unconditional mass functions need to be first averaged over the mass range,
\begin{equation}
\begin{aligned}\label{pbsbiasMGavg}
& b_{MG}^1(M,\delta_{env}) =  \\
& \frac{1}{I_{dM}}\int\left[\frac{\left(\frac{d\delta_{cr}(M,\delta_{env})}{d \delta_{env}}-1\right)}{\delta_{cr}(M,\delta_{env})}\frac{\nu_{cMG}}{M}\frac{\partial f[\nu_{cMG}]}{\partial \nu_{cMG}}\frac{d \ln \nu_{cMG}}{dM}\right]dM, \\
& b_{MG}^2(M,\delta_{env}) = \\
& \frac{1}{I_{dM}}\int\left[\frac{\left(\frac{d\delta_{cr}(M,\delta_{env})}{d \delta_{env}}-1\right)^2}{\delta^2_{cr}(M,\delta_{env})}\frac{\nu^2_{cMG}}{M}\frac{\partial^2 f[\nu_{cMG}]}{\partial \nu^2_{cMG}}\frac{d \ln \nu_{cMG}}{dM}\right]dM \\
& + \frac{1}{I_{dM}}\int\left[\frac{\frac{d^2\delta_{cr}(M,\delta_{env})}{d \delta_{env}^2}}{\delta_{cr}(M,\delta_{env})}\frac{\nu_{cMG}}{M}\frac{\partial f[\nu_{cMG}]}{\partial \nu_{cMG}}\frac{d \ln \nu_{cMG}}{dM}\right]dM, \
\end{aligned}
\end{equation}
with
\begin{equation}
I_{dM} = \int\left[\frac{f[\nu_{cMG}]}{M}\frac{d \ln \nu_{cMG}}{dM}\right]dM.
\end{equation}
The details of the peak-background split derivations, for (\ref{pbsbiasMG}), (\ref{pbsbiasMGavg}), are shown in more detail in Appendix \ref{pbsMG}. 

 We note that another popular method, as an alternative to PBS, for calculating halo biases is the excursion set approach \citep{1991ApJ...379..440B}. Here the universal halo mass function is associated with the Brownian-walk, first crossing distribution of a  collapse threshold. In the GR case, the redshift and scale independence of the collapse barrier leads to an analytical solution that recovers the common PBS biases. Given the potential scale, environment and redshift dependence that $\delta_{cr}$ has in  MG models  however, there is no analytical solution for the excursion set approach in MG and one would need to perform numerical simulations, rather than analytic prediction available for PBS, to determine the predicted biases  \citep{doi:10.1111/j.1365-2966.2011.20404.x,doi:10.1111/j.1365-2966.2012.21746.x,doi:10.1111/j.1365-2966.2012.21592.x}.

In order to make predictions to compare against simulations, given that the correlation statistics sample a distribution of environments, rather than a specific value of $\delta_{env}$, we average all environment-dependent quantities over a probability distribution for environments in which halos form and reside, defining these on a fixed scale $\zeta$ which we set to 8 Mpch/$h$  \citep{Lam:2007qw,doi:10.1111/j.1365-2966.2012.21592.x,doi:10.1111/j.1365-2966.2012.21746.x},
\begin{equation}
\begin{aligned}\label{probenv}
& p_{\zeta}(\delta_{env})= \frac{\beta^{0.5 \omega}}{\sqrt{2 \pi}}\left[1+(\omega-1)\frac{\delta_{env}}{\delta_{cr}}\right]\left(1-\frac{\delta_{env}}{\delta_{cr}}\right)^{-0.5 \omega-1}, \\
& \times \exp\left[-\frac{\beta^{\omega}}{2}\frac{\delta_{env}}{\left(1-\frac{\delta_{env}}{\delta_{cr}}\right)^{\omega}}\right],
\end{aligned}
\end{equation}
where $\beta\equiv {\left(\frac{\zeta}{8}\right)^3}/(\delta_{cr}\sigma_8^{(2/\omega)})$ with $\omega\equiv\delta_{cr}\frac{n_s+3}{3}$. 

The list of environments over which we average the other dependent quantities for the rest of this work, is $\delta_{env}=\left[-1,-0.72,-0.43,-0.15,0.13,0.42,0.7,0.98,1.27,1.55\right]$.

\begin{figure}[!tb]
\bc
{\includegraphics[width=0.425\textwidth]{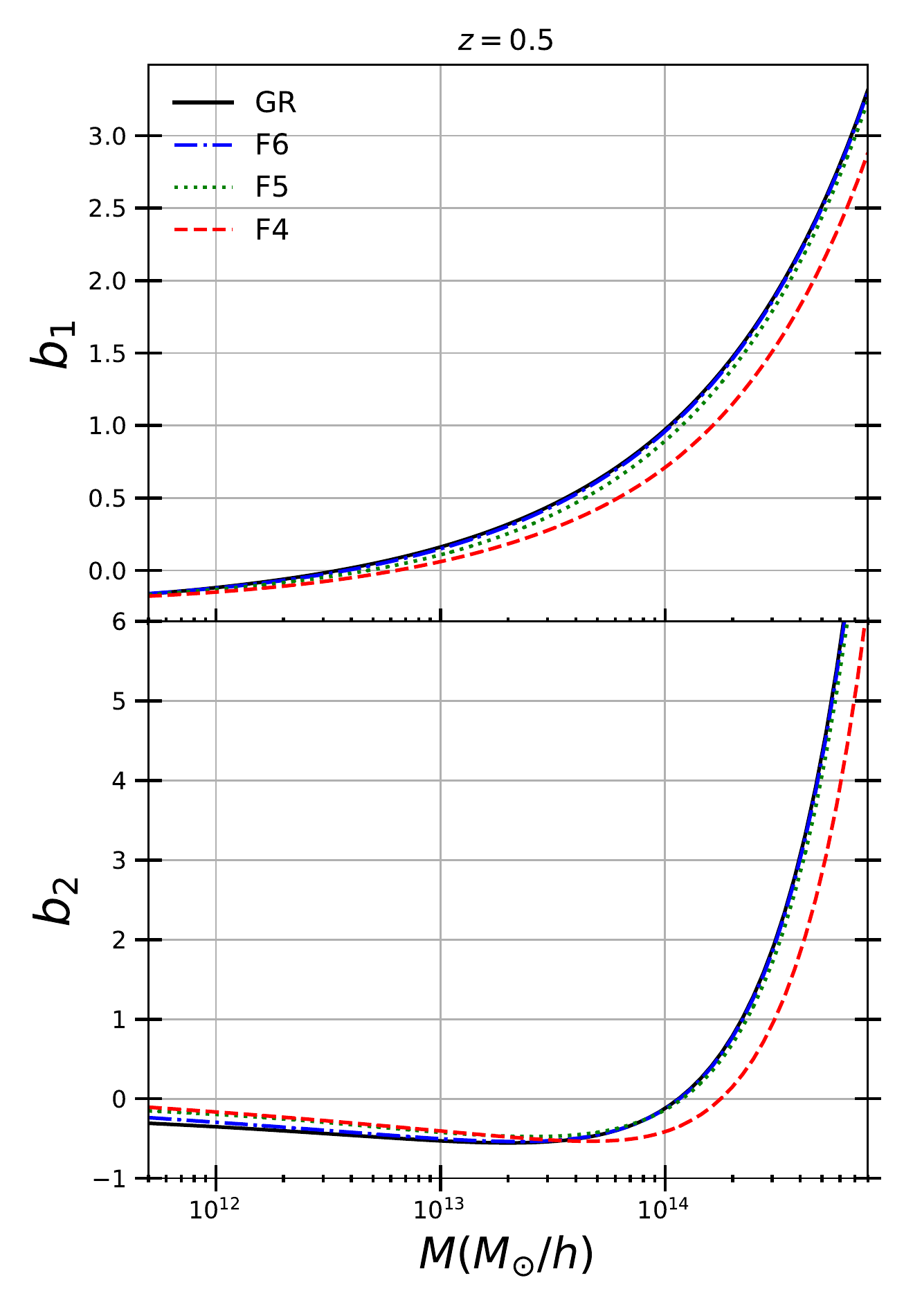}
}
\caption{First and second order Lagrangian bias factors $b_1$ [Top] and $b_2$ [Bottom], as a function of halo mass M, calculated for GR [solid black], F6 [dash-dot blue], F5 [green dotted] and F4 [red dash], through relationships (\ref{pbsbiasGR}) and (\ref{pbsbiasMG}), using the Sheth-Tormen values $(q,p)=(0.707,0.3)$ at $z=0.5$. For the 3 $f(R)$ models, the biases are the environmentally averaged.}
\label{biasvsM}
\ec
\end{figure}

In Figure \ref{biasvsM} we demonstrate the impact of the reduced $\delta_{cr}$ values in modified gravity models, separately from modifications to the halo mass function itself, by plotting the environmentally averaged bias parameters $b_{MG}^1$ and $b_{MG}^2$ for GR and three $f(R)$ models we study calculated from relations (\ref{pbsbiasGR}) and (\ref{pbsbiasMG}), while assuming  {\it the same underlying halo mass model}, with  $(q,p)=(0.707,0.3)$ for $\it{all}$ models. As described earlier, the increased gravitational strength in modified gravity models, parameterized by $G_{eff}$ in (\ref{eq:Geff}), allows haloes of a given mass to  collapse more readily, yielding a lower critical threshold values for $\delta_{cr}$ and a lower bias relative to the background (reduced $b_1$ relative to GR).  The deviations from GR are most pronounced in the models with least screening, like the $F4$ model and are suppressed in the presence of stronger screening.

To accurately predict the biases in the various modified gravity models, we also need to accurately characterize the halo mass function. The values $(q=0.707,p=0.3)$ used to characterize the ST halo mass function in GR were fixed by fitting to $\Lambda$CDM simulations \citep{Sheth:1999mn} and would not be expected to predict the mass function for modified gravity theories, given the different physics involved in the growth rate and  collapse of nonlinear structures. Since the form of the halo mass function is critical to evaluating the biases for the LPT  correlation function and power spectra predictions, we determine the best fit values for $(q,p)$ from the simulated halo mass functions  for each model in the mass ranges considered. These then uniquely determine the predictions for the biased tracer correlation and power spectra. This approach minimizes the errors that would be introduced at the outset of the LPT modeling from an inaccurate halo mass function. To do this, we evaluate the environment averaged Sheth-Tormen halo mass function (\ref{PSfunction}) for various pairs values of $(q,p)$, each using the MG prescription (\ref{peakMG}), over the distribution of environments (\ref{probenv}) and identify the pair of values that best fits the corresponding halo mass functions from the simulations, through the simple criterion that minimizes the quantity
\begin{equation}
\begin{aligned}\label{criterion}
\sum_i \Biggl\lvert\frac{n^{sim}(M_i)}{n^{ST}(M_i,q,p)}-1\Biggr\rvert,
\end{aligned}
\end{equation}
previously used to fit the halo mass function in Galileons  \citep{Barreira:2013xea,Barreira:2014zza}. In (\ref{criterion}), $i$ is the number of mass bins over which the sum is performed, which, 
can be tuned to model a narrow mass range rather than fitting the whole range with a single set of parameters.
 
\begin{figure}[!tb]
\bc
{\includegraphics[width=0.5\textwidth]{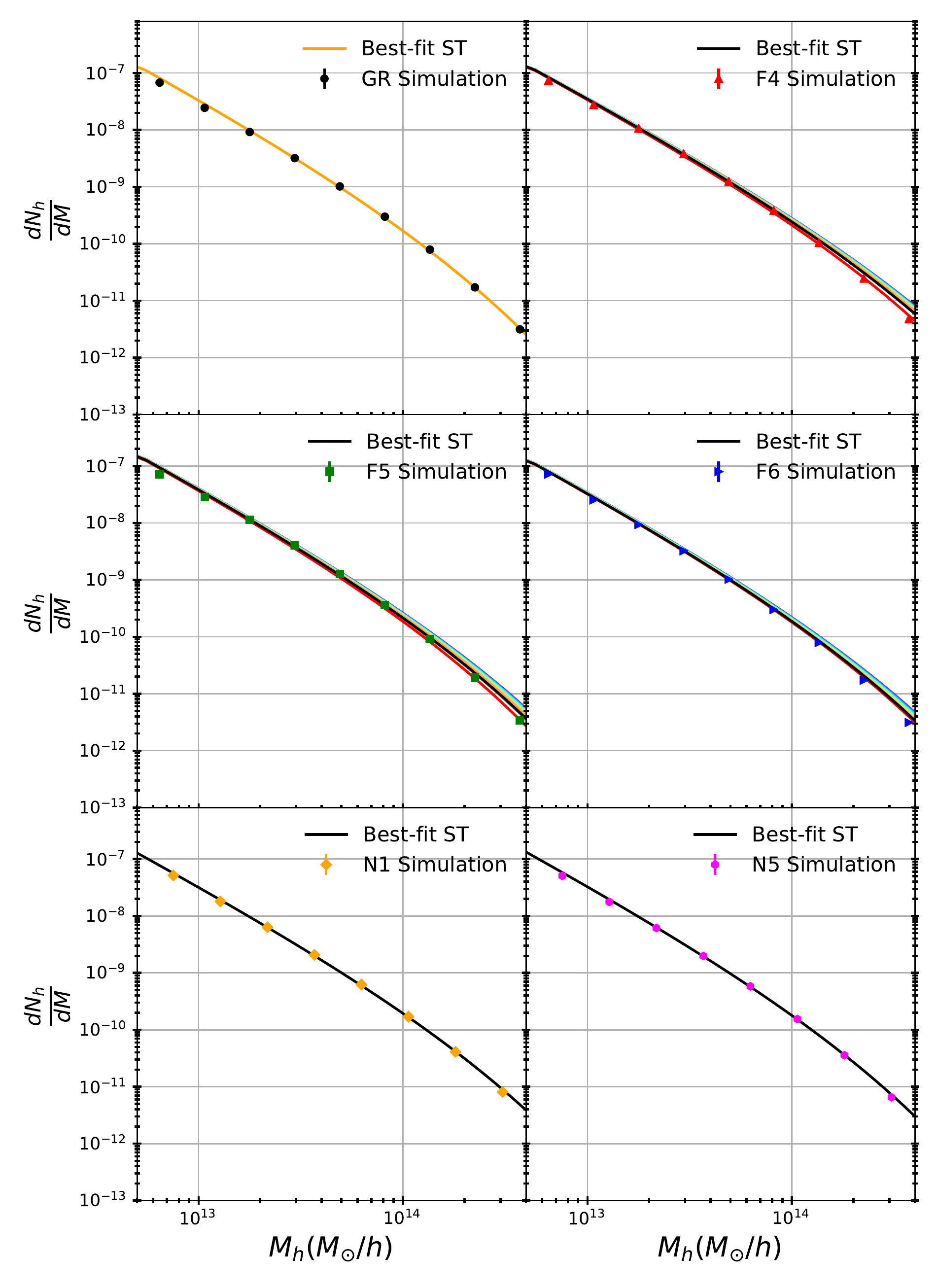}}

\caption{Halo mass function, in the form of number of halos $d N_h$, per mass bin $dM$, as a function of halo mass $M$, calculated from the Group I simulations at $z=0.5$, for GR [black dot] in the upper left panel, F4 [red triangle] in the upper right panel, F5 [green square] in the middle left panel, F6 in the middle right panel and $N1$ and $N5$ $n$DGP models in the lower left and right panels, respectively . Alongside with the simulations, we show the analytical Sheth-Tormen halo mass functions (\ref{PSfunction}), plotted using the best-fit $(q,p)$ values for each model using the criterion (\ref{criterion}). For the $f(R)$ models, the best-fit values correspond to the environmentally averaged halo mass function, through (\ref{probenv}), shown with black curves, while we also plot the halo mass functions for individual values for $\delta_{env}$ for each $f(R)$ model. The color scheme and distribution of values for $\delta_{env}$ is the same as in Figure \ref{collapsevsM}. 
\label{massfR}}
\ec
\end{figure}
  
In Figure \ref{massfR}, we plot the halo mass functions for the GR and three $f(R)$ MG models we consider at  $z=0.5$, together with the best fit ST halo mass functions obtained through (\ref{criterion}). In the case of $f(R)$ gravity, we plot both the mean ST halo mass function, as well as the halo mass function for each value of $\delta_{env}$, to give a sense of the variation among various different environments.
 
In Table \ref{tab1}, we show the best fit values of $(q,p)$ for the different models, mass ranges and simulations considered.  For Group I  simulations, we study predictions at $z=0.5$ for halos in a mass range $(2-3.5)\ \times 10^{12} M_{\odot}/h$, for all models. For the Group II simulations we analyze a $z=1$ snapshot, where we consider halos in three separate mass bins: a lower mass bin of $9\times10^{11}-2\times 10^{12}\ M_{\odot}/h$, an intermediate bin of $5\times10^{12}-1\times 10^{13}\ M_{\odot}/h$ and a higher mass bin, $1.1\times10^{13}-9\times 10^{13}\ M_{\odot}/h$. 

         \begin{table}[t!]
    	\begin{tabular}{ | p{8.2em} | C{3.5em} |C{3.5em} | C{3.5em} |C{3.5em} | }
    		\cline{1-5}
		& \multicolumn{2}{c|}{Best-fit ST} & \multicolumn{2}{c|}{Predicted Biases}
		\\ \hline
    		Models &  $q$ &$p$ &  $b_1$ &$b_2$ 
		 \\ \hline
		$Group\ I: GR$
		& 0.726 & 0.345 
		& 0.301 & -0.501	
		\\ \hline
		$Group\ I: F4$
		& 0.671 & 0.361
		& 0.120 & -0.435
		\\ \hline
		$Group\ I: F5$
		& 0.765 & 0.321 
		& 0.211 & -0.470 
		\\ \hline
		$Group\ I: F6$
		& 0.670 & 0.362 
		& 0.230 & -0.449 
		\\ \hline
	        $Group\ I: N1$
		& 0.701 & 0.369 
		& 0.224 & -0.661
		\\ \hline
	        $Group\ I: N5$
		& 0.702 & 0.357 
		& 0.268 & -0.503 
		\\ \hline	
	        $Group\ II: GR\ Low$
		& 0.674 & 0.362 
		& 0.345 & -0.183 
		\\ \hline	
	        $Group\ II: GR\ Mid.$
		& 0.728 & 0.342 
		& 0.925 & -0.05 
		\\ \hline
	        $Group\ II: GR\ High$
		& 0.806 & 0.594
		& 1.720 & 1.900 
		\\ \hline	
	        $Group\ II: F5\ Low$
		& 0.733 & 0.314 
		& 0.295 & -0.170 
		\\ \hline	
	        $Group\ II: F5\ Mid.$
		& 0.788 & 0.282 
		& 0.909 & -0.033 
		\\ \hline
	        $Group\ II: F5\ High$
		& 0.746 & 0.305 
		& 1.491 & 0.416 
		\\ \hline							
		\end{tabular}
			\caption{The table presents the values for the best-fit Sheth-Tormen parameters $(q,p)$ for the halo mass function (\ref{PSfunction}), with respect to the simulations, through the criterion (\ref{criterion}), as well as the bias factors $b_1$ and $b_2$ evaluated through (\ref{pbsbiasMGavg}) using the best-fit values. All the evaluations for the Group I simulations were performed at redshift $z=0.5$ and for the Group II simulations at $z=1$. The labels low, mid. and high indicate reference to the three mass bins, specified in the text. For all $f(R)$ models, the bias values shown are environmentally averaged as described in the text.}
    	\label{tab1}
	\end{table}

\begin{figure}[t!]
\bc
{\includegraphics[width=0.48\textwidth]{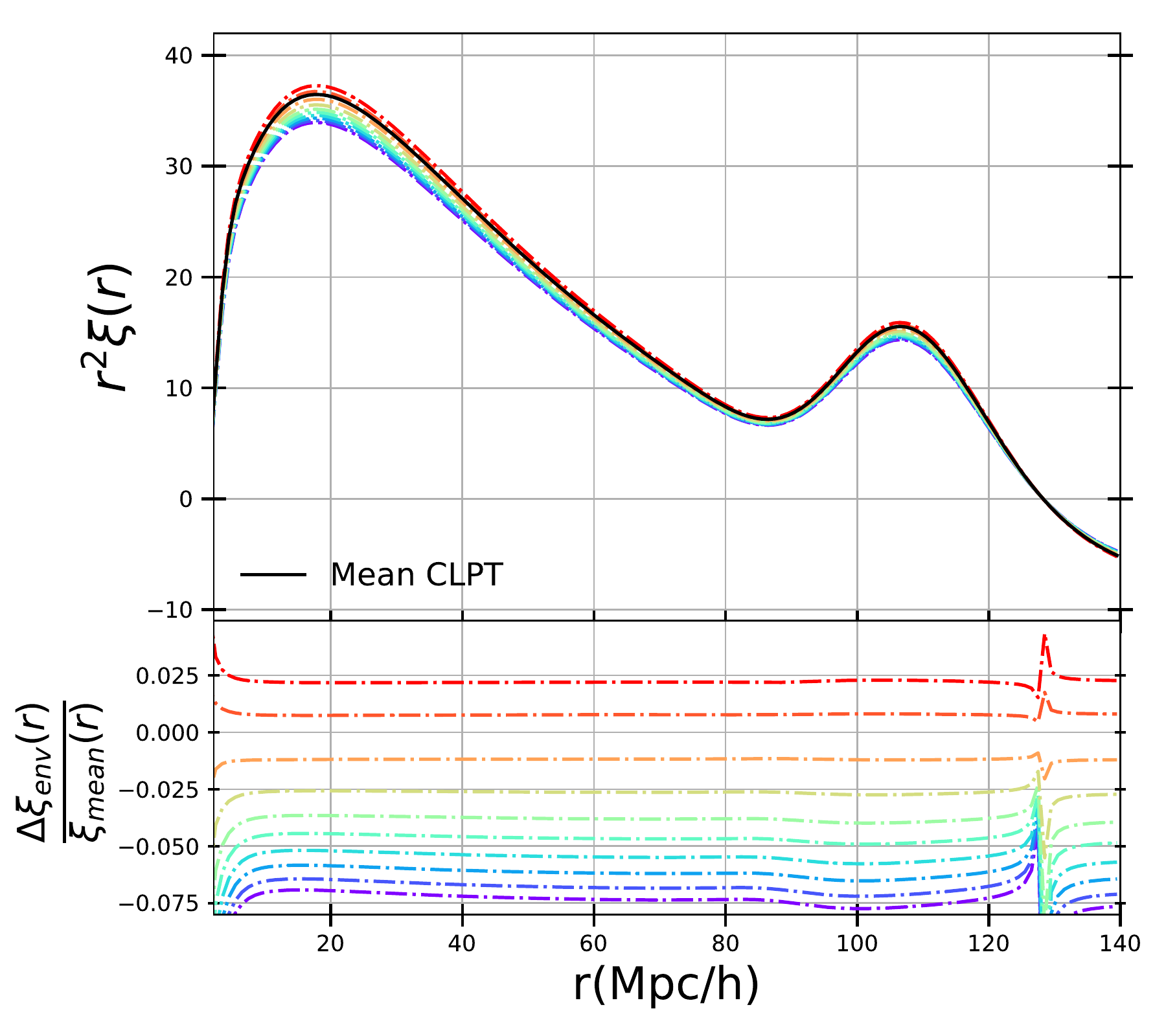}
}
\caption{[Top] Two-point correlation function prediction form CLPT, through (\ref{xiXfinal}), for the F6 model at $z=0.5$ for the various bias values given when $\delta_{env}=\left[-1 (\mathrm{purple}),-0.72,-0.43,-0.15,0.13,0.42,0.7,0.98,1.27,1.55 (\mathrm{red})\right]$ through (\ref{pbsbiasMGavg}) and the result when averaged over environments [black line] using (\ref{probenv}). [Bottom] Fractional deviation, $\frac{\xi_{env}}{\xi_{mean}}-1$, for the CLPT result for each environment in the top panel, with respect to the CLPT curve given by the mean $b_1$ and $b_2$ values (black curve in the top panel).}
\label{xienv}
\ec
\end{figure}

In Figure \ref{xienv} we present the predicted correlation function, $\xi_{env}$ for different environments for the F6 model, along with the correlation function for the environment averaged bias values. Changing the environment can have a notable effect on the predicted correlation function, with variations of $\sim 10\%$ for the F6 model presented in the figure. A decrease in the background environmental density corresponds to a reduction in the correlation function because lower values of $\delta_{env}$ give rise to a lower $\delta_{cr}$ and consequently a lower value for the linear bias $b_1$. 

\subsection{Comparison with simulations}
\label{sec:Results:Compsims}

In this Section, we compare the performance of the various LPT resummation schemes under consideration, combined with our bias model, against the corresponding results from the Group I and Group II simulations, discussed in Section \ref{sim}, with respect to the correlation function and the power spectrum. All correlation functions from the simulations have been calculated employing the publicly available code $\sc{CUTE}$ \citep{2012arXiv1210.1833A}, using 30 linearly spaced bins in the range $0-140$ Mpc/$h$. The power spectra, on the other hand, have been extracted using a Cloud-In-Cell (CIC) mass assignment scheme, on a grid with resolution of $N_{grid}=1024^3$ and $N_{grid}=1200^3$ cells, for the Group I and Group II simulations, respectively. The power was binned in 30 linearly spaced points in the $k$ range of $0.008-0.3$ h/Mpc. 

As discussed in Section \ref{sec:biasparameters}, for Group I  simulations, we study predictions at $z=0.5$ for halos in the mass range $(2-3.5)\ \times 10^{12} M_{\odot}/h$, for all models. For the Group II simulations we analyze a $z=1$ snapshot, where we consider halos in three separate mass bins: a lower mass bin of $9\times10^{11}-2\times 10^{12}\ M_{\odot}/h$, an intermediate bin of $5\times10^{12}-1\times 10^{13}\ M_{\odot}/h$ and a higher mass bin, $1.1\times10^{13}-9\times 10^{13}\ M_{\odot}/h$. 

\subsubsection{Correlation function}
\label{sec:Results:Compsims:xi}

To assess the accuracy and performance of the LPT predictions, we first utilize the Group II simulations as a comparison dataset. Given that we only have 1 realization available for the F5 and GR models, the correlation functions exhibit noise due to the random initial phase, but still facilitate valuable conclusions about the performance of the methods tested given the appropriate combination of large volume and high resolution. These allow us to evaluate the performance of CLPT and its variations simultaneously across a wide range of scales, including both the BAO scales and the region roughly following a power-law scaling relation, down to $r\sim5$ Mpc/$h$. 

The LPT predictions use the PBS biases evaluated from the best fit halo mass function, when fitted over the specific mass ranges, for each bin as summarized inTable \ref{tab1}). Both the LPT and simulation results are compared to the Zel'dovich prediction for the correlation function.

\begin{figure}[!tb]
\bc
{\includegraphics[width=0.5\textwidth]{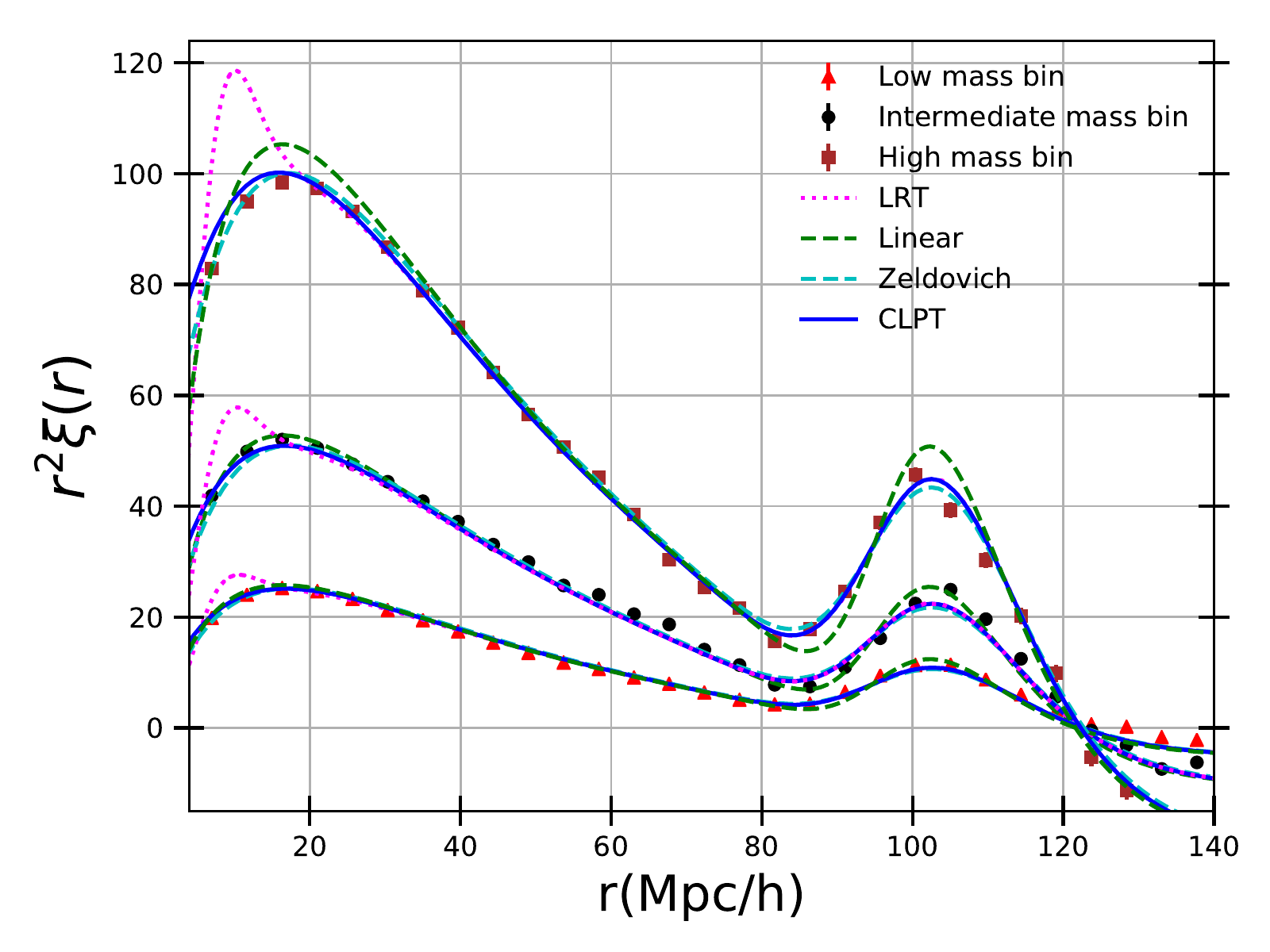}
\includegraphics[width=0.5\textwidth]{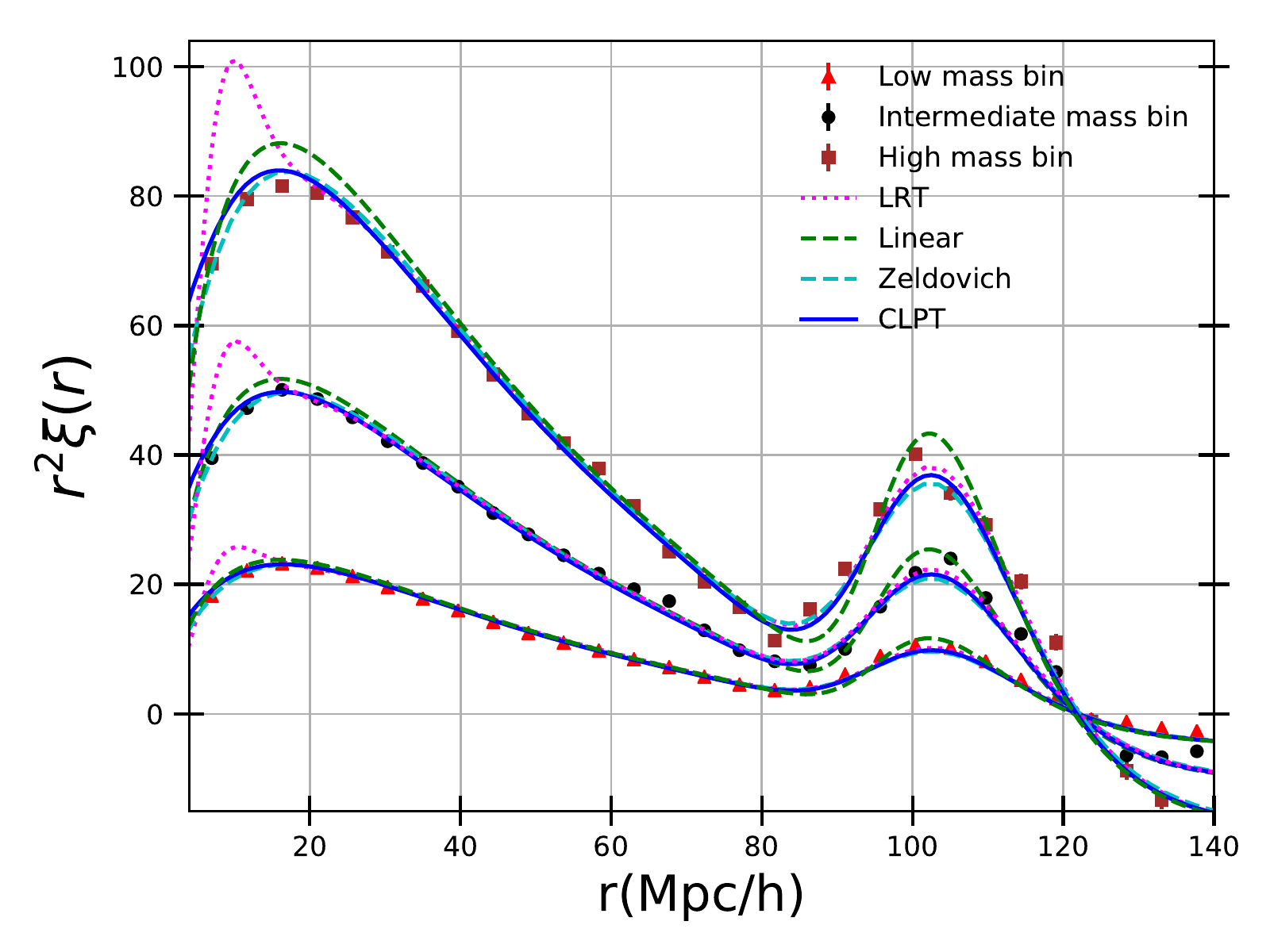}}
\caption{Two-point correlation functions for the Group II simulation snapshots at $z=1$  for GR [top panel] and F5 [lower panel] models. The predictions from CLPT (\ref{xiXfinal}) [solid blue],  the Zel'dovich approximation [dashed cyan],  LRT (\ref{xiXLRT}) [dotted magenta] and the linear theory [dashed green], using the bias values shown in Table \ref{tab1} are compared to the correlation function extracted from simulations, shown with Poisson error bars, for the three mass bins defined in Section \ref{sec:Results}: the low mass [red triangle], intermediate mass [black dot]  and high mass [brown square] bins. }
\label{xiresults}
\ec
\end{figure}

\begin{figure}[!tb]
\bc
{\includegraphics[width=0.5\textwidth]{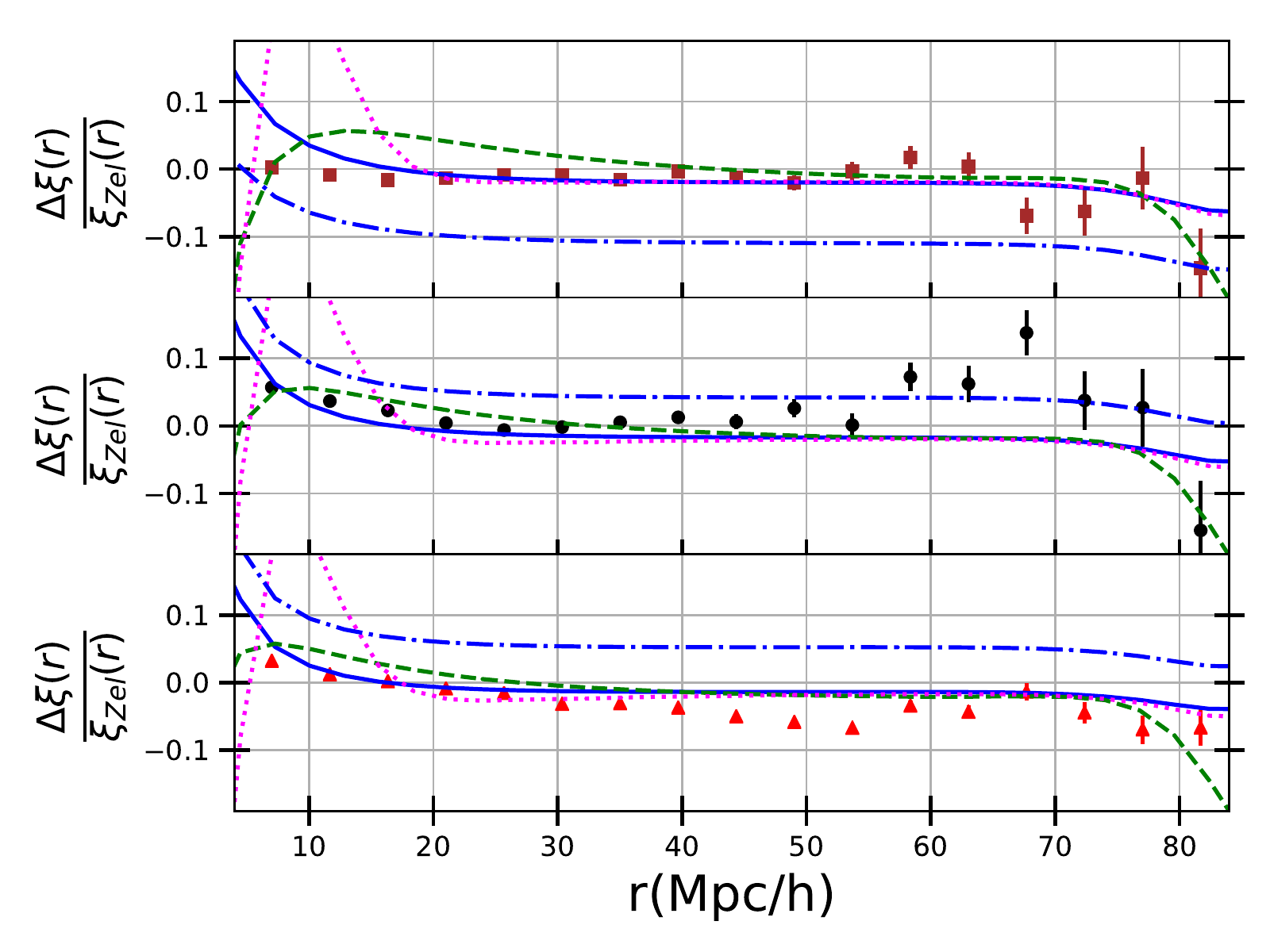}
\includegraphics[width=0.5\textwidth]{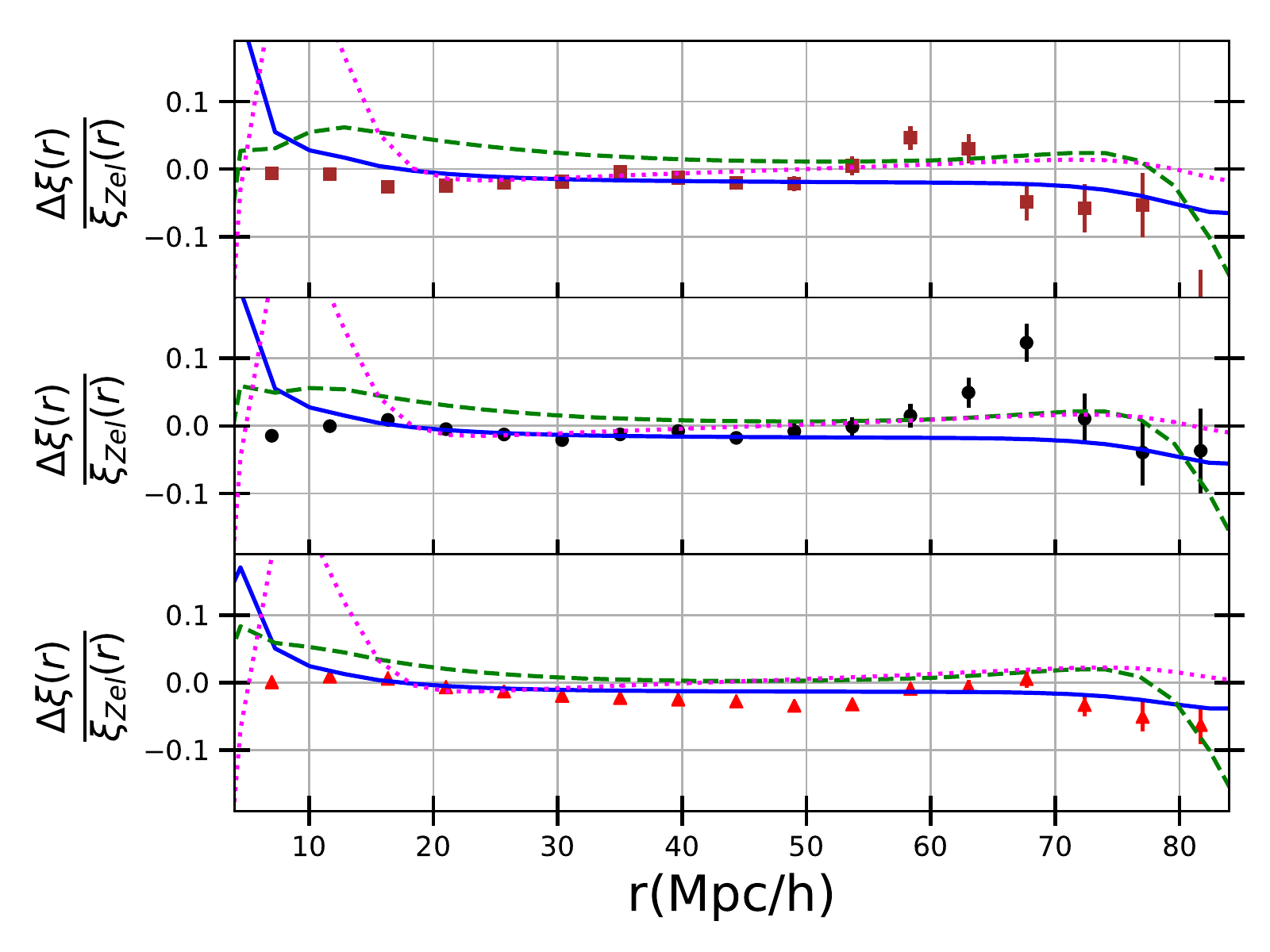}}
\caption{ Fractional deviations in the power-law regime of the correlation function predictions with respect to the Zel'dovich approximation for the results shown in Figure \ref{xiresults}. For the GR analysis [top panel], we also plot the ratio of the CLPT prediction using the standard ST values $(q,p)=(0.75,0.3)$ [dashed-dot blue], rather than the best fit ones in Table \ref{tab1}, divided by the Zel'dovich result. }
\label{xicomp}
\ec
\end{figure}

To benchmark our findings  we  consider both the GR simulations as well as those for the modified gravity model. Figure \ref{xiresults} shows these comparisons for the three mass bin ranges, and  Figure \ref{xicomp} shows the fractional variations with respect to the Zel'dovich component of CLPT. 

For both the GR and modified gravity cases it is important to carefully understand the form of the halo mass function to get accurate LPT predictions. We find that simply adopting the standard ST pair of values,  $(0.75,0.3)$,  gives a poor approximation to the first order bias $b_1$ (for the various values of halo mass), consistent with the findings of the simulation creators \citep{Arnold:2018nmv} when they extracted the bias estimate from the simulations and compared it to a standard ST prediction. For the results with the best fit halo mass parameters, we find that the full CLPT results for both the GR and the F5 model, incorporating the bias parameters evaluated using our PBS model (Table \ref{tab1}) and the environment averaging where necessary, does a very good job,  in describing the power-law correlation function, $20-80$ Mpc/h, for all three mass bins and significantly improves upon linear theory at the BAO peak. For all three mass ranges, shown in  Figure \ref{xicomp}, the simulated correlation function falls below the Zel'dovich approximation and we find that  the CLPT predictions are reflecting this better than both the linear and LRT predictions. The three approaches, CLPT, LRT and Zel'dovich, all perform well in characterizing the BAO peak for the low mass and intermediate mass bins, for the largest differences being in the highest mass bin for F5, where the LRT approach performs slightly better. The LRT performs poorly at the smaller scales, under $20Mpc/h$, significantly overshooting the observed correlation function,  consistent with the results reported in previous studies performed on $\Lambda$CDM cosmologies \citep{Vlah:2014nta,Matsubara:2008wx,2013MNRAS.429.1674C,doi:10.1111/j.1365-2966.2011.19379.x}. The Zel'dovich approximation provides the best agreement at scales below $10Mpch/h$. 

It is also interesting to notice that, in Figure \ref{xiresults}, while the correlation functions have similar values for GR and the F5 model in the lowest mass bins, the F5 result is noticeably lower than the GR one for the highest mass bin. The amplitude of the correlation function depends on the interplay between the dark matter component (which has higher values in MG) and mostly the linear bias factor, $b_1$, which is lower for MG. In the lower mass bins, the combination of the above two is such that the difference between the GR and the F5 curves is almost neutralized, while in the highest mass bins the linear bias factor $b_1$ is, relatively, even lower, causing the F5 two-point function to have clearly lower values than in the GR case.

\begin{figure*}[!tb]
\bc
{\includegraphics[width=0.99\textwidth]{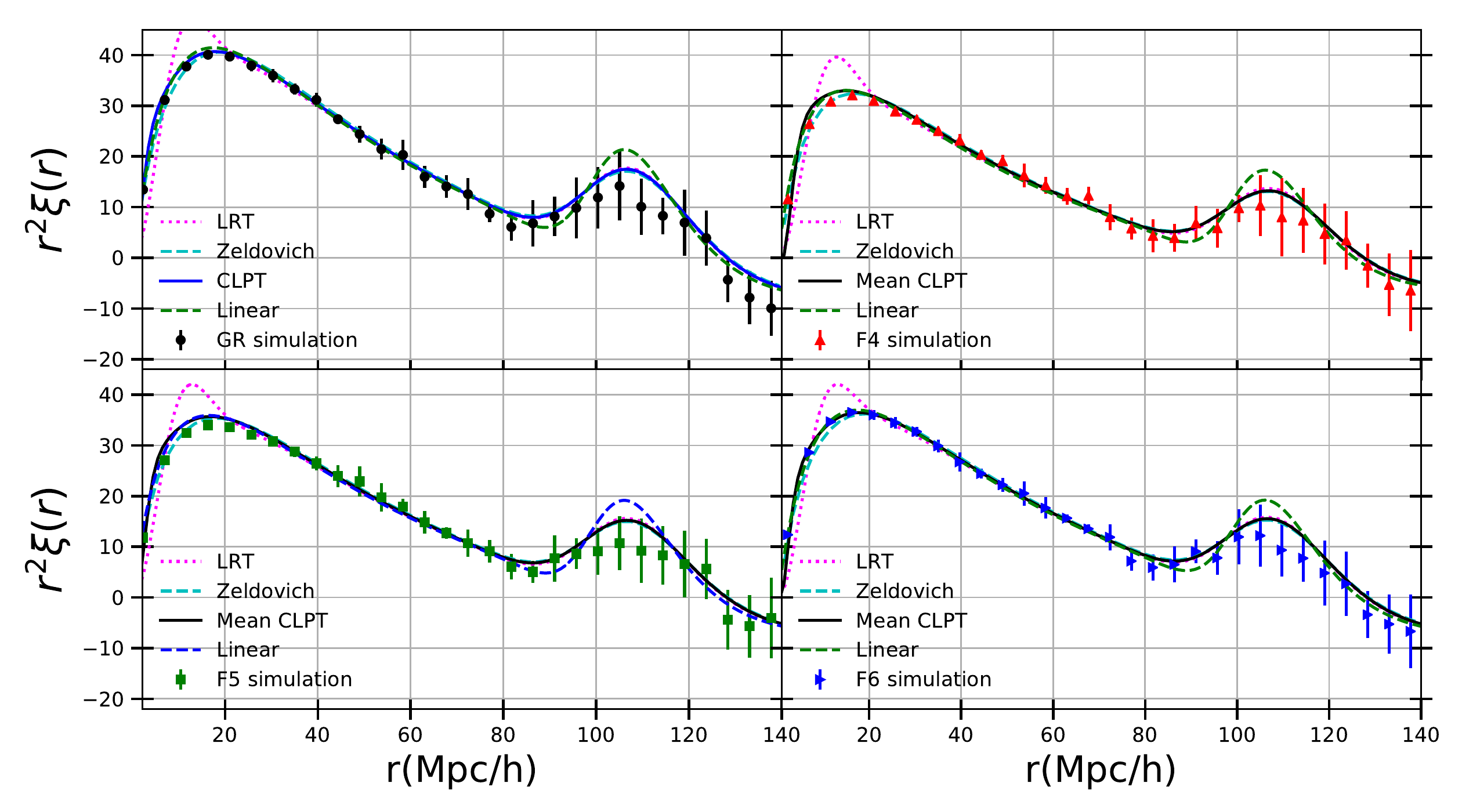}

}
\caption{Two-point correlation functions from the Group I simulations, calculated at $z=0.5$, for GR [black dots] in the upper left panel, for F4 [red triangles] in the upper right panel, for F5 [green squares] in the lower left panel and for F6 [blue right triangles] in the lower right panel. The results are the average over the 5 realizations and the error bars shown are standard deviations. Furthermore, for each model we plot the predictions from CLPT (\ref{xiXfinal}) [solid blue], from the Zel'dovich approximation [dashed cyan], from LRT (\ref{xiXLRT}) [dotted magenta] and from linear theory [dashed green], using the bias values shown in Table \ref{tab1}. The linear theory result for the F5 model is plotted using a blue dashed line instead, for ease of comparison.}
\label{xifR}
\ec
\end{figure*}

To expand on the results from the large volume simulations, we now look into the comparison with the Group I simulations, that, while spanning a smaller volume, allow us to test our schemes for a wider range of models. For each model, five different random realizations are available and the error bars represent the standard deviations over these realizations. Starting with the $f(R)$ family, which is plotted in Figure \ref{xifR}, we see that the picture painted for the $F4$ an $F5$ models here is similar with the one for the $F5$ model (at $z=1$), with CLPT performing the best at scales  $r>20$ Mpc/h and the Zel'dovich result being superior at capturing the smaller scales, while the trend is more pronounced in the F4 case. For the F6 model however, not only does CLPT perform better at these larger scales, but it seems to trace the simulation results more accurately compared to Zel'dovich down to r$\sim7$ Mpc/h.  

To explore this small scale performance sensitivity to screening in more detail note we consider what makes the linear order LPT and its one-loop extensions perform differently. In \citep{Vlah:2014nta} (and also in \citep{Tassev:2013rta}), it was argued, in dark matter-only studies, that LPT does a poor job at estimating the higher order corrections to the linear displacement dispersion, given by (\ref{sigmalin}) and the one-loop correction piece in LPT given by
\begin{equation}\label{sigmaloop}
\sigma_{1loop}^2=\frac{1}{6 \pi^2}\int_0^{\infty}dk\left(\frac{9}{98}Q_1(k)+\frac{10}{21}R_1(k)\right).
\end{equation}
Comparison with simulations in \citep{Tassev:2013rta,Vlah:2014nta}, found CLPT to overestimate the size of this correction at small scales, through (\ref{sigmaloop}), with the true value of the total $\sigma_L^2+\sigma_{1loop}^2$ being closer to $\sigma_L^2$, which is what the Zel'dovich result uses. Because one-loop CLPT strongly depends on these corrections, through its zero-lag terms (as can be seen in Appendix \ref{correlderiv}), it performs worse at the smaller scales compared to its Zel'dovich counterpart. Calculated from our theory prediction for GR at $z=0.5$, the ratio $\frac{\sigma_{1loop}^2}{\sigma_L^2}=0.126$, close to the value in Figure 5 of \citep{Vlah:2014nta}. In comparison,  for the $f(R)$ models at $z=0.5$, the ratio $\frac{\sigma_{1loop}^2}{\sigma_L^2}$ is $(F4, F5, F6)=(0.212, 0.180, 0.152)$. The higher values for F4 and F5 lead to an overestimation in these cases that is responsible for the worse behavior at smaller $r$. It Is worth noting here that if we do not include screening, the ratio $\frac{\sigma_{1loop}^2}{\sigma_L^2}=0.17$ in the $F6$ model, as opposed to the full value of $0.152$. For the $z=1$ Group II simulations, the ratios are $(GR, F5)=(0.08, 0.103)$, which explains the lower discrepancy and better performance of CLPT for F5. This also is consistent with considering that this is an earlier reference in which clustering differences between the theories will be less pronounced. From a physical standpoint, the overestimation reflects an inability in LPT (including the Zel'dovich result), to trap dark matter particles within halos \citep{Vlah:2014nta}, which seems to be more apparent in the LPT approach for stronger MG chameleons. Fortunately, as we said earlier, these models are the ones that violate the observational constraints and are thus less interesting from an astrophysical standpoint. 

\begin{figure}[!t]
\bc
{\includegraphics[width=0.48\textwidth]{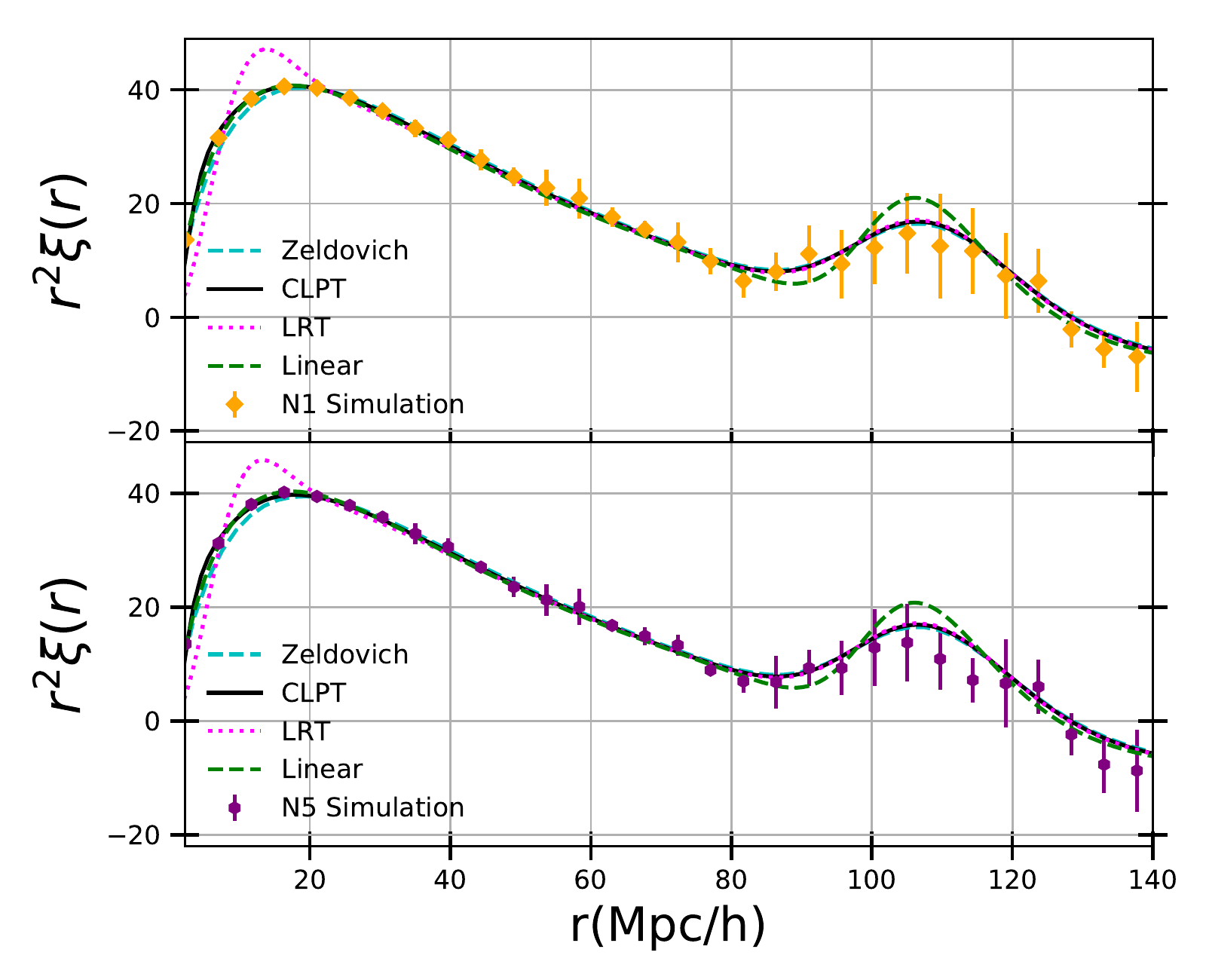}
}
\caption{Two-point correlation functions from the Group I simulations, calculated at $z=0.5$, for N1 [orange diamonds] in the upper panel and for N5 [purple hexagons] in the lower panel. The results are the average over the 5 realizations and the error bars shown are standard deviations. Furthermore, for each model we plot the predictions from CLPT (\ref{xiXfinal}) [solid black], from the Zel'dovich approximation [dashed cyan], from LRT (\ref{xiXLRT}) [dotted magenta] and from linear theory [dashed green], using the bias values shown in Table \ref{tab1}.}
\label{xinDGP}
\ec
\end{figure}

Finally, we test our LPT approaches applied on the $n$DGP models, that represent the Vainshtein screening mechanism, and the correlation functions of which are presented in Figure \ref{xinDGP}, for all LPT schemes and the Group II simulations. Just like in the $f(R)$ models, CLPT does a very good job at describing the correlation function for large scales and beyond that, it even seems to perform at least equally well as the Zel'dovich curve down to scales $r\sim10$ Mpc/h, similar to the F6 and GR cases in Figure \ref{xifR} discussed earlier. The measurement of the 1-loop statistic discussed in the previous paragraph is consistent with this; for the $n$DGP models, the ratio$\frac{\sigma_{1loop}^2}{\sigma_L^2}=(0.129, 0.122)$ for (N1, N5) respectively, very consistent with  the GR value = 0.126. This is the case even in the weaker screening case, the N5 model, and is a very promising sign, given that the Vainshtein mechanism is highly efficient at screening modifications to gravity at smaller scales and contains viable candidates that self-accelerate (even though this particular model does not). The relative performance among the different resummation schemes is very similar to the one observed in the GR and $f(R)$ cases, with all LPT models improving the accuracy at the BAO peak upon linear theory, with the LRT scheme giving more power that CLPT and then the Zel'dovich result. The characterization of the BAO peak on scales $r>100$ Mpc/h is limited in the simulation box with side 1,024 Mpc/h; larger-volume simulations for the $n$DGP model, comparable to the Group II simulations for F5 or GR, will allow us to more clearly trace the region between $100-140$ Mpc/h and draw stronger conclusions about how our models perform at the BAO scales.  

The fact that CLPT performs well for all modified gravity models considered in the power-law and BAO scales  is very encouraging. On the smaller scales, its robustness for highly screened models is also a positive result. If an MG model was ever detected, it would be a highly screened case, given the tight constraints placed on GR; models F4 and F5 are actually ruled out by observations \citep{Burrage:2017qrf}, but we include them in our analysis to fully investigate the chameleon phenomenology with LPT.

\subsubsection{Power spectrum}

\begin{figure}[!t]
\bc
{\includegraphics[width=0.5\textwidth]{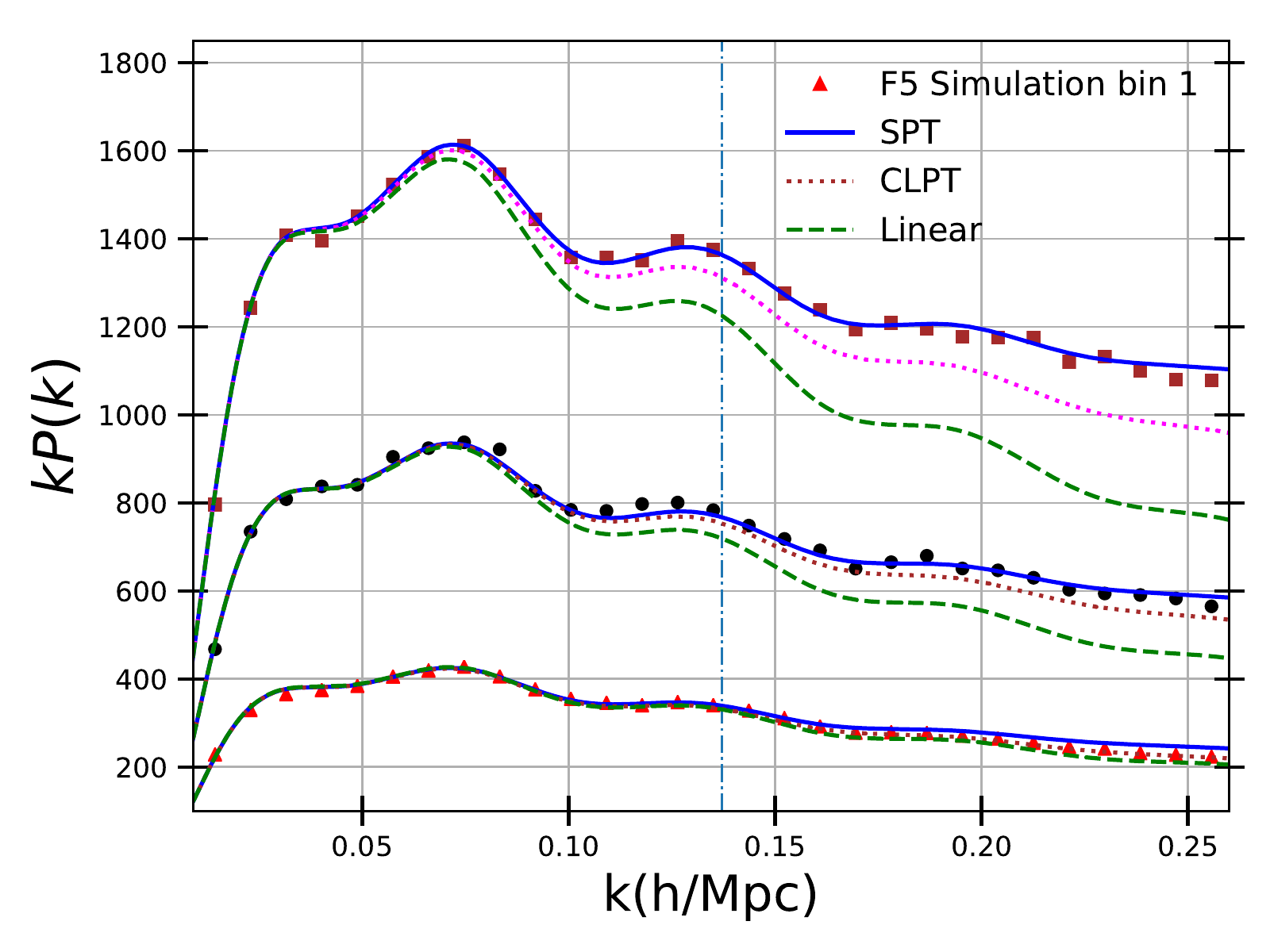}

}
\caption{Power spectra from the Group II simulations, calculated for F5 at $z=1$, in the low mass [red triangle], intermediate mass [black dot]  and high mass [brown square] bins, that were defined in Section \ref{sec:Results}. The error bars shown are Poisson error bars. Furthermore, for each mass bin we plot the predictions from CLPT (\ref{xiXfinal}) [dotted magenta], from SPT (\ref{PkXSPT}) [solid blue] and from linear theory [dashed green], using the bias values shown in Table \ref{tab1}.}
\label{PklightF5}
\ec
\end{figure}

\begin{figure*}[!tb]
\bc
{\includegraphics[width=0.9\textwidth]{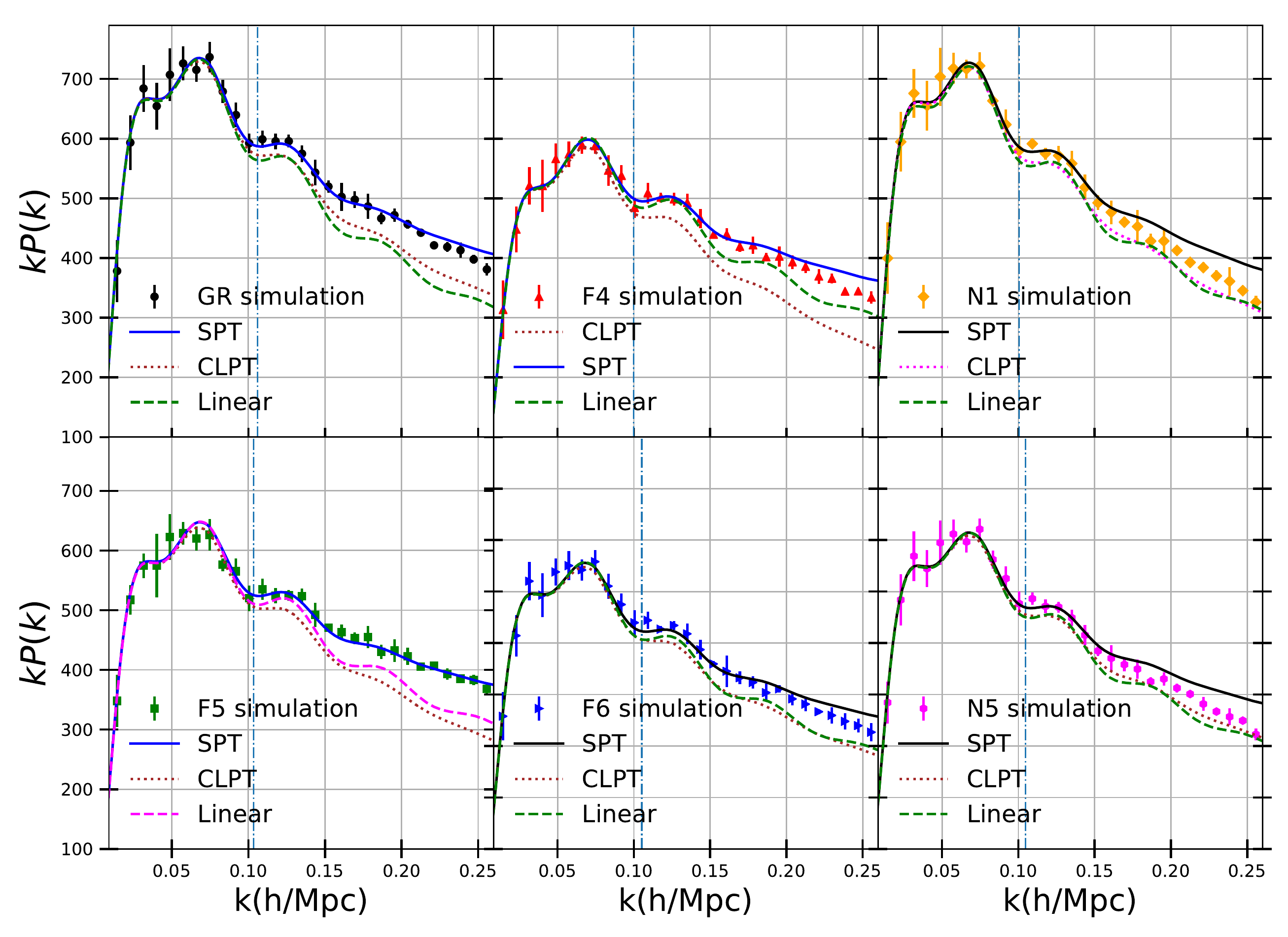}
}
\caption{Power spectra from the Group I simulations, calculated at $z=0.5$, for GR [black dots] in the upper left panel, for F4 [red triangles] in the upper middle panel, for F5 [green squares] in the lower left panel, for F6 [blue right triangles] in the lower middle panel, for N1 [orange diamonds] in the upper right panel and for N5 [purple hexagons] in the lower right panel. The results are the average over the 5 realizations and the error bars shown are standard deviations. Furthermore, for each model we plot the predictions from CLPT (\ref{xiXfinal}) [dotted brown],  from SPT (\ref{PkXSPT}) [solid blue] and from linear theory [dashed green], using the bias values shown in Table \ref{tab1}. The linear theory result for the F5 model is plotted using a pink  dashed line instead, for ease of comparison. }
\label{PkfR}
\ec
\end{figure*}

Complementary to the correlation function, we also perform tests on its Fourier space counterpart, the halo power spectrum. The mass bins and bias values used in the power spectra calculations are exactly the same as the ones presented in the correlation function case, but with the additional step that all power spectra are shot noise corrected \citep{PhysRevLett.103.091303}: $\tilde{P}(k)=P(k)-\frac{1}{n_h}$, where $P(k)$ is the uncorrected power spectrum, and $n_h$ is the number density of halos in each bin. Especially for the higher mass bins that contain less halos, this effect is not negligible, especially at higher $k$. We also identify the scale at which perturbation theory starts to fail, $k_{NL}$, with the vertical dashed-dot blue line, using the definition \citep{Matsubara:2007wj}, $k_{NL} = (2\sigma_L^2)^{-1}$,
with $\sigma_L^2$ the linear power spectrum dispersion defined in (\ref{sigmalin}).

In Figure \ref{PklightF5}, we present  the F5, Group II snapshot at $z=1$. We find the expanded, SPT power spectrum (\ref{PkXSPT}) to perform very well at capturing the small $k$ and to follow the power spectrum until $k\sim0.25h/$Mpc, where it starts to overestimate power compared to the simulations, for all three mass bins. This behavior is consistent with what was found in the GR case in earlier works on \citep{Vlah:2014nta,Matsubara:2008wx} and also for dark matter in MG \citep{Aviles:2017aor}. The linear theory result is only accurate at very large scales and quickly underestimates the power at $k>0.05 h$/Mpc. Unlike in the correlation function comparison, where LPT was found to perform very well at a wide range of scales, here we see that the CLPT power spectrum (\ref{PkXfinal}) decays quickly and performs considerably worse than the SPT expansion. This is not unexpected, since the power spectrum in LPT has been found to receive, unlike in the configuration space, significant contributions from large, nonlinear $k$ modes, where LPT performs poorly and fails to trap particles inside dark matter halos \citep{Vlah:2014nta}. Our results  show that this to also  the case in our MG models. We find that this effect is even more pronounced in the LRT power spectrum (\ref{PkXLRT}), which decays sharply in $k$-space, for this reason we do not include the result in our plots.

In Figure \ref{PkfR} summarizing the GR and $f(R)$ cases from the Group I simulations, at $z=0.5$, while the CLPT consistently underestimates the power spectrum for all models,  the SPT result tracing the simulation points well for F5 and F6 until $k\sim 0.2 h$/Mpc, at which it starts to overestimate the power spectrum. The performance for the the F4 model is slightly worse on small scales. This earlier deviation  is not surprising given that the model has the smallest $k_{NL}$ prediction, resulting from a comparatively higher 1D linear dispersion. 

For the two $n$DGP models, also shown in Figure \ref{PkfR}, we find that the SPT predictions perform well at scales $k<0.15$ Mpc/h but overestimate the power on small scales. The CLPT predictions consistently underestimate the power, and are broadly comparable to the linear prediction. 

\section{Conclusions }
\label{sec:conclusions}

In this work, we modeled the two-point statistics of biased tracers in modified gravity (MG) up to one-loop order in the linear power spectrum, using the Convolution Lagrangian Perturbation Theory (CLPT) framework and its variants. Following standard methods in the literature, the linear piece of the two-point Lagrangian correlator for dark matter is kept exponentiated in the expression for the two-point correlation function, but everything else is expanded, leading to a series of convolution integrals, the expressions of which we derive for scalar-tensor theories. 

The evolution of the underlying dark matter density field is described by the LPT framework for dark matter, suitably extended to study scalar-tensor theories, along with an analytical model for the calculation of the first and second order bias parameters in MG. To perform the bias calculations, we employ the Peak-Background split (PBS) approach, in which biases are modeled rigorously as responses of the universal Sheth-Tormen halo mass function in the presence of a long-wavelength density perturbation. This is extended in MG theories, to account for the dependence of the gravitational collapse on the environment and screening. Our PBS implementation, provides a quantitative prediction for the increased production, and the related lower biases, for haloes of a given mass. We apply this scheme to the $f(R)$ Hu-Sawicki and the $n$DGP braneworld models, that are representatives of the chameleon and Vainshtein screening mechanisms, respectively. We make the code used for the analytic predictions publicly available in \footnote{\url{https://github.com/CornellCosmology/bias_MG_LPT_products}} and evaluate their performance against state-of-the-art cosmological N-body simulations, for a variety of MG models at $z=0.5$ and $z=1$, with respect to the correlation function and the power spectrum in a variety of mass regimes and scales. 

The CLPT implementation, in combination with the analytical bias model, gives good agreement with the simulations, with the only free parameters necessary being those to best-fit the Sheth-Tormen universal halo mass function at the given mass range. The CLPT predicts the  correlation function across scales $20-80$ Mpc/$h$, tracing the simulation results at an accuracy of $2-3\%$ and better. At the BAO scales, that provide a valuable probe of fundamental physics, CLPT was found to improve significantly upon the linear theory and Zel'dovich predictions for the F5 models, just like in GR. The Lagrangian Resummation Theory (LRT) approach improved the accuracy a little further at BAO scales for the highest mass range considered. At scales of $r<20$ Mpc/$h$, the CLPT performed well for the highly screened model $F6$ and for the $n$DGP models, while the Zel'dovich predictions performed better for the weakly screened F5 and F4 models. The reason for this behavior was identified, being an overestimation in these low-screening chameleon models of the one-loop contributions to the zero-lag terms at small scales.

In  Fourier space, consistent with findings for GR, the CLPT power spectrum  was found to underestimate power quickly, compared to the simulations for all MG models. This is due to the power spectrum receiving significant contributions from large $k$, where LPT performs poorly. The Standard Perturbation Theory (SPT) approach, though, which is the low-$k$ expanded version of this power spectrum, performs very well  and remains consistent with the simulation results down to $k\sim0.2 h$/Mpc for the $f(R)$ models and down to $k\sim0.15 h$/Mpc for the two $n$DGP models. Beyond these scales, the SPT curve overestimates the power spectrum, as has been found for GR previously.

While  we have focused our analysis on LPT predictions for real space, our model can be expanded to capture the redshift-space distortions required for upcoming LSS surveys. Furthermore, even though we focused on a local in matter density bias scheme in the Lagrangian space, in which the bias is purely a function of the local density, one can extend this to include other factors determining bias into the formalism, such as curvature bias, and model them successfully by this PBS scheme. The same applies for potential extensions to include EFT corrections to our LPT model, as in \citep{Vlah:2016bcl}, which could also be used to calculate the components of the Gaussian Streaming Model for MG theories. Finally, our CLPT MG framework can be used to analytically predict marked statistics in MG and assess their ability to boost the MG signals carried in cosmic density fields, as in \citep{White:2016yhs,Valogiannis:2017yxm}. We leave these natural extensions to future work.

In the coming decade, a wide array of cosmological surveys will span a large part of the observable universe, searching for hints of new physics beyond $\Lambda$CDM. In this work we demonstrate that semi-analytical approaches, extensively employed in the context of standard GR, can serve as invaluable tools to predict structure formation in cosmologies with an extra degree of freedom in the gravitational sector. A next step for these approaches are to confront them in comparison to realistic simulations of galaxies and clusters that will be observed with  surveys coming online in the coming year or two and assess survey ability to identify and constrain potential deviations from GR.

\section*{Acknowledgments}
We wish to thank Wojciech Hellwing for kindly making available the $n$DGP simulations, on behalf of \citep{Hellwing:2017pmj} and Baojiu Li for kindly providing the $\sc{ELEPHANT}$ simulations, on behalf of \citep{Cautun:2017tkc}, for numerous discussions and for making us awaref of the Lightcone simulations. We are also grateful to Christian Arnold, as well as to all the other authors of \citep{Arnold:2018nmv}, for kindly providing their Lightcone simulations and for numerous discussions. We also wish to thank Martin White for useful discussions on available simulations of $\Lambda$CDM cosmologies. The work of Georgios Valogiannis and Rachel Bean is supported by DoE grant DE-SC0011838, NASA ATP grants NNX14AH53G and 80NSSC18K0695,  and NASA ROSES grant 12-EUCLID12-0004.

\appendix
\label{sec:app}
\section*{Appendix}
\section{One loop corrections for biased statistics in MG} 
\label{App:AppendixA}

In section \ref{sec:twopointMG} of the main text, it was stated that the two point statistics for biased tracers, up to one loop in CLPT, are given by equations (\ref{xiXfinal}) and (\ref{PkXfinal}), in the configuration space and the Fourier space, respectively. The expressions are convolutional integrals over a sum of individual terms that depend on the Lagrangian correlators (\ref{eq:correlMG}), which are essentially the fundamental blocks of CLPT. These expressions (\ref{eq:correlMG}), however, are functions of the Lagrangian coordinates $\bold{q}$, while the LPT solutions for the displacement fields up to various orders are found in the Fourier space (through the growth factors (\ref{growth1st}), (\ref{eq:D2sources}) and (\ref{D3MG})). As was also stated in \ref{sec:twopointMG}, substituting the LPT solutions (\ref{psifourmain}) into (\ref{eq:correlMG}) gives the integral expressions for the MG Lagrangian correlators (\ref{qfuncsmain}), that depend on the functions (\ref{QRformmain}), which are also the building blocks of the SPT and LRT power spectra (\ref{PkXSPT}) and (\ref{PkXLRT}). We start with the innermost layer of integration, deriving the expressions for the k-dependent functions (\ref{QRformmain}) in section \ref{polyderiv}, before showing how to get to the correlators (\ref{qfuncsmain}) in section \ref{correlderiv}. Finally, in section \ref{spectraderiv}, we show how the SPT and LRT expressions for the two-point statistics are derived. The notation and index structure is the one adopted in \citep{Matsubara:2007wj,Matsubara:2008wx,2013MNRAS.429.1674C}. These results are consistent with those recently presented in \citep{Aviles:2018saf}.
\subsection{Polyspectra and k-functions in MG}\label{polyderiv}

In LPT, we solve for the displacement field $\bold{\Psi}(\bold{q})$ across various orders in perturbation theory, as
\begin{equation}\label{eq:psiexpappend}
\bold{\Psi}(\bold{q},t) = \sum_{n=1}^{\infty}\bold{\Psi}^{(n)}(\bold{q},t) = \bold{\Psi}^{(1)}(\bold{q},t)+\bold{\Psi}^{(2)}(\bold{q},t)+\bold{\Psi}^{(3)}(\bold{q},t)...
\end{equation}
In the Fourier space representation, $\tilde{\bold{\Psi}}(\bold{p})$, the solutions are expanded as
\begin{equation}
\begin{aligned}\label{psifour}
&\tilde{\Psi}^{(n)}_j(\bold{p}) = \frac{i}{n!}\int\frac{d^3p_1}{(2\pi)^3}..\frac{d^3p_{n}}{(2\pi)^3} \delta_D^3\left(\sum_{j=1}^{n}p_j-p\right) \\
&\times L_j^{(n)}(\bold{p}_1,..,\bold{p}_n)\tilde{\delta}_{L}(\bold{p}_1)..\tilde{\delta}_{L}(\bold{p}_n), 
\end{aligned}
\end{equation}
where $\tilde{\delta}_{L}(\bold{p}_n)$ are the linear density fields in the Fourier space at the time of evaluation. For gravitational evolution governed by GR, the growth factors are only functions of time, and under the additional assumption of an Einstein-De Sitter evolution, the Kernels $L_j^{(n)}(\bold{p}_1,..,\bold{p}_n)$ admit simple scale-independent expressions and (\ref{psifour}) can be simply evolved in time by powers of the linear growth factor \citep{Matsubara:2007wj}. This assumption gives results accurate at the sub-percent level for $\Lambda$CDM cosmologies \citep{Bouchet:1994xp}.

In MG, however, the simple description presented above does not hold, in principle, because the growth factors depend on both space and time, as we saw in Section \ref{LPTdarkmatter}. Following \citep{Aviles:2017aor}, we define
\begin{equation}
\begin{aligned}\label{LPTKernels}
&L_j^{(1)}(\bold{p}) = \frac{p^j}{p^2} \\
& L_j^{(2)}(\bold{p}_1,\bold{p}_2) = \frac{p^j}{p^2} \frac{D^{(2)}(\bold{p_1},\bold{p_2})}{D^{(1)}(\bold{p_1})D^{(1)}(\bold{p_2})}  \\
& (L_j^{(3)})_{sym}(\bold{p}_1,\bold{p}_2,\bold{p}_3) = i\frac{p^j}{p^2} \frac{D^{(3)}_{sym}(\bold{p_1},\bold{p_2},\bold{p}_3)}{D^{(1)}(\bold{p_1})D^{(1)}(\bold{p_2})D^{(1)}(\bold{p_3})} ,
\end{aligned}
\end{equation}
where the MG growth factors are calculated through the prescription described in Section \ref{LPTdarkmatter} and their time arguments have been dropped for notational simplicity. Furthermore, the subscript in the third order Kernel is meant to emphasize on the fact that the configuration that enters the 2-point statistics should be symmetrized \citep{Matsubara:2007wj,Aviles:2017aor}. The symmetrized third order growth factor that enters the two-point statistics is given by \citep{Aviles:2017aor}:
\begin{equation}
D^{(3)}_{sym}(\bold{k},-\bold{p},\bold{p})=D^{(3)}(\bold{k},-\bold{p},\bold{p})+D^{(3)}(\bold{k},\bold{p},-\bold{p}),
\end{equation}
with $D^{(3)}(\bold{k},-\bold{p},\bold{p})$ given by:
\begin{equation}
\begin{aligned}\label{D3MG}
&\left(\mathcal{\hat{T}}-A(k)\right)D^{(3)}(\bold{k},-\bold{p},\bold{p})= D^{(1)}(p)\left(A(p)+\mathcal{\hat{T}}-A(k)\right)D^{(2)}(\bold{p},\bold{k})\times \\
&  \left[1-\frac{\left(\bold{p}\cdot\left(\bold{k}+\bold{p}\right)\right)^2}{p^2|\bold{p}+\bold{k}|^2}\right]- D^{(1)}(p)\left(A(p)+A(|\bold{p}+\bold{k}|)-2A(k)\right)D^{(2)}(\bold{p},\bold{k}) \\
&+ \left(2A(k)-A(|\bold{p}+\bold{k}|)-A(p)\right)D^{(1)}(k)\left(D^{(1)}(p)\right)^2\frac{\left(\bold{k}\cdot\bold{p}\right)^2}{k^2p^2}\\
& -\left(A(|\bold{p}+\bold{k}|)-A(k)\right)D^{(1)}(k)\left(D^{(1)}(p)\right)^2 - D^{(1)}(k)\left(D^{(1)}(p)\right)^2 \times \Biggl[ \\
&\frac{M_1(\bold{p}+\bold{k})}{3\Pi(|\bold{p}+\bold{k}|)}K^{(2)}_{FL}(\bold{p},\bold{k})-\left(\frac{2A(0)}{3}\right)^2\frac{M_2(\bold{p},\bold{k})|\bold{p}+\bold{k}|^2}{6 a^2 \Pi(|\bold{p}+\bold{k}|)\Pi(k)\Pi(p)} \Biggr]+\\
&\frac{M_1(k)}{3\Pi(k)}\Biggl[\left(2\frac{\left(\bold{p}\cdot\left(\bold{k}+\bold{p}\right)\right)^2}{p^2|\bold{p}+\bold{k}|^2}-\frac{\bold{p}\cdot\left(\bold{k}+\bold{p}\right)}{p^2}\right)\left(A(p)-A(0)\right)D^{(2)}(\bold{p},\bold{k})D^{(1)}(p) \\
&+\left(2\frac{\left(\bold{p}\cdot\left(\bold{k}+\bold{p}\right)\right)^2}{p^2|\bold{p}+\bold{k}|^2}-\frac{\bold{p}\cdot\left(\bold{k}+\bold{p}\right)}{|\bold{k}+\bold{p}|^2}\right)\left(A(|\bold{k}+\bold{p}|)-A(0)\right)D^{(2)}_{\phi}(\bold{p},\bold{k})D^{(1)}(p) \\
&+3\frac{\left(\bold{k}\cdot\bold{p}\right)^2}{k^2p^2}\left(A(p)+A(k)-2A(0)\right)D^{(1)}(k)\left(D^{(1)}(p)\right)^2 \Biggr] \\
&-\frac{1}{2}\frac{k^2}{6 a^2 \Pi(k)}K^{(3)}_{\delta \mathcal{I} sym}(\bold{k},-\bold{p},\bold{p})D^{(1)}(k)\left(D^{(1)}(p)\right)^2.  \\
\end{aligned}
\end{equation}
In (\ref{D3MG}), we additionally defined 
\begin{equation}
\begin{aligned}\label{D2phi}
&D^{(2)}_{\phi}(\bold{p},\bold{k})= D^{(2)}(\bold{p},\bold{k})+\left(1+\frac{\left(\bold{k}\cdot\bold{p}\right)^2}{k^2p^2}\right)D^{(1)}(k)D^{(1)}(p) \\
&-\frac{2A(0)}{3}\frac{M_2(\bold{p},\bold{k})+2\left(\frac{3}{2A(0)}\right)^2K^{(2)}_{FL}(\bold{p},\bold{k})\Pi(k)\Pi(p)}{3\Pi(k)\Pi(p)}D^{(1)}(k)D^{(1)}(p)  \\
\end{aligned}
\end{equation}
and also used the kernels $K^{(2)}_{FL}$ and $K^{(3)}_{\delta \mathcal{I} sym}$, the forms of which were shown in \citep{Aviles:2017aor}. The second and third order LPT kernels (\ref{LPTKernels}) need to be evaluated numerically, now, for each value of $\bold{p}_1,\bold{p}_2$ and $z$ and so should (\ref{psifour}), which is the main point of divergence between the MG implementation and the corresponding one in GR. 

Now, following \citep{Matsubara:2008wx}, we define the mixed polyspectra $C_{i_1...i_N}$, as
\begin{equation}
\begin{aligned}\label{mixedploy}
& \langle \tilde{\delta}_{L}(\bold{k}_1)...\tilde{\delta}_{L}(\bold{k}_l)\tilde{\Psi}_{i_1}(\bold{p}_1)...\tilde{\Psi}_{i_N}(\bold{p}_N) \rangle_c = \\
& = (2\pi)^3 \delta_D^3(\bold{k}_1+..+\bold{k}_l+..\bold{p}_1+..+\bold{p}_N)(-i)^{N}C_{i_1...i_N}(\bold{p}_1,..,\bold{p}_N).
\end{aligned}
\end{equation}
It is useful to decompose the various polyspectra into the various constituents, order by order in perturbation theory, as e.g.
\begin{equation}
\begin{aligned}\label{polyorders}
& C_{ij}(\vec{p})=C_{ij}^{(11)}(\vec{p})+C_{ij}^{(22)}(\vec{p})+C_{ij}^{(13)}(\vec{p})+C_{ij}^{(31)}(\vec{p})+.... \\
& C_{ijk}(\vec{p}_1,\vec{p}_2,\vec{p}_3)=C_{ijk}^{(112)}(\vec{p}_1,\vec{p}_2,\vec{p}_3)+\\
& C_{ijk}^{(121)}(\vec{p}_1,\vec{p}_2,\vec{p}_3)+C_{ijk}^{(211)}(\vec{p}_1,\vec{p}_2,\vec{p}_3)+...,
\end{aligned}
\end{equation}
where the additional notation has been adopted
\begin{equation}
\begin{aligned}\label{mixedployorder}
& \langle \tilde{\delta}_{L}(\bold{k}_1)...\tilde{\delta}_{L}(\bold{k}_l)\tilde{\Psi}_{i_1}^{(r)}(\bold{p}_1)...\tilde{\Psi}_{i_N}^{(s)}(\bold{p}_N) \rangle_c = \\
& = (2\pi)^3 \delta_D^3(\bold{k}_1+..+\bold{k}_l+..\bold{p}_1+..+\bold{p}_N)(-i)^{N}C_{i_1...i_N}^{(r...s)}(\bold{p}_1,..,\bold{p}_N)
\end{aligned}
\end{equation}
and as previously, the bracketed numbers in the superscripts indicate the orders of contribution in each $\tilde{\Psi}_i(\bold{p})$. The various polyspectra can be expressed as functions of the Lagrangian kernels (\ref{LPTKernels}), by identifying the different contributions across each order in LPT, as in (\ref{polyorders}) and plugging the solutions (\ref{psifour}) into equation (\ref{mixedployorder}). The ones relevant for the two-point statistics of biased tracers are \citep{Matsubara:2008wx}:
\begin{equation}
\begin{aligned}\label{mixedployorderspef}
 &C_{ij}^{(11)}(\bold{p}) =  L_i^{(1)}(\bold{p})L_j^{(1)}(\bold{p}')P_L(p) \\
 &C_{ij}^{(22)}(\bold{p}) =\frac{1}{2}\int \frac{d^3p'}{(2\pi)^3}L_i^{(2)}(\bold{p}',\bold{p}-\bold{p}')L_j^{(2)}(\bold{p}',\bold{p}-\bold{p}')\times \\
 &\times P_L(p) P_L(|p-p'|)  \\
&C_{ij}^{(13)}(\bold{p}) = C_{ji}^{(31)}(\bold{p}) =   -\frac{1}{2}L_i^{(2)}(\bold{p}) P_L(p)\times \\
&\times \int \frac{d^3p'}{(2\pi)^3}(L_j^{(3)})_{sym}(\bold{p},-\bold{p}',\bold{p}')P_L(p') \\
&C_{i}^{(2)}(\bold{p}_1,\bold{p}_2;\bold{p}_3)  =   -L_i^{(2)}(\bold{p}_1,\bold{p}_2) P_L(p_1)  P_L(p_2)\\
&C_{ij}^{(2)}(\bold{p}_1;\bold{p}_2,\bold{p}_3)  =   C_{ji}^{(21)}(\bold{p}_1;\bold{p}_3,\bold{p}_2) = \\
&= -L_i^{(1)}(\bold{p}_2)L_j^{(2)}(\bold{p}_1,\bold{p}_2) P_L(p_1)  P_L(p_2)\\
&C_{ijk}^{(112)}(\bold{p}_1,\bold{p}_2,\bold{p}_3)  =   C_{kij}^{(211)}(\bold{p}_3,\bold{p}_1,\bold{p}_2) = C_{jki}^{(121)}(\bold{p}_2,\bold{p}_3,\bold{p}_1)=\\
&= L_i^{(1)}(\bold{p}_1)L_j^{(1)}(\bold{p}_2) P_L(p_1)  P_L(p_2),\\
\end{aligned}
\end{equation}
where by $P_L(p)$ we denote the MG linear power spectrum.

The scalar functions $Q_n(k)$ and $R_n(k)$ that contribute to the SPT power spectrum (and as we shall see in the next section, to the Lagrangian correlators (\ref{eq:correlMG})), are expressed as functions of the polyspectra (\ref{mixedployorder}) in GR \citep{Matsubara:2008wx}. Fortunately, since in MG the above picture is only modified through the modified Kernels in (\ref{LPTKernels}), the relationships that give the various scalar functions are of the same form as the ones presented in \citep{Matsubara:2008wx} (in particular, equations (A50)-(A67) in Appendix A). However, one should be cautious at this point, because certain symmetries that are present in the GR solutions, are not preserved anymore. In particular, the integral in the l.h.s of eq. (A59) in \citep{Matsubara:2008wx} will not be equal to $R_1(k)+R_2(k)$ anymore, because of the scale and redshift dependence of the MG growth factors. In a similar manner, the functions resulting from eq. (A57) and (A61), that used to be equal to $R_1(k)$ and $Q_1(k)$, respectively, in GR, will differ for our MG models and should be additionally calculated. We denote these by $\left[R_1(k)+R_2(k)\right]_{MG}$, $\left[R_1(k)\right]_{MG}$ and $\left[Q_1(k)\right]_{MG}$ to emphasize on their GR limit. In order to illustrate how these calculations are performed, we show the derivations for $\left[R_1(k)+R_2(k)\right]_{MG}$ and $\left[Q_1(k)\right]_{MG}$, that are both new in MG and serve as a representative example of each category. For the former, we have:
\begin{equation}
\begin{aligned}\label{R12plus}
 &\left[R_1(k)+R_2(k)\right]_{MG} = -\frac{7}{3} k_i k_j \int \frac{d^3 p}{8 \pi^3}C_{ji}^{(21)}(-\bold{p};\bold{p}-\bold{k},\bold{k})= \\
 & = \frac{7}{3} \int \frac{d^3 p}{8 \pi^3}k_i k_j L_j^{(1)}(\bold{k})L_j^{(2)}(-\bold{p},\bold{k}) P_L(p)  P_L(k)=\\
  & = P_L(k) \frac{7}{3} \int \frac{dr dx}{4 \pi^2} k^2 k r^2 \frac{k^2 -\bold{k}\cdot\bold{p}}{|k-p|^2} \frac{D^{(2)}(-\bold{p},\bold{k})}{D^{(1)}(p)D^{(1)}(p-k)} P_L(kr)  =\\
  & = \frac{k^3}{4 \pi^2 }P_L(k) \int_0^{\infty} drP_L(kr)\int_{-1}^{1}dx \frac{r^2 \left(1-rx\right)}{1+r^2-2rx} \bar{D}^{(2)}(-\bold{p},\bold{k}), 
\end{aligned}
\end{equation}
where we defined the quantities $x=\hat{\bold{k}}\cdot\hat{\bold{p}}$, $p=kr$ and
\begin{equation}
\begin{aligned}\label{growthdef}
 & \bar{D}^{(2)}(-\bold{p},\bold{k}) = \frac{7}{3}\frac{D^{(2)}(-\bold{p},\bold{k})}{D^{(1)}(p)D^{(1)}(k)} = \\
&=\frac{7}{3}\frac{D^{(2)}\left(k\sqrt{1+r^2-2rx},k,kr\right)}{D^{(1)}(kr)D^{(1)}(k)} = \\
& \bar{D}_a-\bar{D}_b x^2+\bar{D}_{FL}-\bar{D}_{\delta \mathcal{I}}, \\
\end{aligned}
\end{equation}
as was done in \citep{Aviles:2017aor}. Similarly, for $\left[Q_1(k)\right]_{MG}$ we will have:
\begin{equation}
\begin{aligned}\label{Q1mod}
&\left[Q_1(k)\right]_{MG}= \frac{7}{3} (k_i k_j k_l - k^2 k_i \delta_{jl}) \int \frac{d^3 p}{8 \pi^3}C_{ijl}^{(211)}(\bold{k},-\bold{p},\bold{p}-\bold{k})= \\
& = \frac{7}{3} (k_i k_j k_l - k^2 k_i \delta_{jl}) \times \\
& \int \frac{d^3 p}{8 \pi^3}L_l^{(1)}(\bold{p})L_l^{(1)}(\bold{p}-\bold{k})L_i^{(2)}(-\bold{p},\bold{p}-\bold{k}) P_L(p)  P_L(|k-p|)=\\
& = \frac{k^3}{4 \pi^2 } \int dr dx \frac{(\bold{k}\cdot\bold{p})(\bold{k}\cdot\bold{p}-k^2)-k^2\bold{p}(\bold{k}-\bold{p})}{p^2 |k-p|^2} \times \\
& \bar{D}^{(2)}(\bold{p},\bold{k}-\bold{p}) P_L(kr)P_L(|k-p|)  =\\
& \frac{k^3}{4 \pi^2 }\int_0^{\infty} drP_L(kr)\int_{-1}^{1}dx \frac{r^2 (1-x^2)}{(1+r^2-2rx} \bar{D}^{(2)}(\bold{p},\bold{k}-\bold{p})P_L(|k-p|), 
\end{aligned}
\end{equation}
where $|k-p|=k\sqrt{1+r^2-2rx}$ is used and similarly as before, we defined
\begin{equation}
\begin{aligned}\label{growthdef2}
 & \bar{D}^{(2)}(\bold{p},\bold{k}-\bold{p}) = \frac{7}{3}\frac{D^{(2)}(\bold{p},\bold{k}-\bold{p})}{D^{(1)}(p)D^{(1)}(|k-p|)} = \\
&=\frac{7}{3}\frac{D^{(2)}\left(k,kr,k\sqrt{1+r^2-2rx}\right)}{D^{(1)}(p)D^{(1)}(k\sqrt{1+r^2-2rx})} = \\
& \bar{D}_a-\bar{D}_b \frac{x^2+r^2-2rx}{1+r^2-2rx}+\bar{D}_{FL}-\bar{D}_{\delta \mathcal{I}}. \\
\end{aligned}
\end{equation}
The above two Kernels, as well as all of the rest that we need, can be compactly written in the form \citep{Matsubara:2008wx}:
\begin{equation}
\begin{aligned}\label{Qform}
Q_n(k) &= \frac{k^3}{4 \pi^2 }\int_0^{\infty} drP_L(kr) \\
& \times \int_{-1}^{1}dx P_L(k\sqrt{1+r^2-2rx})\tilde{Q}_n(r,x) 
\end{aligned}
\end{equation}
and 
\begin{equation}
\begin{aligned}\label{Rform}
R_n(k) &= \frac{k^3}{4 \pi^2 }P_L(k) \int_0^{\infty} drP_L(kr)\int_{-1}^{1}dx \tilde{R}_n(r,x).
\end{aligned}
\end{equation}
Using similar methods as the one presented above we get that the various $Q_n(k)$, and after some algebra, are given by:
\begin{equation}
\begin{aligned}\label{Qs}
&\tilde{Q}_1 = r^2\left(\bar{D}_a^{(2)}-\bar{D}_b^{(2)}\frac{x^2+r^2-2rx}{1+r^2-2rx}+\bar{D}_{FL}^{(2)}-\bar{D}_{\delta \mathcal{I}}^{(2)}\right)^2 \\
&\tilde{Q}_2 = \frac{rx(1-rx)}{1+r^2-2rx}\left(\bar{D}_a^{(2)} - \bar{D}_b^{(2)}\frac{x^2+r^2-2rx}{1+r^2-2rx}+\bar{D}_{FL}^{(2)}-\bar{D}_{\delta \mathcal{I}}^{(2)}\right) \\
&\tilde{Q}_3 = \frac{x^2(1-rx)^2}{(1+r^2-2rx)^2} \\
&\tilde{Q}_5 = rx\left(\bar{D}_a^{(2)}-\bar{D}_b^{(2)}\frac{x^2+r^2-2rx}{1+r^2-2rx}+\bar{D}_{FL}^{(2)}-\bar{D}_{\delta \mathcal{I}}^{(2)}\right) \\
&\tilde{Q}_7 = \frac{x^2(1-rx)}{(1+r^2-2rx)} \\
&\tilde{Q}_8 = r^2\left(\bar{D}_a^{(2)}-\bar{D}_b^{(2)}\frac{x^2+r^2-2rx}{1+r^2-2rx}+\bar{D}_{FL}^{(2)}-\bar{D}_{\delta \mathcal{I}}^{(2)}\right) \\
&\tilde{Q}_9 = \frac{rx(1-rx)}{1+r^2-2rx} \\
&\tilde{Q}_{11} = x^2 \\
&\tilde{Q}_{12} = rx \\
&\tilde{Q}_{13} = r^2 \\
&\left[\tilde{Q}_1\right]_{MG} = \frac{r^2 (1-x^2)}{1+r^2-2rx}\left(\bar{D}_a^{(2)}-\bar{D}_b^{(2)}\frac{x^2+r^2-2rx}{1+r^2-2rx}+\bar{D}_{FL}^{(2)}-\bar{D}_{\delta \mathcal{I}}^{(2)}\right).\\ 
\end{aligned}
\end{equation}
Similarly, in accordance with equation (\ref{Rform}), we will have the $R_n(k)$ functions:
\begin{equation}
\begin{aligned}\label{Rs}
&\tilde{R}_{1} = \frac{21}{10} r^2 \frac{D^{(3)}_{sym}(\bold{k},-\bold{p},\bold{p})}{D^{(1)}(k)\left(D^{(1)}(p)\right)^2} \\
&\tilde{R}_{2} = \frac{rx(1-rx)}{1+r^2-2rx}\left(\bar{D}_a^{(2)}-\bar{D}_b^{(2)}x^2+\bar{D}_{FL}^{(2)}-\bar{D}_{\delta \mathcal{I}}^{(2)}\right) \\
&\left[\tilde{R_1}(k)+\tilde{R_2}(k)\right]_{MG} = \frac{r^2(1-rx)}{1+r^2-2rx}\left(\bar{D}_a^{(2)}-\bar{D}_b^{(2)}x^2+\bar{D}_{FL}^{(2)}-\bar{D}_{\delta \mathcal{I}}^{(2)}\right)\\
&\left[\tilde{R}_1\right]_{MG} = \frac{r^2(1-x^2)}{1+r^2-2rx}\left(\bar{D}_a^{(2)}-\bar{D}_b^{(2)}x^2+\bar{D}_{FL}^{(2)}-\bar{D}_{\delta \mathcal{I}}^{(2)}\right).\\ 
\end{aligned}
\end{equation}
The functions $Q_1$-$Q_3$, $R_1$ and $R_2$ are the only ones that are necessary to calculate LPT statistics for pure dark matter considerations in MG, with the rest of them that we present here, being the additional functions needed for statistics of biased tracers in MG (in the configuration space). It should be emphasized at this point, that even the functions that have the same integral structure as in GR (for example, $Q_9$-$Q_{13}$), do differ from their GR values, but this difference is manifested in the MG linear power spectra that appear in the integral relations (\ref{Qform}) and (\ref{Rform}). 

Let us finish this section, by noting that in the GR limit, the above functions can be rather easily shown to recover their standard GR forms given in \citep{Matsubara:2008wx}, if one keeps in mind that $\bar{D}_{\delta \mathcal{I}}^{(2)}=\bar{D}_{FL}^{(2)}=0$ and $\bar{D}_a^{(2)}=\bar{D}_b^{(2)}=1$ (for Einstein-De Sitter) in this limit. 

\subsection{Lagrangian correlators and q-functions in MG}\label{correlderiv}

Having derived the expressions for the scalar functions $Q_n(k)$ and $R_n(k)$ in MG cosmologies, we will now derive the integral formulas for the Lagrangian correlators (\ref{eq:correlMG}), that are the building blocks of the 2-point statistics in CLPT. We will adopt the notation of \citep{2013MNRAS.429.1674C} in this section and will show how the functions in their Appendix B will change for our MG models. To illustrate how the connection between the functions (\ref{eq:correlMG}) and the ones presented in the previous section is drawn and also to show how these calculations are performed, we pick a reprsentative example of one these functions, $U_{11}^{(2)}$ and show the derivation below. Starting with the definition:
\begin{equation}
\begin{aligned}\label{Udef}
U_{11}^{(2)}(\bold{q}) =  \hat{q}^i\langle \delta_1^{(1)} \delta_2^{(1)} \Delta_i^{(2)} \rangle_c,
\end{aligned}
\end{equation}
we plug in the Fourier space representation of the field $\Delta_i=\Psi_{i}(\bold{q_2})-\Psi_{i}(\bold{q_1})$, as well of the linear overdensities and get:
\begin{equation}
\begin{aligned}\label{Urel}
& U_{11}^{(2)}(\bold{q}) =   \hat{q}^i \times\\
&\bigg \langle \int\frac{d^3 p}{\left(2 \pi\right)^3} \frac{d^3 p_1}{\left(2 \pi\right)^3} \frac{d^3 p_2}{\left(2 \pi\right)^3} e^{i \bold{p}_2\cdot\bold{q}_2} e^{i \bold{p}_1\cdot\bold{q}_1}\left(e^{i \bold{p}\cdot\bold{q}_2}-e^{i \bold{p}\cdot\bold{q}_1}\right)\tilde{\delta}(\bold{p}_2)\tilde{\delta}(\bold{p}_1)\tilde{\Psi}^{(2)}(\bold{p})\bigg \rangle_c \\
& = \hat{q}^i \int\frac{d^3 p}{\left(2 \pi\right)^3} \frac{d^3 p_1}{\left(2 \pi\right)^3} \frac{d^3 p_2}{\left(2 \pi\right)^3} e^{i \bold{p}_1\cdot\bold{q}_1}e^{i (\bold{p}+\bold{p}_2)\cdot\bold{q}_2}\bigg \langle \tilde{\delta}(\bold{p}_2)\tilde{\delta}(\bold{p}_1)\tilde{\Psi}^{(2)}(\bold{p})\bigg \rangle_c \\
& - \hat{q}^i \int\frac{d^3 p}{\left(2 \pi\right)^3} \frac{d^3 p_1}{\left(2 \pi\right)^3} \frac{d^3 p_2}{\left(2 \pi\right)^3} e^{i \bold{p}_2\cdot\bold{q}_2}e^{i (\bold{p}+\bold{p}_1)\cdot\bold{q}_1}\bigg \langle \tilde{\delta}(\bold{p}_2)\tilde{\delta}(\bold{p}_1)\tilde{\Psi}^{(2)}(\bold{p})\bigg \rangle_c. \\
\end{aligned}
\end{equation}
This expression contains two terms, which turn out to be equal. For this reason, we focus on the first one and notice that the cumulant can be expressed as a polyspectrum, through (\ref{mixedployorder}). The substitution gives:
\begin{equation}
\begin{aligned}\label{Urelpart}
&\hat{q}^i \int\frac{d^3 p}{\left(2 \pi\right)^3} \frac{d^3 p_1}{\left(2 \pi\right)^3} \frac{d^3 p_2}{\left(2 \pi\right)^3} e^{i \bold{p}_1\cdot\bold{q}_1}e^{i (\bold{p}+\bold{p}_2)\cdot\bold{q}_2}\bigg \langle \tilde{\delta}(\bold{p}_2)\tilde{\delta}(\bold{p}_1)\tilde{\Psi}^{(2)}(\bold{p})\bigg \rangle_c = \\
&-i \int\frac{d^3 p_1}{\left(2 \pi\right)^3} \frac{d^3 p}{\left(2 \pi\right)^3} e^{i \bold{p}_1\cdot\bold{q}} \hat{q}^i C_{i}^{(2)}(\bold{p}_1, \bold{p};\bold{p}_1-\bold{p}) = \\
&-i \int\frac{d^3 p_1}{\left(2 \pi\right)^3} \hat{q}^i e^{i \bold{p}_1\cdot\bold{q}} \underbrace{\int \frac{d^3 p}{\left(2 \pi\right)^3} C_{i}^{(2)}(\bold{p}_1, \bold{p};\bold{p}_1-\bold{p})}_\text{$I_{C}$}. \\
\end{aligned}
\end{equation}
In the last line, we defined the integral $I_{C}$, that can be calculated by using the definitions (\ref{mixedployorderspef}):
\begin{equation}
\begin{aligned}\label{Urelpart2}
& I_{C} = \int \frac{d^3 p}{\left(2 \pi\right)^3} C_{i}^{(2)}(\bold{p}_1, \bold{p};\bold{p}_1-\bold{p}) = \\
& -\int \frac{d^3 p}{\left(2 \pi\right)^3} L_i^{(2)}(\bold{p}_1-\bold{p})P_L(p_1)P_L(p) =\\
& = -\frac{3}{7}\frac{p_{1i}}{p_1^2}\int \frac{d^3 p}{\left(2 \pi\right)^3} p_{1i} \frac{(p_{1i}-p_i)}{|\bold{p}_{1}-\bold{p}|^2}\bar{D^{(2)}}P_L(p_1)P_L(p) \\
& = -\frac{3}{7}\frac{p_{1i}}{p_1^2}\underbrace{\frac{p_{1}^3}{4 \pi^2 }P_L(p_{1}) \int_0^{\infty} drP_L(p_{1}r)\int_{-1}^{1}dx \frac{r^2 \left(1-rx\right)}{1+r^2-2rx} \bar{D}^{(2)}}_\text{$\left[R_1(p_1)+R_2(p_1)\right]_{MG}$} \\
& = -\frac{3}{7}\frac{p_{1i}}{p_1^2} \left[R_1(p_1)+R_2(p_1)\right]_{MG},\\
\end{aligned}
\end{equation}
where we made use of the previous result (\ref{R12plus}). Plugging the result (\ref{Urelpart2}) into (\ref{Urelpart}) and relabelling $p_1$ as $k$, we have
\begin{equation}
\begin{aligned}\label{Urelpart3}
&\hat{q}^i \int\frac{d^3 p}{\left(2 \pi\right)^3} \frac{d^3 k}{\left(2 \pi\right)^3} \frac{d^3 p_2}{\left(2 \pi\right)^3} e^{i \bold{k}\cdot\bold{q}_1}e^{i (\bold{p}+\bold{p}_2)\cdot\bold{q}_2}\bigg \langle \tilde{\delta}(\bold{p}_2)\tilde{\delta}(\bold{k})\Psi^{(2)}(\bold{p})\bigg \rangle_c = \\
&i \int\frac{dk dx}{4\pi^2} e^{i x(kq)}x k \frac{3}{7}\left[R_1(k)+R_2(k)\right]_{MG} =\\
&\frac{1}{2\pi^2}\int dk k \left(-\frac{3}{7}\right)\left[R_1(k)+R_2(k)\right]_{MG} j_1(kq), \\
\end{aligned}
\end{equation}
where we made use of the Bessel function identity $\frac{1}{2}\int_{-1}^{1}dxxe^{ixkq}=i j_1(kq)$. In exactly the same way, the second term in (\ref{Urelpart}) is equal to the first, which finally gives:
\begin{equation}
\begin{aligned}\label{Urelfinal}
&U_{11}^{(2)}(q)=\frac{1}{2\pi^2}\int dk k \left(-\frac{6}{7}\right)\left[R_1(k)+R_2(k)\right]_{MG} j_1(kq). \\
\end{aligned}
\end{equation}
This is the MG equivalent of equation (B$28$) in Appendix B of \citep{2013MNRAS.429.1674C}, which is obviously recovered in the GR limit. In a similar manner, but after lengthy calculations, we get the expressions for all the correlators:
\begin{equation}
\begin{aligned}\label{qfuncs}
&V_1^{(112)}(q)=\frac{1}{2\pi^2}\int \frac{dk}{k}  \left(-\frac{3}{7}\right)\left[R_1(k)\right]_{MG} j_1(kq), \\
&V_3^{(112)}(q)=\frac{1}{2\pi^2}\int \frac{dk}{k}  \left(-\frac{3}{7}\right)\left[Q_1(k)\right]_{MG} j_1(kq), \\
&S^{(112)}=\frac{3}{14\pi^2}\int \frac{dk}{k} \left[2\left[R_1\right]_{MG}+4R_2 +\left[Q_1\right]_{MG} +2Q_2\right]\frac{j_2(kq)}{kq}, \\
&T^{(112)}=\frac{-3}{14\pi^2}\int \frac{dk}{k} \left[2\left[R_1\right]_{MG}+4R_2 +\left[Q_1\right]_{MG} +2Q_2\right]j_3(kq), \\
&U^{(1)}(q)=\frac{1}{2\pi^2}\int dk k \left(-1\right)P_L(k) j_1(kq), \\
&U^{(3)}(q)=\frac{1}{2\pi^2}\int dk k \left(-\frac{5}{21}\right)R_1(k) j_1(kq), \\
&U_{20}^{(2)}(q)=\frac{1}{2\pi^2}\int dk k \left(-\frac{6}{7}\right)Q_8(k) j_1(kq), \\
&U_{11}^{(2)}(q)=\frac{1}{2\pi^2}\int dk k \left(-\frac{6}{7}\right)\left[R_1(k)+R_2(k)\right]_{MG} j_1(kq), \\
&X_{10}^{(12)}(q)=\frac{1}{2\pi^2}\int dk \frac{1}{14}\Biggl(2\left(\left[R_1\right]_{MG}-R_2(k)\right) +3\left[R_1\right]_{MG} j_0(kq) \\
& -3\left[ \left[R_1\right]_{MG} + 2R_2+2\left[R_1(k)+R_2(k)\right]_{MG} +2Q_5 \right]\frac{j_1(kq)}{kq} \Biggr), \\
&Y_{10}^{(12)}(q)=\frac{1}{2\pi^2}\int dk \left(-\frac{3}{14}\right)\Biggl( \left[R_1\right]_{MG} + 2R_2  \\
& +2\left[R_1(k)+R_2(k)\right]_{MG} +2Q_5\Biggr)\times \left[j_0(kq)-3\frac{j_1(kq)}{kq}\right] , \\
&X(q)=\frac{1}{2\pi^2}\int dk\ a(k) \left[\frac{2}{3}-2\frac{j_1(kq)}{kq}\right], \\
&Y(q)=\frac{1}{2\pi^2}\int dk\ a(k) \left[-2j_0(kq)+6\frac{j_1(kq)}{kq}\right], \\
\end{aligned}
\end{equation}
where we defined $a(k)=P_L(k)+\frac{9}{98}Q_1(k)+\frac{10}{21}R_1(k)$ and, following \citep{2013MNRAS.429.1674C}, we decomposed the matter terms as 
\begin{equation}
\begin{aligned}\label{decomp}
&A_{ij}^{mn}(q)=X_{mn}(q)\delta_{ij}+Y_{mn}\hat{q}_{i}\hat{q}_{j}\\
&W_{ijk}(q)=V_{1}(q)\hat{q}_{i}\delta_{jk}+V_{2}(q)\hat{q}_{j}\delta_{ki}+V_{3}(q)\hat{q}_{k}\delta_{ij}+T(q)\hat{q}_{i}\hat{q}_{j}\hat{q}_{k}.  \\
\end{aligned}
\end{equation}

Now that the basic framework has been laid out, let us finish this section by briefly summarizing the steps followed to calculate the two point statistics for a given model: after calculating the necessary MG growth factors, (\ref{growth1st}), (\ref{eq:D2sources}) and (\ref{D3MG}), using our $\sc{Mathematica}$ notebook, we feed our modified version of the code in \footnote{\url{https://github.com/martinjameswhite/CLEFT_GSM}} 
with tabulated values of the growth factors for the various values of $k$, $r$ and $x$ needed at a given cosmological redshift $z$, as well as with the MG linear power spectrum given by:
\begin{equation}\label{Plin}
P_{MG}^{L}(k,z) = \left(\frac{D^1_{MG}(k,z)}{D^1_{GR}(k,0)}\right)^2P_{GR}^{L}(k,0).
\end{equation}
The linear power spectrum for the background $\Lambda$CDM cosmology is calculated using the publicly available code CAMB \citep{Lewis:1999bs}.
 The PYTHON module computes the various $Q_n(k)$ and $R_n(k)$ functions through equations (\ref{Qs}) and (\ref{Rs}), which are then used to calculate the various components of the CLPT power spectrum $P_{X}(k)$, through the integrations (\ref{qfuncs}) and (\ref{PkXfinal}). To calculate (\ref{qfuncs}), the $q-$function module is modified accordingly. The k functions can then be simply combined to give the SPT and LRT power spectra, by equations (\ref{PkXSPT}) and (\ref{PkXLRT}), respectively. Finally, the modified C$++$ counterpart follows a similar procedure to compute the configuration space two-point correlation function given by CLPT, through (\ref{xiXfinal}).

\subsection{SPT and LRT Power spectra}\label{spectraderiv}
In the main text, it was stated that the SPT and LRT power spectra, given by (\ref{PkXSPT}) and (\ref{PkXLRT}), correspondingly, are produced when one expands the resummed terms in the exponent of relation (\ref{PkXfinal}). Here we show how this derivation takes place, a process that once again shares a lot of similarities with the corresponding one in GR. When fully expanding the resummed Lagrangian correlator $A^L_{ij}$ in (\ref{PkXfinal}), one gets:
\begin{equation}
\begin{aligned}\label{PkXinterm}
&P_{X}(k) = \int d^3q e^{i\bold{k}\cdot\bold{q}} \\
&   \times \Biggl[ 1 - \frac{1}{2}k_ik_j A_{ij} - \frac{i}{6}k_ik_jk_k W_{ijk} + b_1 \left(2i k_i U_i  - k_ik_jA^{10}_{ij}\right)  \\
&+ b_2\left(i k_i U^{20}_i - k_ik_jU^{(1)}_iU^{(1)}_j\right)+b_1^2\left(\xi_L + ik_iU^{11}_i -k_ik_jU^{(1)}_iU^{(1)}_j\right) \\
& +\frac{1}{2}b_2^2\xi_L^2 + 2b_1b_2\xi_L i k_iU^{(1)}_i\Biggr].
\end{aligned}
\end{equation}
As done previously, we pick one of terms that is modified, that is the term $b_1^2 ik_iU^{11}_i$, and perform the integration as an example. Plugging in $U^{11}_i$ from (\ref{qfuncs}):
\begin{equation}
\begin{aligned}\label{integrSPT}
&  i\ b_1^2\int d^3q e^{i\bold{k}\cdot\bold{q}}\hat{q}^ik_iU^{11} = \\
&  \frac{-i\ b_1^2}{2 \pi^2}\frac{6}{7}\int d^3q dp e^{i kq x}k x p \left[R_1(p)+R_2(p)\right]_{MG} j_1(pq)=\\
&  \frac{-i\ b_1^2}{\pi}\frac{6}{7}\int dq dp k  p q^2 \left[R_1(p)+R_2(p)\right]_{MG} j_1(pq) \int^{1}_{-1} dxe^{i kq x}x=\\
&  \frac{b_1^2}{\pi}\frac{12}{7}\int dq\ dp\ k  p q^2 \left[R_1(p)+R_2(p)\right]_{MG} j_1(pq)j_1(kq) =\\
&  \frac{b_1^2}{\pi}\frac{12}{7}\int dp\ k  p \left[R_1(p)+R_2(p)\right]_{MG} \int_0^{\infty} dq q^2 j_1(pq)j_1(kq) =\\
&  b_1^2\frac{6}{7}\int dp\ \frac{k  p}{k\ p} \delta_D(p-k) \left[R_1(p)+R_2(p)\right]_{MG} =\\
&  b_1^2 \left[R_1(k)+R_2(k)\right]_{MG}, \\
\end{aligned}
\end{equation}
where we also used the identity $ \int_0^{\infty} dq q^2 j_1(pq)j_1(kq)=\frac{\pi}{2pk}\delta_D(p-k)$. Similar computations for the rest of the terms give (\ref{PkXSPT}). In the LRT case, we expand everything but the $q$-independent, ``zero-lag" piece of $A^L_{ij}$, which is equal to $2 \sigma^2_L \delta_{ij}$, with $\sigma_L^2=\frac{1}{6 \pi^2}\int_0^{\infty}dkP_L(k)$. Since this term is scale-independent, it can be pulled out of the q integral and all the other integrations can be performed in the same manner as above, resulting in (\ref{PkXLRT}), that differs from (\ref{PkXSPT}) only in terms of the resummed exponential factor. 

\section{PBS biases in MG}\label{pbsMG}
In this section, we will explain the derivation of the Lagrangian PBS biases (\ref{pbsbiasMG}) in MG models with environmentally dependent gravitational collapse. The PBS argument is commonly employed in conjunction with the halo approach \citep{Mo:1995cs,Mo:1996cn,Sheth:1998xe}, where one states that the conditional halo mass function $\bar{n}_h(M,\Delta)$, modulated by a long-wavelength density perturbation $\Delta$, is modeled by the universal mass function prescription (\ref{PSfunction}), with the collapse threshold shifted as $\delta_{cr}\rightarrow\delta_{cr}-\Delta$. The same result was also derived in \citep{PhysRevD.88.023515}, based on the rigorous definition (\ref{biasrig}), following a separate universe approach: with regards to galaxy clustering, a large-scale density perturbation, $\Delta$, can be viewed as a modification of the mean physical density $\bar{\varrho}_m$ by an offset, that is \citep{Desjacques:2016bnm}
\begin{equation}\label{shift}
\bar{\varrho}'_m = (1+\Delta)\bar{\varrho}_m, 
\end{equation}
where $\bar{\varrho}_m$ should not be confused with the mean comoving density $\bar{\rho}_m$ and in a similar manner, the fractional overdensity at a point $\bold{x}$, $\delta_m(\bold{x})$, is shifted as:
\begin{equation}\label{delshift}
\delta'_m(\bold{x}) = \delta_m(\bold{x})+\Delta.
\end{equation}
This reasoning can be employed to calculate the conditional halo mass function $\bar{n}_h(\Delta)$, by noticing that (\ref{PSfunction}) depends solely on comoving quantities (that will not change), with the only exception of the density threshold $\delta_{cr}$, through the peak significance 
\begin{equation}\label{peakapp}
\nu_c= \frac{\delta_{cr}}{D^{(1)}(z)\sigma(M)}=\frac{\varrho_{cr}-\bar{\varrho}_m}{D^{(1)}(z)\sigma(M)\bar{\varrho}_m},
\end{equation}
that quantifies how rare a fluctuation above the density barrier $\varrho_{cr}=(1+\delta_{cr})\bar{\varrho}_m$ is, given an RMS amplitute at the time of collapse z, $\delta\varrho_{RMS}=D^{(1)}(z)\sigma(M)\bar{\varrho}_m$. Now following Birkhoff's theorem, the critical density for collapse, $\varrho_{cr}$, will be unaffected in the presence a density perturbation $\Delta$, since a collapsing overdensity will not depend on the external spacetime. Combining this fact with equation (\ref{shift}), relation (\ref{peakapp}) gives \citep{PhysRevD.88.023515}
\begin{equation}
\begin{aligned}\label{peakGRsep}
\nu'_{c}(M)=\frac{(1+\delta_{cr})\bar{\varrho}_m-(1+\Delta)\bar{\varrho}_m}{D^{(1)}(z)\sigma(M)\bar{\varrho}_m}=\frac{\delta_{cr}-\Delta}{D^{(1)}(z)\sigma(M)}.
\end{aligned}
\end{equation}
The conditional halo mass function $\bar{n}_h(M,\Delta)$ is now given by the universal prescription (\ref{PSfunction}), but with the peak significance $\nu'_{c}(M)$ in (\ref{peakGRsep}). Combining this fact with the rigorous definition of the bias (\ref{biasrig}), gives the known PBS biases (\ref{biasgeneral}) for GR.

Let us now turn to the MG case. Following the discussion in Section \ref{sec:biasparameters}, it is now clear how the density threshold will not be scale and redshift independent anymore, but will depend on the comoving halo mass $M$, as well as the environmental density $\delta_{env}$, that is, a function $\delta_{cr}(M,\delta_{env})$. As a consequence, the peak significance in MG will now become:
\begin{equation}\label{peakappMG}
\nu_{cMG}= \frac{\delta_{cr}(M,\delta_{env})}{D^{(1)}(z)\sigma(M)}=\frac{\varrho_{cr}(M,\delta_{env})-\bar{\varrho}_m}{D^{(1)}(z)\sigma(M)\bar{\varrho}_m}.
\end{equation}
In the presence of a long-wavelength density perturbation $\Delta$, $\bar{\varrho}_m$ will again change according to (\ref{shift}), but also $\delta_{env}$ will now change as dictated by (\ref{delshift}), and so will $\delta_{cr}$ that depends on it, which will become $\delta'_{cr}(M,\delta'_{env})=\delta_{cr}(M,\delta_{env}+\Delta)$. From the equivalent of (\ref{peakGRsep}), the MG peak significance will now become:
\begin{equation}
\begin{aligned}\label{peakGRsep2}
\nu'_{cMG}(M)=\frac{(1+\delta'_{cr})\bar{\varrho}_m-(1+\Delta)\bar{\varrho}_m}{D^{(1)}(z)\sigma(M)\bar{\varrho}_m}=\frac{\delta_{cr}(M,\delta_{env}+\Delta)-\Delta}{D^{(1)}(z)\sigma(M)}.
\end{aligned}
\end{equation}
As in the GR case, the conditional halo mass function $\bar{n}_h(M,\Delta)$ for MG is now given by the universal prescription (\ref{PSfunction}), but with the peak significance $\nu'_{cMG}(M)$ given by (\ref{peakGRsep2}). Calculating, now, the first and second order derivatives that we need in (\ref{biasrig}), we will have, starting with the linear order:
\begin{equation}
\begin{aligned}\label{deriv1}
& \frac{d\bar{n}_h(M,\Delta)}{d \Delta}\Biggr|_{\substack{\Delta=0}} = \frac{d\nu'_{cMG}}{d \Delta}\Biggr|_{\substack{\Delta=0}} \frac{d\bar{n}_h(M,\Delta)}{d \nu'_{cMG}}\Biggr|_{\substack{\Delta=0}} = \\
&\frac{1}{D^{(1)}(z)\sigma(M)}\frac{d \left[\delta_{cr}(M,\delta_{env}+\Delta)-\Delta\right]}{d \Delta}\Biggr|_{\substack{\Delta=0}} \frac{d\bar{n}_h(M,0)}{d \nu_{cMG}} = \\
&\frac{\left[\frac{d \delta_{cr}(M,\delta_{env})}{d \delta_{env}}-1\right]}{D^{(1)}(z)\sigma(M)}\frac{d\bar{n}_h(M,0)}{d \nu_{cMG}},
\end{aligned}
\end{equation}
where we used the chain rule a few times and also the fact that $\nu'_{cMG}(M)=\nu_{cMG}(M)$ at $\Delta=0$. The definition (\ref{biasrig}), combined with the result (\ref{deriv1}) and the universal prescription (\ref{PSfunction}), gives:
\begin{equation}\label{bias1st}
b_{MG}^1(M, \delta_{env})= \frac{\left[\frac{d \delta_{cr}(M,\delta_{env})}{d \delta_{env}}-1\right]}{D^{(1)}(z)\sigma(M)}\frac{1}{\nu_{cMG}f\left[\nu_{cMG}\right]}\frac{d \left(\nu_{cMG}f\left[\nu_{cMG}\right]\right)}{d\nu_{cMG}}, 
\end{equation}
Similarly, the second order derivative will be:
\begin{equation}
\begin{aligned}\label{deriv2}
& \frac{d^2\bar{n}_h(M,\Delta)}{d \Delta^2}\Biggr|_{\substack{\Delta=0}} = \frac{d}{d\Delta}\frac{d\bar{n}_h(M,\Delta)}{d \Delta}\Biggr|_{\substack{\Delta=0}}\\
&=\frac{d}{d\Delta} \left[\frac{\left[\frac{d \delta_{cr}(M,\delta_{env})}{d \delta_{env}}-1\right]}{D^{(1)}(z)\sigma(M)}\frac{d\bar{n}_h(M,0)}{d \nu_{cMG}}\right],\\
& =\frac{d^2 \delta_{cr}(M,\delta_{env})}{d \delta^2_{env}} \frac{1}{D^{(1)}(z)\sigma(M)}\frac{d\bar{n}_h(M,0)}{d \nu_{cMG}} + \\
& \frac{\left[\frac{d \delta_{cr}(M,\delta_{env})}{d \delta_{env}}-1\right]^2}{\left[D^{(1)}(z)\sigma(M)\right]^2}\frac{d\bar{n}^2_h(M,0)}{d \nu^2_{cMG}},\\
\end{aligned}
\end{equation}
which will give the expression for the second order bias factor:
\begin{equation}\label{bias2st}
\begin{aligned}
&b_{MG}^2(M, \delta_{env})= \frac{\frac{d^2 \delta_{cr}(M,\delta_{env})}{d \delta^2_{env}}}{D^{(1)}(z)\sigma(M)}\frac{1}{\nu_{cMG}f\left[\nu_{cMG}\right]}\frac{d \left(\nu_{cMG}f\left[\nu_{cMG}\right]\right)}{d\nu_{cMG}} + \\
& \frac{\left[\frac{d \delta_{cr}(M,\delta_{env})}{d \delta_{env}}-1\right]^2}{\left[D^{(1)}(z)\sigma(M)\right]^2}\frac{1}{\nu_{cMG}f\left[\nu_{cMG}\right]}\frac{d^2 \left(\nu_{cMG}f\left[\nu_{cMG}\right]\right)}{d\nu^2_{cMG}}.\\
\end{aligned}
\end{equation}
Equations (\ref{bias1st}) and (\ref{bias2st}) give the Lagrangian PBS biases of first and second order in MG, for any universal mass function $f\left[\nu_{cMG}\right]$. Applying these to the particular ST form (\ref{stmult}), we arrive at the relationships (\ref{pbsbiasMG}) and (\ref{pbsbiasMGavg}) in the main text. The derivatives of the form $\frac{d \delta_{cr}(M,\delta_{env})}{\delta_{env}}$ can be easily calculated numerically, as soon as the function $\delta_{cr}(M,\delta_{env})$ is known from integrating equation (\ref{collapseh}).
In MG models that still possess a scale and redshift $\it{independent}$ barrier, like the $n$DGP model (and possibly other models in the Vainshtein family), these derivatives will vanish and we recover the standard GR expressions for the PBS biases (\ref{biasgeneral}), but with a different $\delta_{cr}$.

\newpage
\nocite{*}
\bibliographystyle{apsrev}


\end{document}